%% file: dqpaper-ALLTMP.tex
\documentclass[twocolumn,showpacs,preprintnumbers,amsmath,amssymb,
               floatfix,superscriptaddress]{revtex4}

\usepackage{bm}             % Bold math
\usepackage{dcolumn}        % Align table columns on decimal point
\usepackage{graphicx}       % Include figure files
\usepackage{mathrsfs}
\newcommand{\spar}{{\stackrel{\rightarrow}{\Rightarrow}}}
\newcommand{\sant}{{\stackrel{\rightarrow}{\Leftarrow}}}

\newcommand{\pap}{{\spar(\sant)}}

\newcommand{\MeV}{{\mathrm{MeV}}}
\newcommand{\GeV}{{\mathrm{GeV}}}
\newcommand{\mrad}{{\mathrm{mrad}}}

\newcommand{\Tm}{\mathrm{Tm}}

\newcommand{\m}{\mathrm{m}}
\newcommand{\cm}{\mathrm{cm}}
\newcommand{\mm}{\mathrm{mm}}
\newcommand{\mum}{\mathrm{\mu m}}

\newcommand{\Kelvin}{\mathrm{K}}

\newcommand{\minutes}{\mathrm{min}}
\newcommand{\seconds}{\mathrm{s}}

\newcommand{\Hz}{\mathrm{Hz}}

\newcommand{\labeq}{\stackrel{\mathrm{lab}}{=}}

\begin{document}
\graphicspath{{figs/}}

\preprint{HERMES/DC--42}

\title{Quark helicity distributions in the nucleon for \textit{up},
  \textit{down}, and \textit{strange} quarks from semi--inclusive
  deep--inelastic scattering}

%
%  List of authors for the long delta-q paper
%
%  Derived from the file rec-dqlong.tex by Naomi, updated 16-Jun-2003
%  Naomi's original files are located on
%   worf:/user/makins/transfer/authors/rec-dqlong.tex
%

% List of Institute Addresses 

\def\groupalberta{\affiliation{Department of Physics, University of
    Alberta, Edmonton, Alberta T6G 2J1, Canada}}
\def\groupargonne{\affiliation{Physics Division, Argonne National
    Laboratory, Argonne, Illinois 60439-4843, USA}}
\def\groupbari{\affiliation{Istituto Nazionale di Fisica Nucleare,
    Sezione di Bari, 70124 Bari, Italy}}
\def\groupcolorado{\affiliation{Nuclear Physics Laboratory, University
    of Colorado, Boulder, Colorado 80309-0446, USA}}
\def\groupdesy{\affiliation{DESY, Deutsches Elektronen-Synchrotron,
    22603 Hamburg, Germany}}
\def\groupzeuthen{\affiliation{DESY Zeuthen, 15738 Zeuthen, Germany}}
\def\groupdubna{\affiliation{Joint Institute for Nuclear Research,
    141980 Dubna, Russia}}
\def\grouperlangen{\affiliation{Physikalisches Institut, Universit\"at
    Erlangen-N\"urnberg, 91058 Erlangen, Germany}}
\def\groupferrara{\affiliation{Istituto Nazionale di Fisica Nucleare,
    Sezione di Ferrara and Dipartimento di Fisica, Universit\`a di
    Ferrara, 44100 Ferrara, Italy}}
\def\groupfrascati{\affiliation{Istituto Nazionale di Fisica Nucleare,
    Laboratori Nazionali di Frascati, 00044 Frascati, Italy}}
\def\groupfreiburg{\affiliation{Fakult\"at f\"ur Physik, Universit\"at
    Freiburg, 79104 Freiburg, Germany}}
\def\groupgent{\affiliation{Department of Subatomic and Radiation
    Physics, University of Gent, 9000 Gent, Belgium}}
\def\groupgiessen{\affiliation{Physikalisches Institut, Universit\"at
    Gie{\ss}en, 35392 Gie{\ss}en, Germany}}
\def\groupglasgow{\affiliation{Department of Physics and Astronomy,
    University of Glasgow, Glasgow G12 8QQ, United Kingdom}}
\def\groupillinois{\affiliation{Department of Physics, University of
    Illinois, Urbana, Illinois 61801-3080, USA}}
\def\groupmit{\affiliation{Laboratory for Nuclear Science,
    Massachusetts Institute of Technology, Cambridge, Massachusetts
    02139, USA}}
\def\groupmichigan{\affiliation{Randall Laboratory of Physics,
    University of Michigan, Ann Arbor, Michigan 48109-1120, USA }}
\def\groupmoscow{\affiliation{Lebedev Physical Institute, 117924
    Moscow, Russia}}
\def\groupmunich{\affiliation{Sektion Physik, Universit\"at M\"unchen,
    85748 Garching, Germany}}
\def\groupnikhef{\affiliation{Nationaal Instituut voor Kernfysica en
    Hoge-Energiefysica (NIKHEF), 1009 DB Amsterdam, The Netherlands}}
\def\groupstpetersburg{\affiliation{Petersburg Nuclear Physics
    Institute, St. Petersburg, Gatchina, 188350 Russia}}
\def\groupprotvino{\affiliation{Institute for High Energy Physics,
    Protvino, Moscow region, 142281 Russia}}
\def\groupregensburg{\affiliation{Institut f\"ur Theoretische Physik,
    Universit\"at Regensburg, 93040 Regensburg, Germany}}
\def\grouprome{\affiliation{Istituto Nazionale di Fisica Nucleare,
    Sezione Roma 1, Gruppo Sanit\`a and Physics Laboratory, Istituto
    Superiore di Sanit\`a, 00161 Roma, Italy}}
\def\groupsimonfraser{\affiliation{Department of Physics, Simon Fraser
    University, Burnaby, British Columbia V5A 1S6, Canada}}
\def\grouptriumf{\affiliation{TRIUMF, Vancouver, British Columbia V6T
    2A3, Canada}}
\def\grouptokyo{\affiliation{Department of Physics, Tokyo Institute of
    Technology, Tokyo 152, Japan}}
\def\groupamsterdam{\affiliation{Department of Physics and Astronomy,
    Vrije Universiteit, 1081 HV Amsterdam, The Netherlands}}
\def\groupwarsaw{\affiliation{Andrzej Soltan Institute for Nuclear
    Studies, 00-689 Warsaw, Poland}}
\def\groupyerevan{\affiliation{Yerevan Physics Institute, 375036
    Yerevan, Armenia}}

% Set Institute Order 

\groupalberta
\groupargonne
\groupbari
\groupcolorado
\groupdesy
\groupzeuthen
\groupdubna
\grouperlangen
\groupferrara
\groupfrascati
\groupfreiburg
\groupgent
\groupgiessen
\groupglasgow
\groupillinois
\groupmit
\groupmichigan
\groupmoscow
\groupmunich
\groupnikhef
\groupstpetersburg
\groupprotvino
\groupregensburg
\grouprome
\groupsimonfraser
\grouptriumf
\grouptokyo
\groupamsterdam
\groupwarsaw
\groupyerevan

% List of Authors 

\author{A.~Airapetian}  \groupyerevan
\author{N.~Akopov}  \groupyerevan
\author{Z.~Akopov}  \groupyerevan
\author{M.~Amarian}  \groupzeuthen \groupyerevan
\author{V.V.~Ammosov}  \groupprotvino
\author{A.~Andrus}  \groupillinois
\author{E.C.~Aschenauer}  \groupzeuthen
\author{W.~Augustyniak}  \groupwarsaw
\author{R.~Avakian}  \groupyerevan
\author{A.~Avetissian}  \groupyerevan
% ...

\collaboration{The HERMES Collaboration} \noaffiliation

%
% ================================ End of file ================================
%

%%%\date{\today{} --- Version 3.3} % It is always \today, today,

\begin{abstract}
Polarized deep--inelastic scattering data on longitudinally polarized
hydrogen and deuterium targets have been used to determine double spin
asymmetries of cross sections.  Inclusive and semi--inclusive 
asymmetries for the production of positive and negative pions 
from hydrogen were
obtained in a re--analysis of previously published data.  Inclusive
and semi--inclusive asymmetries for the production of negative and
positive pions and kaons were measured on a polarized deuterium
target.
The separate helicity densities for the up and down quarks and the
anti--up, anti--down, and strange sea quarks were computed from these
asymmetries in a ``leading order'' QCD analysis.  The polarization of the
up--quark is positive and that of the down--quark is negative.  All
extracted sea quark polarizations are consistent with zero, and 
the light quark sea helicity densities
are flavor symmetric within the
experimental uncertainties.  First and second
moments of the extracted quark helicity densities 
in the measured range are consistent with
fits of inclusive data.

%%% Local Variables: 
%%% mode: latex
%%% TeX-master: t
%%% End: 
\end{abstract}

%
% PACS-numbers (the Physics and Astronomy Classification Scheme.)
%    See http://publish.aps.org/eprint/gateway/pacslist for a list
%    of PACS numbers
%
% -> Needs to be checked and/or augmented!
%

%
% Use showkeys class option if keyword display desired
%

%\keywords{Suggested keywords}%

% -------------------------------------------------------------------------- 
%   Here we go...  Make the title
% --------------------------------------------------------------------------
\maketitle

% -------------------------------------------------------------------------- 
%   Include the standardized HERMES list of authors
% --------------------------------------------------------------------------

% ==========================================================================
%   Start the main text body:
%   Include separate files for the individual sections
% ==========================================================================

%% ----------------------------------------------------------------------- %%
%%
%%   File        : sect-I.tex
%%
%%   Description : Source file for section I of the second delta-q paper
%%                 (DC number 42)
%%   
%%   Main Author : Mike Vetterli
%%
%%   Date        : 19-Aug-2002
%%
%%   Remarks     : 
%%
%%   Modified    : 28-May-2003 - Added Ref spin:filippone-ji and paragraph
%%                               on the organization of the paper //MB
%%
%% ----------------------------------------------------------------------- %%

\section{Introduction}
\label{sect:intro}

Understanding the internal structure of the nucleon remains a fundamental
challenge of contemporary hadron physics. From studies of deep-inelastic
lepton-nucleon scattering(DIS), much has been learned
about the quark-gluon structure
of the nucleon, but a clear picture of the origins 
of its spin has yet to emerge. The pioneering experiments 
to explore the spin structure of the nucleon performed at SLAC
\cite{e80:g1p,e130:g1p} were measurements of inclusive spin asymmetries,
in which only the scattered lepton is observed.
Until recently, inclusive measurements have provided most of the current
knowledge of nucleon spin structure. The objective of these studies was to
determine the fraction of the spin of the nucleon which is carried by the 
quarks. The nucleon spin can be decomposed conceptually into the angular
momentum contributions of its constituents according to the equation
\begin{equation}
\langle s^{N}_{z}\rangle = \frac{1}{2}  =\frac{1}{2}{\Delta\Sigma
}+{L_{q}}+{J_{g}},
\label{nuclspin}
\end{equation}
where the three terms give the contributions to the nucleon spin from the
quark spins, the quark orbital angular momentum, and the total angular 
momentum of the gluons, respectively. Early calculations based on 
relativistic quark models \cite{Jaffe:1990jz,th:suzuki-weise} suggested
$\Delta\Sigma\approx 2/3$, while more precise experiments on DIS at 
CERN, performed by the European Muon Collaboration (EMC)
\cite{emc:g1,Ashman:1989ig}, led to the conclusion that 
$\Delta\Sigma \approx 0.1-0.2$.

With these indications of the complexity of the spin structure, it was
quickly realized that a simple leading order (LO)
analysis that did not include
contributions from gluons
was incomplete. More recent next-to-leading order (NLO)
treatments provide a picture more appropriate to our present understanding
of QCD. The focus has been on the polarized structure function 
$g_{1}(x,Q^{2})$ for the proton, given by \cite{ph:altarelli}
\begin{equation}
\begin{split}
g_{1}(x,Q^{2})&=\frac{\langle
e^{2}\rangle}{2}[C_{NS}(x,\alpha_{s}(Q^{2}))\otimes\Delta q_{NS}(x,Q^{2}) \\ 
&+C_{S}(x,\alpha_{s}(Q^{2}))\otimes\Delta\Sigma(x,Q^{2}) \\ 
&+2n_{f}C_{g}(x,\alpha_{s}(Q^{2}))\otimes\Delta g(x,Q^{2})],
\end{split}
\label{g1}
\end{equation}
where $\langle e^{2}\rangle =n_{f}^{-1}\Sigma_{i=1}^{n^{f}}e^{2}_i$,
$e_i$ is the electric charge of the quark
of flavor $q$, the operator $\otimes$ denotes convolution
over $x$, $\Delta q_{NS}$ and $\Delta\Sigma$ are
respectively the nonsinglet and
singlet quark helicity distributions, and $\Delta g$ is the gluon helicity
distribution.
Here $x$ is the usual Bjorken scaling variable, $-Q^2$ is the squared
four--momentum transfer, and $n_f$ is the number of active quark flavors.
The coefficient functions
$C_{NS}$, $C_S$, and $C_g$ have
been computed up to next-to-leading order \cite{ph:mertig,ph:vogelsang} 
in $\alpha_{s}$.
At NLO they as well as their associated parton distributions depend on
the renormalization and factorization schemes. While the physical
observables are scheme independent, parton distributions will be
strongly scheme dependent, but connected from scheme to scheme by well-defined
relationships. In a recent NLO analysis \cite{smc:NLO-g1} of available data
for $g_{1}$, the SMC group presented results for the first moment of $g_{1}$,
which is given by
\begin{equation}
\begin{split}
\int_{0}^{1} dx\: g_{1}(x,Q^{2})&=\frac{\langle e^2\rangle}{2}
[C_{NS}(Q^{2},\alpha_{s}(Q^{2})){\Delta}q_{NS}(Q^{2}) \\
+&C_{S}(Q^{2},\alpha_{s}(Q^{2}))a_0(Q^{2})],
\end{split}
\label{1stmom}
\end{equation}
where the $Q^{2}$ dependent quantities $C_{NS}$, $C_S$, and  
${\Delta}q_{NS}$ are the first moments over $x$.
%%where we have followed the notation of Ref.\cite{ph:altarelli} in denoting
%%moments of coefficient functions and parton densities as
%%$f(N)=\int_{0}^{1}dx\,x^{N-1}f(x)$.
In the Adler--Bardeen scheme used by the SMC 
group the singlet axial charge $a_{0}$ is
\begin{equation}
a_{0}(Q^{2})= \Delta\Sigma -3\frac{\alpha _{s}
    (Q^{2})}{2\pi}\Delta g(Q^{2}),
\label{a0}
\end{equation}
where $\Delta\Sigma $ is the first moment of the 
singlet quark distribution, and
$\Delta g(Q^{2})$ the gluonic first moment.
The SMC group finds that the analysis of the $Q^2$
evolution of the world data base
gives 
$\Delta\Sigma = 0.38^{+0.03}_{-0.03}$
(stat)$^{+0.03}_{-0.02}$(syst)$^{+0.03}_{-0.05}$(th)
and $\Delta g(1\GeV^{2})=0.99^{+1.17}_{-0.31}$
(stat)$^{+0.42}_{-0.22}$(syst)$^{+1.43}_{-0.45}$(th).
The resulting value of
the singlet axial charge is $a_{0}=0.23\pm 0.07$(stat)$\pm 0.19$(syst).
While this result strongly constrains the total quark spin contribution to the
nucleon spin, the limited information it provides
on the flavor structure
of $\Delta\Sigma$ is critically dependent on the assumptions 
of $SU(3)$ flavor symmetry in the interpretation of hyperon
beta--decay which are made to constrain $\Delta q_{NS}$.
%%However, the
%%large uncertainties in $\Delta g$ preclude a precise determination of
%%$\Delta\Sigma$ in the absence of direct measurements of $\Delta g$
%%\cite{ph:adams}.  
A central issue in the analysis of the inclusive
data from these experiments is their sensitivity to $SU(3)$ symmetry
breaking, and the reliability of estimates of the contributions to the
first moments coming from the unmeasured low $x$ region.

With rare exceptions, the experiments listed above have studied
\textit{inclusive} polarized DIS where only the scattered lepton is
detected.  Their sensitivity is limited to the polarization of the
\textit{combination} of quarks and antiquarks  because
the scattering cross section depends on the square of the charge of
the target parton. The key to further progress is more specific probes of the
individual contributions of Eq.~(\ref{nuclspin}) to the proton spin.
Determination of the
polarization of the gluons is clearly of high priority, and 
a more precise measurement
will eliminate a major current ambiguity in the implications
of existing inclusive data. A more
direct determination of the strange quark polarization will avoid
the need for the use of data from hyperon decay and the assumption of
$SU(3)$ flavor symmetry. Measurements which are sensitive to quark flavors will
allow the separation of quark and antiquark polarizations. The HERMES
experiment attempts to achieve these objectives by emphasizing
semi-inclusive DIS, in which a $\pi ,K,$ or $p$ is observed in
coincidence with the scattered lepton. The added dimension of flavor in
the final hadron provides a valuable probe of the  flavor dependence
and other features of parton helicity distributions. With the advanced
state of inclusive measurements and the HERMES data with its added
dimension in the flavor sector, important issues such as measurements of
moments of matrix elements and their accessibility to measurement can be 
revisited. Indeed, the results reported here, which 
address the issue of the flavor dependence of quark helicity densities,
mark the logical next step in unraveling the spin structure of the proton.

This paper begins with a brief development of the formalism required
to describe semi--inclusive DIS. It is followed by a description of 
the HERMES experiment and the analysis procedures for flavor tagging
which produce a comprehensive set of spin asymmetries and a detailed 
flavor decomposition of the quark helicity densities in the nucleon.
The formalism and experiment are described in sections~\ref{sect:dis}
and \ref{sect:exp}. Sections~\ref{sect:data} and \ref{sect:asyms} 
detail the analysis procedures and the resulting cross section 
asymmetries. The extraction of the helicity distributions is explained
in section~\ref{sect:deltaq}, while section~\ref{sect:ds} summarizes
an alternative approach to measuring strange quark distributions.
%%The organization of this paper is as follows: After a brief review of
%%the formalism of polarized DIS in Sect.~\ref{sect:dis}, a detailed
%%description of the HERMES experiment will follow in
%%Sect.~\ref{sect:exp}.  Sect.~\ref{sect:data} covers the relevant
%%details of the data analysis leading to the inclusive and
%%semi--inclusive cross--section asymmetries presented in
%%Sect.~\ref{sect:asyms}.  From these asymmetries, helicity
%%distributions for the various quark and anti--quark flavors are
%%extracted, as explained in Sect.~\ref{sect:deltaq}.  Section
%%\ref{sect:ds} presents a second, alternative approach for the
%%extraction of the particularly interesting helicity distribution of
%%the strange quarks in the nucleon.  
Partial first and second moments
of the extracted helicity distributions and of their singlet and
non--singlet combinations in the measured kinematic range are given in
section~\ref{sect:moments}, where they are also compared to other
existing measurements and to results from global QCD fits.  The
conclusions from these results are discussed in
section~\ref{sect:conclusions}.  The formalism used for the QED
radiative and detector smearing corrections is presented in some
detail in App.~\ref{sect:app:radcorr} and tables with the numerical
results of the present analysis are given in
App.~\ref{sect:app:results}.

%%
%% ======================= End of file 'sect-I.tex' ====================== %%
%%

%%% Local Variables: 
%%% mode: latex
%%% TeX-master: t
%%% End: 
%% ----------------------------------------------------------------------- %%
%%
%%   File        : sect-II.tex
%%
%%   Description : Source file for section II of the second delta-q paper
%%                 (DC number 42)
%%   
%%   Main Author : Mike Vetterli
%%
%%   Date        : 19-Aug-2002
%%
%%   Remarks     : 
%%
%%   Modified    : JW 2003-04-29
%%
%% ----------------------------------------------------------------------- %%

\section{Polarized DIS}
\label{sect:dis}

%%
%% ----------------------------------------------------------------------- %%
%%

\subsection{Polarized Inclusive DIS Formalism}
\label{subsec:formalism}

\begin{figure}[htbp]
  \includegraphics[width=.95\columnwidth]{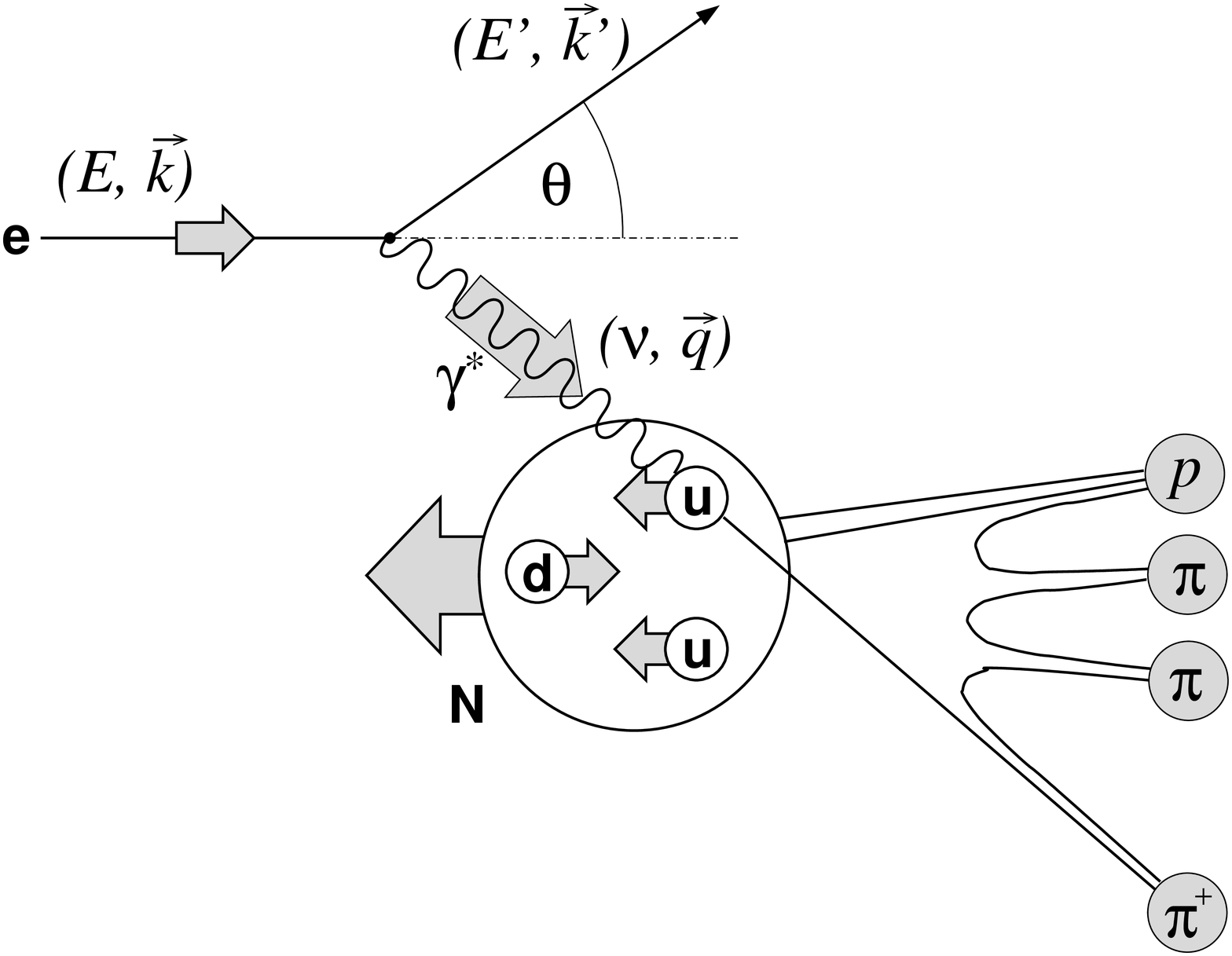}
  \caption{\label{fig:DIS} Diagram of the deep-inelastic scattering process.
    The incoming lepton emits a virtual photon which is absorbed by
    one of the quarks in the nucleon. In the case depicted, the struck
    quark fragments into a pion in the final state.  In
    semi--inclusive processes, the scattered lepton and part of the
    hadronic final state are detected in coincidence.}
\end{figure}

The main process studied here is depicted in
Fig.~\ref{fig:DIS}.  An incoming positron or electron emits 
a spacelike virtual
photon, which is absorbed by a quark in the nucleon.  The nucleon is
broken up, and the struck quark and the target remnant fragment into
hadrons in the final state. Only the lepton is detected in inclusive
measurements while detection of one or more hadrons in the final state
in semi--inclusive measurements adds important information on the
scattering process.  Contributions from Z$^0$ exchange can be
neglected at the energy of the present experiment.

The kinematic variables relevant for this process are listed in
Tab.~\ref{tab:DIS}.  The formalism for DIS is developed in many texts
on particle physics~\cite{halzen,roberts,thomas}.  Here, the formalism
for polarized DIS is briefly summarized in order to introduce
the various measured quantities.

The inclusive DIS cross--section can be written as follows:
\begin{equation}
 \label{eqn:lw}
  \frac{d^2\sigma}{dx\,dQ^2} \propto \, L_{\mu\nu}\,W^{\mu\nu} \, ,
\end{equation}
where $L_{\mu\nu}$ is a tensor that describes the emission of the
virtual photon by the lepton and other radiative processes; 
it can be calculated 
in Quantum Electro Dynamics (QED).  The tensor $W^{\mu\nu}$ describes
the absorption of the virtual photon by the target; it contains all of
the information related to the structure of the target.  Symmetry
considerations and conservation laws determine the form of
$W^{\mu\nu}$ ({\it cf.}~\cite{roberts,thomas}), which for a spin--$1/2$
target and pure electromagnetic interaction reads:

\begin{equation}
  \begin{split}
    \label{eqn:wmunu}
%%% Everything I saw in the literature was different ....
%   W^{\mu\nu} =&  -g^{\mu\nu}\; \mathbf{F_1}
%   + p^\mu p^\nu / \nu \;\mathbf{F_2}\\
%   &+ i/\nu\;\epsilon^{\mu\nu\lambda\sigma}\;q_{\lambda}
%   s_{\sigma}\; \mathbf{g_1} \\
%   &+i/\nu^2\,\epsilon^{\mu\nu\lambda\sigma}\,q_{\lambda} 
%   (p\cdot q\,s_{\sigma}-s\cdot q\,p_{\sigma})\;\mathbf{g_2}.
%%% like so in fact:
    \!W_{\mu\nu}&=
    \left( -g_{\mu\nu} - \frac{q_\mu\,q_\nu}{Q^2}\right) F_1\\
    +&\left( P_\mu + \frac{P\cdot q}{Q^2}\, q_\mu \right) \left( P_\nu
      + \frac{P\cdot q}{Q^2}\, q_\nu \right)
    \frac{F_2}{P\cdot q}\\
    +&\,i\epsilon_{\mu\nu\alpha\beta}\,q^\alpha\frac{M}{P\cdot q}
    \left[ S^\beta \, g_1 +\left( S^\beta - \frac{S\cdot q}{P\cdot q}
        \, P^\beta \right) g_2 \right] \, .
  \end{split}
\end{equation}
\begin{table}[htbp]
  \caption{Kinematic variables in deep inelastic
    scattering}\label{tab:DIS}
  \begin{ruledtabular}
    \begin{tabular}{ll}
      \parbox{0.4\columnwidth}{\raggedright 
        $k=(E,\vec{k})$, $k'=(E',\vec{k'})$}
      &  \parbox{0.55\columnwidth}{\raggedright 
        4--momenta of the initial and final state leptons}\\[3mm]
      \parbox{0.4\columnwidth}{\raggedright 
        $\theta,\; \phi$} &
      \parbox{0.55\columnwidth}{\raggedright 
        Polar and azimuthal angle of the scattered lepton}\\[3mm]
      \parbox{0.4\columnwidth}{\raggedright 
        $P\labeq(M,0)$}
      &  \parbox{0.55\columnwidth}{\raggedright
        4--momentum of the initial target nucleon}\\[3mm]
      \parbox{0.4\columnwidth}{\raggedright 
        $q=k-k'$ }
      &\parbox{0.55\columnwidth}{\raggedright
        4--momentum of the virtual photon }\\[3mm]
      \parbox{0.4\columnwidth}{\raggedright 
        $Q^2\equiv -q^2$\\\hspace*{1em}$\labeq 4EE'\sin^2\frac{\theta}{2}$ }
      &\parbox{0.55\columnwidth}{\raggedright
        Negative squared 4--momentum transfer}\\[3mm]
      \parbox{0.4\columnwidth}{\raggedright 
        $ \nu\equiv \frac{P\cdot q}{M}\labeq E-E'$ }
      &\parbox{0.55\columnwidth}{\raggedright
        Energy of the virtual photon }\\[3mm]
      \parbox{0.4\columnwidth}{\raggedright 
        $ x=\frac{Q^2}{2\,P\cdot q}=\frac{Q^2}{2\,M\nu}$ }
      & \parbox{0.55\columnwidth}{\raggedright
        Bjorken scaling variable }\\[3mm]
      \parbox{0.4\columnwidth}{\raggedright 
        $ y\equiv \frac{P\cdot q}{P\cdot k}\labeq\frac{\nu}{E}$ }
      &\parbox{0.55\columnwidth}{\raggedright
        Fractional energy of the virtual photon }\\[3mm]
      \parbox{0.4\columnwidth}{\raggedright 
        $W^2=(P+q)^2$\\\hspace*{1em} $= M^2+2M\nu-Q^2$} 
      & \parbox{0.55\columnwidth}{\raggedright 
        Squared invariant mass of the photon--nucleon system }\\[3mm]
      \parbox{0.4\columnwidth}{\raggedright 
        $ p=(E_h,\vec{p})$ }
      & \parbox{0.55\columnwidth}{\raggedright 
        4--momentum of a hadron in the final state }\\[3mm]
      \parbox{0.4\columnwidth}{\raggedright 
        $ z=\frac{P\cdot p}{P\cdot q}\labeq \frac{E_h}{\nu}$
      }&\parbox{0.55\columnwidth}{\raggedright 
        Fractional energy of the observed final state hadron}\\[3mm]
      \parbox{0.4\columnwidth}{\raggedright       
        $ x_F=\frac{p_{CM}^\|}{|\vec{q}|}
        \stackrel{\mathrm{lab}}{\simeq}\frac{2\,p_{CM}^\|}{W}$
      }&\parbox{0.55\columnwidth}{\raggedright 
        Longitudinal momentum fraction of the hadron}
    \end{tabular}
  \end{ruledtabular}
\end{table}

In this expression, $F_1$ and $F_2$ are \textit{unpolarized}
structure functions, while $g_1$ and $g_2$ are
\textit{polarized} structure functions that contribute to the
cross section only if both the target and the beam are polarized.  The
usual Minkowski metric is given by $g_{\mu\nu}$, and
$\epsilon_{\mu\nu\alpha\beta}$ is the totally anti--symmetric
tensor.  The four--vector $S$ is the spin of the nucleon, and $q$ and
$P$ are defined in Tab.~\ref{tab:DIS}. In general, the structure
functions depend on $\nu$ and $Q^2$. They can also be defined in terms
of the dimensionless scaling variables $y$, the fractional
energy transfer to the nucleon, and $x$, the Bjorken 
scaling variable. The latter is equal to the fraction of the nucleon's
light-cone momentum carried by the struck quark.  

%%However, if the scattering
%%occurs from point--like free constituents, they depend on only one
%%variable $x\labeq Q^2/2M\nu$.  This phenomenon which is known as
%%scaling was the first experimental confirmation that the nucleon is
%%composed of point like particles.  In the infinite momentum
%%frame, where the nucleon is moving very fast, $x$ corresponds to the
%%fraction of the nucleon's momentum carried by the struck quark.  
The structure functions are given in the quark--parton model (QPM) by:
\begin{align}
  \label{eqn:f1}
  F_1(x)&=\frac{1}{2}\sum_{q}e_q^2\;(q^+(x)+q^-(x))=
  \frac{1}{2}\sum_{q}e_q^2\;q(x) \, , \\
  \label{eqn:g1} 
  g_1(x)&=\frac{1}{2}\sum_{q}e_q^2\;(q^+(x)-q^-(x))=
  \frac{1}{2}\sum_{q}e_q^2\;\Delta q(x) \, ,
\end{align}
where the sum is over quark \emph{and} antiquark flavors, and $e_q$ is
the charge of the quark (or antiquark) in units of the elementary
charge $e$. The functions $q^+(x)$ ($q^-(x)$) are the number densities
of quarks or antiquarks with their spins in the same (opposite)
direction as the spin of the nucleon.  The structure function $F_1(x)$
measures the total quark number density in the nucleon, whereas
$g_1(x)$ is the helicity difference quark number density.  Both
densities are measured as a function of the momentum fraction carried
by the quark.  
                              %
%%\begin{equation}
%%  F_2(x) = x \sum_q e_q^2 \: q(x) \, .
%%\end{equation}  
                              %
The structure functions $F_1(x)$ and $F_2(x)$ are related by the equation 
\begin{equation} \label{eq:FL-contrib}
  2 x\: F_1(x)=\frac{1+\gamma^2}{1+R}\, F_2(x) \, ,
\end{equation}
which reduces to the well--known Callan--Gross relation
\cite{th:callangross} in the Bjorken limit.  
$R(x,Q^{2})$ is the ratio of longitudinal to transverse
DIS cross sections, and $\gamma\equiv \sqrt{Q^2/\nu^2}$. 
%%The factor $C_R \equiv
%%(1+R)/(1+\gamma^2)$ connects tabulations of $q(x)$ with
%%the parton distributions $q1(x)$ required here. In the present analysis the
%%parameterization for $R(x,Q^2)$ given in Ref.~\cite{R1998} was used.
The structure function $g_2(x)$ vanishes in the quark--parton model since
it is related to $Q$ suppressed longitudinal--transverse
interference, which is absent in the
simple QPM.  

In typical
experiments the polarized cross sections are not measured
directly. Rather, their asymmetry
\begin{equation}
  \label{eq:a1}
  A_1 = \frac{\sigma_{1/2} - \sigma_{3/2}}{\sigma_{1/2} + \sigma_{3/2}}
\end{equation}
is measured, where $\sigma_{1/2}$ is the photo--absorption cross section for
photons whose spin is antiparallel to the target nucleon spin,
while $\sigma_{3/2}$ is the corresponding cross section for photons
whose spin is parallel to the target nucleon spin.
Angular momentum conservation requires
that in an infinite momentum frame, the spin-1
photon be absorbed by only quarks whose spin is oriented in the opposite
direction of the photon spin. Consequently, a measurement of 
the difference of these two cross sections is related to the
polarized structure function $g_1$:
\begin{equation}
  g_1 \propto \sigma_{1/2}-\sigma_{3/2} \, .
\end{equation}
The structure function $F_1$ is proportional to the sum of
the cross sections, $\sigma_{1/2}+\sigma_{3/2}$, with the result that
the spin structure function $g_1$ can be deduced from 
$A_1$ by using a parameterization of $F_1$ based on world data.

The picture of the structure functions described to this point is based on
the quark--parton model of point--like constituents
in the nucleon. The model can be extended to a more general picture
that includes quark interactions through gluons in the framework of
quantum chromodynamics (QCD). In this QCD inspired parton model,
scaling is violated and the quark densities become $Q^2$ dependent.
However, in leading order of the strong coupling constant,
Eqs.~\eqref{eqn:f1} and \eqref{eqn:g1} still hold if the replacements
$F_1(x)\longrightarrow F_1(x,Q^2)$ etc.\ are made.  To this order the
structure functions describe the nucleon structure in any hard
interaction involving nucleons; they are universal.

%%
%% ----------------------------------------------------------------------- %%
%%

%%
%%  Measurements in practice
%%

\subsection{Relation to the Inclusive Asymmetries}
\label{subsec:practice}

While the spin orientation of the nucleon and the virtual photon
is the configuration of primary interest, in any experiment only 
the polarizations of the target and the beam
can be controlled and measured directly.  The measured asymmetry
$A_{\mathrm{meas}}$ of count rates in the anti--aligned and aligned
configuration of beam and target polarizations is related to the
asymmetry $A_\|$ of cross sections via
\begin{equation}
  A_{\mathrm{meas}} = p_B\,p_T\,f_D \; A_\| \, ,
\end{equation}
where the kinematic dependencies on $x$ and $Q^2$ have been dropped
for clarity.  The factors $p_B$ and $p_T$ are the beam and
target polarizations, and $f_D$ is the target dilution factor.  This
quantity $f_D$ is the cross section fraction that is due to
polarizable nucleons in the target (1 for H, 0.925 for D, 
and $\sim 1/3$ for $^3$He in gas targets; 
generally smaller for other commonly used polarized
targets).  In this experiment the dilution factor $f_D$ is not further 
reduced by extraneous unpolarized
materials in the target such as windows, etc.

The asymmetry in the lepton--nucleon system, $A_\|$, is related to the
physically significant asymmetry $A_1$ for photo--absorption on the
nucleon level by
\begin{equation}
  \label{eqn:apar1}
  A_\| = D(1 + \eta \gamma)\;A_1 \, ,
\end{equation}
where $\eta\equiv\epsilon\,\gamma\, y/[1-(1-y)\,\epsilon]$ is
a kinematic factor,
and $g_{2}\approx 0$ is assumed.  The factor
$D$ depends on $x$ and $Q^2$, and accounts for the degree of
polarization transfer from the lepton to the virtual photon.  It is
called the depolarization factor and is given by
\begin{equation}
  D=\frac{1-(1-y)\,\epsilon}{1+\epsilon\, R} \, ,
\end{equation}
where $\epsilon$ is the polarization parameter of the virtual photon,
\begin{equation}
  \label{eq:epsilon}
  \epsilon=\left[1+\frac{2\,\vec{q}^{\,2}}{Q^2}\,
    \tan^2\frac{\theta}{2}\right]^{-1}
  = \frac{1-y-\frac{1}{4}\,\gamma^2y^2}{1-y+\frac{1}{4} \, y^2 \,
    (\gamma^2+2)} \, .
\end{equation}

The photon--nucleon asymmetry $A_1$ is related to the structure
function $g_1$ by
\begin{equation}
  \label{eq:A1g1F1}
  A_1 = \frac{g_1}{F_1} \, ,
\end{equation}
when $g_2 = 0$.  This
approximation is justified in view of the small measured values of
$g_2(x)$ \cite{e143:g2_A2,e154:g2n,e155x:g2p_g2d} and the
kinematic suppression of its contributions in the present experiment.  
The residual effect of the
small non--zero value of $g_2(x)$ is included in the systematic
uncertainty on $A_1$ as described in section~\ref{sect:asyms}.
% See Refs.~\cite{hermes:marc,hermes:juergen} for a more detailed
% treatment.

%%
%% ----------------------------------------------------------------------- %%
%%

%%
%%  SIDIS
%%

\subsection{Polarized Semi--Inclusive DIS Formalism}
\label{subsec:si-formalism}

As noted in section~\ref{sect:intro}, inclusive polarized DIS is
sensitive only to the sum of the quark and antiquark distribution
functions because the scattering cross section depends on the squared charge of
the (anti--)quarks. 
The polarizations of the individual flavors and
anti--flavors are accessible in fits to only the inclusive data, where
additional assumptions are used; e.g.~the Bjorken sum rule is imposed
and the quark sea is assumed to be $SU(3)$ symmetric
\cite{pdf:grsv2000}.

The contributions from the various quarks and antiquarks can be
separated more directly if hadrons in the final state are detected in
coincidence with the scattered lepton. Measured fragmentation functions
reveal a statistical correlation between the flavor of the struck quark
and the hadron type formed in the fragmentation process. This reflects
the enhanced probability that the hadron will contain the flavor of the
struck quark.  
For example, the presence of a
$\pi^+$ in the final state indicates that it is likely that a
$u$--quark or a $\bar{d}$--quark was struck in the scattering because
the $\pi^+$ is a $(u\bar{d})$ bound state.  The technique of
detecting hadrons in the final state to isolate contributions to the
nucleon spin by specific quark and antiquark flavors is called
\textit{flavor tagging}. Note that in this case
scattering from a $u$--quark is favored both by its charge ($2e/3$) and
by the fact that the $\bar{d}$--quark is a sea quark and hence has a
reduced probability of existing in the proton in the $x$ range covered
in the analysis presented here ($0.023<x<0.6$).
%%The relevant kinematic properties of the hadron are described by $z$
%%defined as the fraction of the energy transfer carried by the
%%hadron and the Feynman variable $x_F\stackrel{\mathrm{lab}}{\simeq}
%%2\,p^\|_{CM}/W$, where $p^\|_{CM}$ is the longitudinal momentum of the
%%hadron with respect to the virtual photon direction in the
%%photon--nucleon center--of--mass frame.

While the cross section asymmetry $A_1$ is of interest for inclusive
polarized DIS, the relevant quantity for polarized semi--inclusive DIS
(SIDIS) is the asymmetry in the cross sections of produced hadrons in
the final state:
\begin{equation}
   A_1^{h} = \frac{\sigma_{1/2}^h - \sigma_{3/2}^h}
   {\sigma_{1/2}^h + \sigma_{3/2}^h}
\end{equation}
in analogy to Eq.~\eqref{eq:a1}, but where $\sigma^h$ now refers to
the semi--inclusive cross section of produced hadrons of type $h$
instead of the inclusive cross--section.

In analogy to Eq.~\eqref{eqn:lw} , the semi-inclusive DIS cross section
can be written as                                
\begin{equation}
 \label{eqn:lwh}
  \frac{d^5\sigma}{dx\,dQ^2\,dz\,dp^{2}_{T}\,d\phi} 
\propto \, L_{\mu\nu}\,W^{\mu\nu}_{h} \, ,
\end{equation}
where the hadron tensor, $W^{\mu\nu}_{h}$ now contains additional
degrees of freedom corresponding to the fractional energy $z$ of the final
state hadron, the component $p_T$ of the final hadron three momentum transverse
to that of the virtual photon, and the azimuthal angle $\phi$ of the hadron 
production plane relative to the lepton scattering plane. 
Integration over $\phi$ and $p_{T}^{2}$ produces the cross section
relevant to the present experiment. The assumption of factorization permits
the separation of the hadron degrees of freedom from the variables associated
with the lepton vertex. Consequently, kinematic factors depending only on
$x$ and $Q^2$, {\it e.g.} the depolarization factor $D$, 
are carried over directly from inclusive scattering in
relating the semi-inclusive asymmetry $A^{h}_{\|}$ to $A^{h}_{1}$.

In leading order, the resulting cross sections $\sigma_{1/2}^h$ and
$\sigma_{3/2}^h$ can be written in terms of the quark distribution
functions and fragmentation functions $D_q^h(z,Q^2)$:
\begin{equation}
  \label{eq:dsigma}
   \frac{d^3\sigma_{1/2(3/2)}^h}{dx\,dQ^2\,dz}
   \propto \sum_q e_q^2 \, q^{+(-)}(x,Q^2) \, D_q^h(z,Q^2)\, ,
\end{equation}
where the dependence on the kinematics is made explicit. The
fragmentation function $D_q^h$ is a measure of the probability that a
quark of flavor $q$ will fragment into a hadron of type $h$. 
%% The
%% quantity $z$ is the fraction of the energy transfer carried by the
%% hadron, as defined in Tab.~\ref{tab:DIS}.

A procedure identical to that described in the previous subsection
relates the measured quantity $A_{\mathrm{meas}}^{h}$ to the
photon--nucleon asymmetry $A_1^{h}$.  The latter can be expressed in
terms of the quark helicity densities $\Delta q$ and the fragmentation
functions:
\begin{equation}
  \label{eq:A-QPM}
  A_1^{h}(x,Q^2\!,z) =
  \frac{\sum_q e_q^2 \, \Delta q(x,Q^2)\, D_q^h(z,Q^2)} 
  {\sum_{q'} e_{q'}^2 \, q'(x,Q^2)\, D_{q'}^h(z,Q^2)} \, .
\end{equation}
This equation can be rewritten as follows:
\begin{equation}
  \label{eq:a1si}
  A_1^{h}(x,Q^2\!,z)= 
  \sum_q {\cal P}_q^h(x,Q^2\!,z) \, \frac{\Delta q(x,Q^2)}{q(x,Q^2)} \, ,
\end{equation}
where the quark polarizations ($\Delta q / q$) are factored out and
\textit{purities} $\mathcal{P}_q^h$ are introduced.  The purity is 
the conditional probability that a hadron of type $h$ observed in the
final state originated from a struck quark of flavor $q$
in the case that the beam/target is unpolarized. It is
related to the fragmentation functions by:
\begin{equation}
  \label{eq:puridef}
  {\cal P}_q^h(x,Q^2\!,z) = \frac{e_q^2 \, q(x,Q^2) \, D_q^h(z,Q^2)}
  {\sum_{q'} \, e_{q'}^2 \, q'(x,Q^2) \, D_{q'}^h(z,Q^2)} \, .
\end{equation}
This concept of a purity is generalized to inclusive scattering by
setting the fragmentation functions to unity in
Eq.~\eqref{eq:puridef}.  This allows the inclusion of the inclusive
data in the same formalism.

The determination of the quark polarizations using flavor tagging
based on Eq.~\eqref{eq:a1si} trades the assumptions used in global
fits to inclusive data for the modeling of the fragmentation process.
The purity formalism based on Eq.~\eqref{eq:dsigma} additionally implies
the factorization of the hard scattering reaction and the
fragmentation process. In the analysis presented here the
purities were calculated from a Monte Carlo simulation of the entire
scattering process. 
%%However, in the present analysis, factorization
%%was not assumed, as the purities were calculated from a Monte Carlo
%%simulation of the entire scattering process. 
The determination of the
purities and the extraction of the quark polarizations on the basis of
Eq.~\eqref{eq:a1si} are explained in detail in
section~\ref{sect:deltaq}.

% In order to account for the contribution from the longitudinal
% structure function $F_L$ to the unpolarized parton densities,
% Eqs.~\eqref{eq:a1si} and \eqref{eq:A-QPM} are augmented by an
% additional factor (cf.~Eq.~\eqref{eq:FL-contrib}).

%%
%% ====================== End of file 'sect-II.tex' ====================== %%
%%
%% ----------------------------------------------------------------------- %%
%%
%%   File        : sect-III.tex
%%
%%   Description : Source file for section III of the second delta-q paper
%%                 (DC number 42)
%%   
%%   Main Author : Mike Vetterli
%%
%%   Date        : 19-Aug-2002
%%
%%   Remarks     : 
%%
%%   Modified    : May 1, 2003; jw
%%
%% ----------------------------------------------------------------------- %%

\section{Experiment}
\label{sect:exp}

The HERMES experiment is located in the East Hall of the HERA facility
at DESY (see Fig.~\ref{fig:hera}).  Although HERA accelerates both
electrons (or positrons) and protons, only the lepton beam is used by
HERMES in a fixed--target configuration.  The proton beam passes
through the mid--plane of the experiment.  The target is a gas cell
internal to the lepton ring.  There are three major components to the
HERMES experiment: the polarized beam, the polarized target, and the
spectrometer.  All three are described in detail elsewhere.  As this
paper reports on data collected from the years 1996 until 2000, the
following describes the experimental status during this time.

%%
%%  The Polarized Beam
%%
\subsection{Polarized Beam}
\label{sect:beam}

Detailed descriptions of the polarized beam, the beam polarimeters,
and the spin rotators are given in Refs.~\cite{hera:spinrot,
  hermes:tpol1, hermes:tpol2, hermes:lpol}.
\begin{figure}[htb]
  \centering
  \includegraphics[width=\columnwidth]{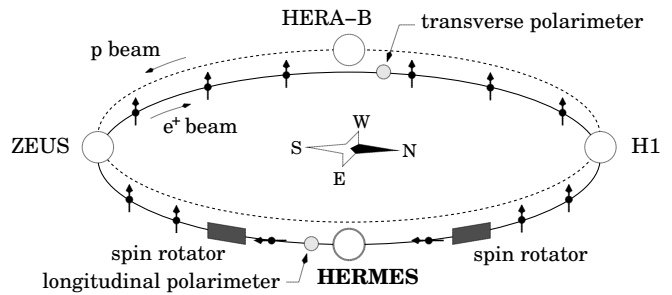}
  \caption{\label{fig:hera}Schematic diagram of the HERA accelerator
    layout until 2000 with the location of the four experiments. Also
    shown are the locations of the spin--rotators and the two
    polarimeters.}
\end{figure}
The electron/positron beam at HERA is self--polarized by the
Sokolov--Ternov mechanism \cite{beampol:sokolov}, which exploits a
slight asymmetry in the emission of synchrotron radiation, 
depending on whether the spin of the
electron/positron in the spin flip associated with the emission 
is parallel or anti--parallel to the magnetic guide
field.  This very small asymmetry (one part in $10^{10}$
\cite{hermes:dueren}), causes the polarization of the beam to grow
asymptotically, with a time constant that depends on the final
polarization of the beam.  This time constant is used in dedicated runs
to verify the calibration of the polarimeters that measure the
degree of polarization of the beam.

Polarizing the lepton beam at HERA is therefore a matter of minimizing
depolarizing effects rather than one of producing an {\it a priori}
polarized beam and keeping it polarized.  An unpolarized beam is
injected into the storage ring and polarization builds up over time,
typically in 30--40 minutes.

The Sokolov--Ternov mechanism polarizes the beam in the transverse
direction, {\it i.e.}~the beam spin orientation is perpendicular to the
momentum.  The beam spin orientation is rotated into the longitudinal
direction just upstream of HERMES, and is rotated back into the
transverse direction downstream of the spectrometer. The locations of
the spin rotators are indicated in Fig.~\ref{fig:hera}.

The beam polarization is measured continuously by two instruments,
both based on asymmetries in the Compton backscattering of polarized
laser light from the lepton beam. 
The transverse polarimeter \cite{hermes:tpol1,hermes:tpol2}, measures
the polarization of the lepton beam 
at a point where it is polarized in the
transverse direction.  The interaction point (IP) of the polarized
light with the lepton beam is located about $120\,\m$ downstream of
the HERA West Hall (see Fig.~\ref{fig:hera}).  The polarimeter uses a
spatial (up--down) asymmetry in the back--scattering of laser light
from the polarized lepton beam.  Back--scattered photons are measured
in a split lead--scintillator sampling calorimeter, where the change
in the position of the photons with initial circular polarization
determines the polarization of the lepton beam. The calorimeter is
located $70\,\m$ downstream of the IP.

A second polarimeter \cite{hermes:lpol} some $90\,\mathrm{m}$
downstream of the HERMES target (and just before the spin is rotated
back to the transverse direction)
measures the polarization of the beam when it is in the longitudinal
orientation.  This polarimeter is also based on Compton
back--scattering of laser light, but in this case the asymmetry is in
the total cross--section, and not in the spatial distribution.  The
larger asymmetry in this case allows a more precise measurement of the
beam polarization.  The higher precision is reflected in the smaller
systematic uncertainties of the polarization measurements in the years
1999 and 2000.  Additionally, this second polarimeter provides the
possibility to measure the polarization of each individual positron
bunch in HERA.  This feature is particularly useful for the
optimization of the beam polarization when the HERA lepton beam is in
collision with the HERA proton beam.  The existence of two
polarimeters also allows a cross--check of the polarization
measurement to be made.

The beam polarization was typically greater than $50\,\%$ in the later
years of the experiment, attaining values near $60\,\%$ for many fills
of the storage ring.  Average beam polarizations, the precision of the
polarization measurement, as well as the charge of the HERA lepton
beam for each year are given in Tab.~\ref{tab:beampol}.

\begin{table}[ht]
  \caption{\label{tab:beampol}Beam polarization and HERA lepton beam
    charge for each year of HERMES running covered in this paper.  The
    numbers are weighted by the luminosity so that the value at the
    beginning of the fill dominates.  Polarization values were larger
    at the end of the fill. The fractional uncertainties quoted are
    the ones used in the data analysis.}
  \renewcommand{\extrarowheight}{1pt}
  \renewcommand{\tabcolsep}{6pt}
  \begin{ruledtabular}
    \begin{tabular}{l|ccc}
           & Lepton      & Average       & Fractional   \\
      Year & beam charge & Polarization  & Uncertainty  \\[.5ex] \hline
      1996 \rule{0mm}{3ex} & $e^+$       & $52.8 \, \%$ & $ 3.4\, \% $ \\
      1997 & $e^+$       & $53.1 \, \%$ & $ 3.4\, \% $ \\
      1998 & $e^-$       & $52.1 \, \%$ & $ 3.4\, \% $ \\
      1999 & $e^+$       & $53.3 \, \%$ & $ 1.8\, \% $ \\
      2000 & $e^+$       & $53.3 \, \%$ & $ 1.9\, \% $    
    \end{tabular}
  \end{ruledtabular}
  \renewcommand{\extrarowheight}{0pt} % default
  \renewcommand{\tabcolsep}{2pt}      % default (at least in revtex)
\end{table}

%%
%%  The Target
%%
\subsection{Polarized Target}
\label{sec:target}

\begin{figure*}[htb]
  \centering
  \includegraphics[clip,width=0.9\textwidth]{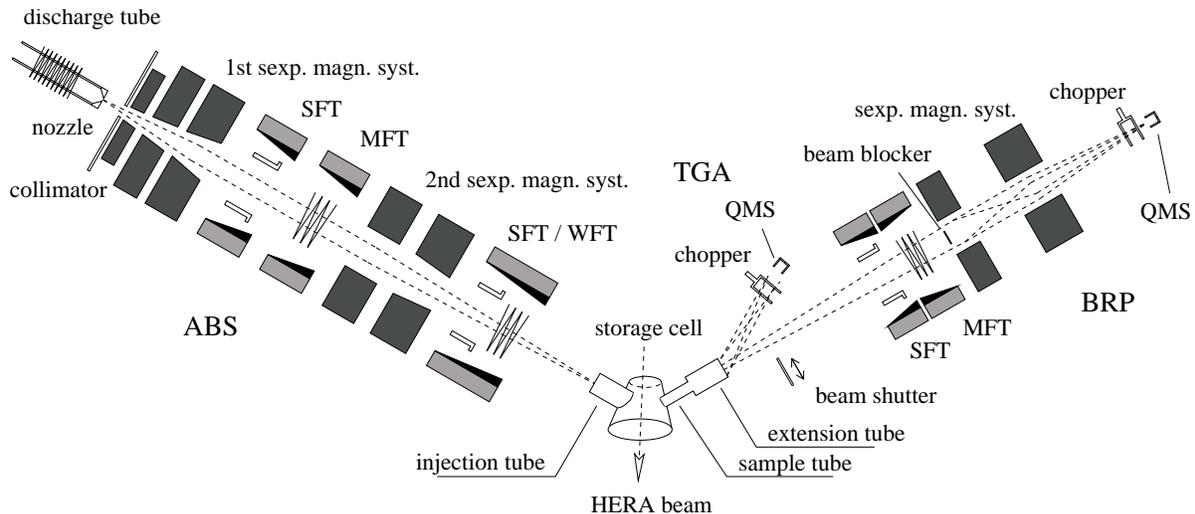} 
  \caption{\label{fig:poltarget}Diagram of the HERMES polarized target.
    Shown are the atomic beam source (ABS), the target gas analyzer (TGA)
    and the Breit--Rabi polarimeter (BRP). SFT, MFT, and WFT label the
    strong, medium, and weak field transitions in the ABS and the
    BRP.}
\end{figure*}

HERMES has used two types of polarized targets over the years.  In
1995 an optically pumped polarized $^3$He target was installed
\cite{hermes:targ3he}.  Since these data are not used in the present
analysis, no description of this target is given here.
In 1996--97, polarized hydrogen was used, while in 1998--2000 the
target was polarized deuterium.  In both cases, the source of
polarized atoms was an atomic beam source (ABS).  The ABS and the
Breit--Rabi polarimeter (BRP) used to monitor the degree of
polarization are described in \cite{hermes:ABS, hermes:BRP,
  hermes:BRP_TOF}. A schematic diagram of the polarized target is
shown in Fig.~\ref{fig:poltarget}.
Briefly, the atomic beam source is based on the Stern--Gerlach effect.
Neutral atomic hydrogen or deuterium is produced in a dissociator and
is formed into a beam using a cooled nozzle, collimators and a series
of differential pumping stations.  A succession of magnetic sextupoles
and radio--frequency (RF) fields are used to select one (or two)
particular atomic hyperfine states that have a given nuclear
polarization.

The ABS feeds a storage cell \cite{hermes:storagecell} which serves to
increase the density by two orders of magnitude.  This storage cell is
located in the HERA ring vacuum and it is cooled to a temperature
between $70\,\Kelvin$ (deuterium) and $100\,\Kelvin$ (hydrogen).  The
cell is $40\,\cm$ long and had elliptical cross--sectional dimensions
of $29 \times 9.8\,\mm^2$ in 1996--1999 and $21 \times 8.9\,\mm^2$ in
2000.

The polarization and the atomic fraction of the target were monitored
by sampling the gas in the target cell using the target gas analyzer
(TGA) and the BRP.  The TGA is a quadrupole mass spectrometer
(QMS) which measures the relative fluxes of atomic and molecular hydrogen
or deuterium and thereby determines the molecular fraction of the
target gas.  The BRP works essentially in reverse to the ABS.  Single
atomic hyperfine states are isolated using magnetic and RF fields and
the atoms in each hyperfine state are counted using again a QMS.  The
electromagnetic fields are varied in a sequence such that atoms in
each hyperfine state are counted in succession.  More details are
given in \cite{hermes:BRP,hermes:BRP_TOF,hermes:simani,hermes:TGA}.
Tab.~\ref{tab:tar} lists the target type, average polarization, and
uncertainty for each data set. The differences in the systematic
uncertainties are largely due to varying running conditions and the
quality of the target cell in use.
\begin{table}[ht]
  \caption{\label{tab:tar}Target type and polarization for each year
    of HERMES running. These values are weighted by the luminosity.
    The uncertainty quoted is the one used in the data analysis.}
  \renewcommand{\extrarowheight}{1pt}
  \renewcommand{\tabcolsep}{6pt}
  \begin{ruledtabular}
    \begin{tabular}{l|ccc}
           &      & Average      & Fractional \\
      Year & Type & Polarization & Uncertainty \\[.5ex] \hline
      1996 \rule{0ex}{3ex} &   H & $  75.9 \,\% $&    $ 5.5 \,\% $    \\
      1997 &   H & $  85.0 \,\% $&    $ 3.8 \,\% $    \\
      1998 &   D & $  85.6 \,\% $&    $ 7.5 \,\% $    \\
      1999 &   D & $  83.2 \,\% $&    $ 7.0 \,\% $    \\
      2000 &   D & $  +85.1,-84.0 \,\% $&    $ 3.5 \,\% $    
    \end{tabular}
  \end{ruledtabular}
  \renewcommand{\extrarowheight}{0pt} % default
  \renewcommand{\tabcolsep}{2pt}      % default (at least in revtex)
\end{table}

%%
%%  The Spectrometer
%%
\subsection{The HERMES Spectrometer}
\label{sec:spectr}

\begin{figure*}[th]
  \centering \includegraphics[width=\textwidth]{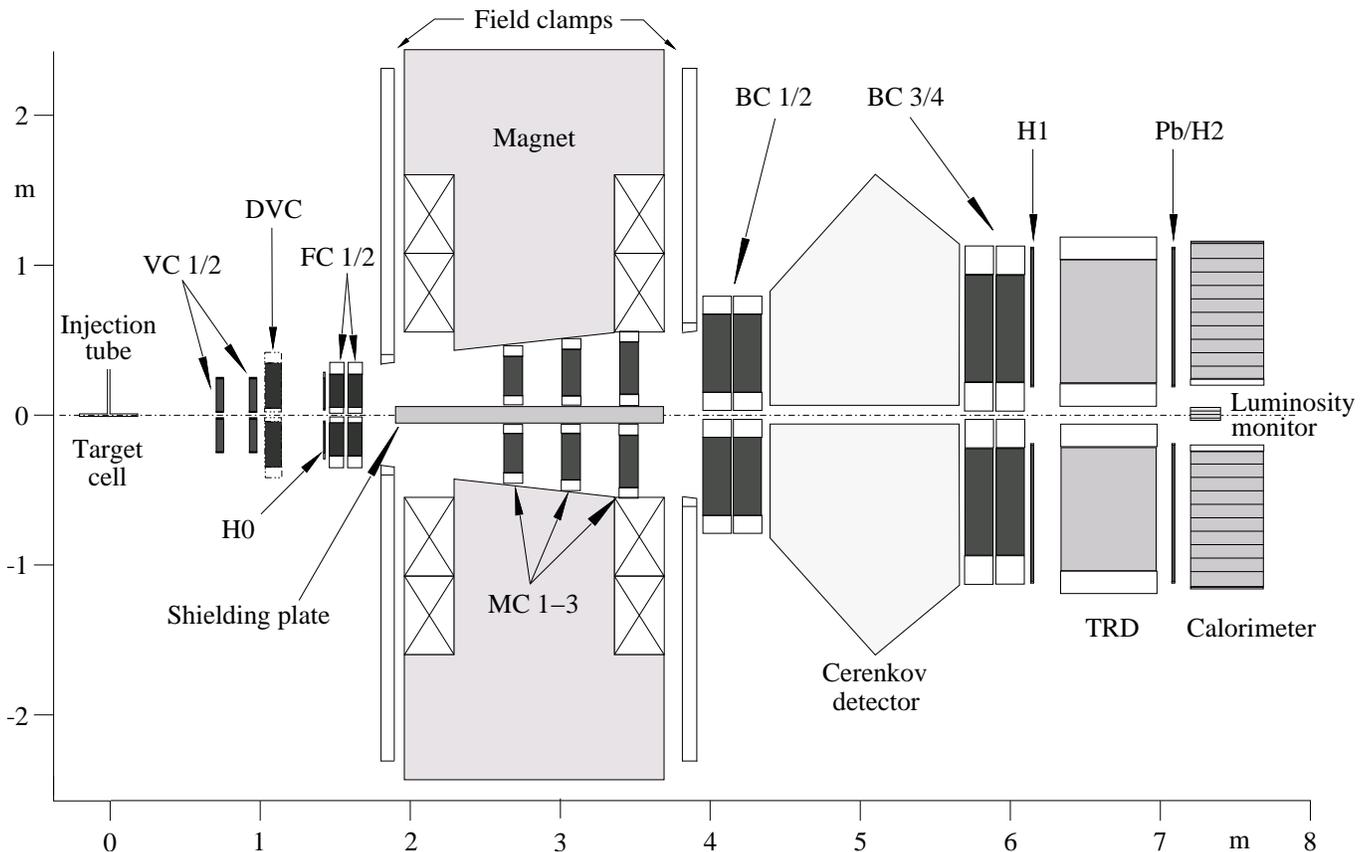}
  \caption{\label{fig:spect}Side view of the HERMES spectrometer.
    The positron beam enters from the left.  The spectrometer is split
    into two halves, one above the beam and one below, by a flux
    exclusion plate to protect the beams from the magnetic field.  
    See the text for further details on the detectors.}
\end{figure*}

The HERMES spectrometer is described in detail in
\cite{hermes:spectr}.  It is a forward spectrometer with large
acceptance that can detect the scattered electron/positron as well as
hadrons in coincidence.  This allows semi--inclusive measurements of
the polarized DIS process, which are the focus of this paper.  A
diagram of the spectrometer is shown in Fig.~\ref{fig:spect}.

Briefly, the HERMES spectrometer consists of multiple tracking stages
before and after a $1.3\,\Tm$ dipole magnet, as well as extensive
particle identification.  The geometrical acceptance of the
spectrometer is $\pm170\,\mrad$ in the horizontal direction and
between $\pm40\,\mrad$ and $\pm140\,\mrad$ in the vertical.  The range
of scattering angles is therefore $40\,\mrad$ to $220\,\mrad$.  The
spectrometer is split into two halves (top/bottom) due to the need for
a flux shielding plate in the mid--plane of the magnet to eliminate
deflection of the primary lepton and proton beams, which pass through the
spectrometer. The lepton beam passes along the central axis of the 
spectrometer. The proton beam traverses the spectrometer 
parallel to the lepton beam but displaced horizontally by 71.4 cm.

\subsubsection{Particle Tracking}
\label{sec:tracking}

The tracking system serves several functions:
\begin{itemize}
\item Determine the event vertex to ensure the event came from the
  target gas, not from the walls of the target cell or from the
  collimators upstream of the target.
\item Measure the scattering angles of all particles.
\item Measure the particle momentum from the deflection of the track
  in the spectrometer magnet.
\item Identify hits in the PID detectors associated with each track.
\end{itemize}

The tracking system consists of 51 planes of wire chambers and six
planes of microstrip gas detectors.  Because of the width of the
tracking detectors in the rear section of the spectrometer, it was not
possible to use horizontal wires in these chambers.  Instead, wires
tilted $ \pm30^\circ$ from the vertical were used ($U$ and $V$
planes), together with vertical wires ($X$ planes).  All chambers have
this geometry to simplify the tracking algorithm (see below).

The majority of the tracking detectors are horizontal drift chambers
with alternating anode and cathode wires between two cathode foils.
The chambers are assembled in modules of six layers in three
coordinate doublets ($XX'$, $UU'$, and $VV'$).  The primed planes are
offset by a half--cell to resolve left--right ambiguities.

\noindent
In order starting at the target, the tracking chambers are:
\paragraph{\underline{Vertex Chambers (VC1/2):}}
The main purpose of the vertex chambers \cite{hermes:VC1,hermes:VC2}
is to measure the scattering angle to high precision and determine the
vertex position of the interaction.  Because of severe geometrical
constraints and the high flux of particles in the region so close to
the target, microstrip gas chambers (MSGC) were chosen for the VCs.
Each of the upper and lower VC detectors consist of six planes grouped
into two modules (VUX and XUV for VC1 and VC2 respectively).  The
pitch of the strips is $0.193\,\mm$.

\paragraph{\underline{Drift Vertex Chambers (DVC):}}
The drift vertex chambers have a cell size of $6\,\mm$ in each module
in the $XX'UU'VV'$ geometry.

\paragraph{\underline{Front Chambers (FC1/2):}}
The front chambers \cite{hermes:FCs} are drift chambers mounted on the
front face of the spectrometer magnet.  The cell size is $7\,\mm$ in
the $XX'UU'VV'$ geometry.

\paragraph{\underline{Magnet Chambers (MC1--3):}}
The magnet chambers \cite{hermes:MCs} are located in the magnet gap.
The MCs are proportional wire chambers with a cell width of $2\,\mm$.
Each module consists of three planes in the $UXV$ geometry.

\paragraph{\underline{Back Chambers (BC1--4):}}
The back chambers \cite{hermes:BCs} are large drift chambers located
behind the spectrometer magnet.  The cell width is $15\,\mm$ in the
$UU'XX'VV'$ geometry.

A tracking algorithm \cite{hermes:wander} defines tracks in front of
and behind the magnet and the momentum of the scattered particles can
therefore be determined.  
The MSGCs contained in the VCs were not available after 1998 due to 
radiation damage. In their place, vertex determination was accomplished
by a refined tracking algorithm that used data from the FCs 
together with the point defined by the the intersection of the track 
in the rear of
the spectrometer (the back--track) with the mid--plane of the magnet as
an additional tracking parameter.
%% used the point defined by the intersection of the track in the rear of
%% the spectrometer (the back--track) with the mid plane of the magnet as
%% an additional point to refine the front tracking.  
%% As the MSGCs contained in the VCs were not
%% available after 1998 due to radiation damage, the tracking algorithm
%% used the point defined by the intersection of the track in the rear of
%% the spectrometer (the back--track) with the mid plane of the magnet as
%% an additional point to refine the front tracking.  
The tracking
algorithm is described in more detail in section~\ref{sect:data}.

Note that the magnet chambers are used only to track particles that do
not reach the back of the spectrometer.  They are useful for the
measurement of partial tracks (mostly low--energy pions) that can,
under certain conditions, increase the acceptance for the
reconstruction of short--lived particles, such as $\Lambda$ particles.
However, these chambers as well as the vertex and the drift
vertex chambers are not used in the analysis reported in this paper.

Multiple scattering, and bremsstrahlung in the case of electrons or
positrons, in the windows and other detector and target cell material
which the particle tracks traverse
limit the momentum resolution of the spectrometer. After its installation
in 1998, the RICH detector because of its aerogel radiator assembly and
heavy gas radiator increased this limit significantly.
Plots of the momentum and angular resolution are shown in
section~\ref{sect:data}.

\subsubsection{Particle Identification}
\label{sec:pid}

There are several particle identification (PID) detectors in the
HERMES spectrometer. 
Electrons and positrons are identified by the combination of a
lead--glass calorimeter, a scintillator hodoscope preceded by two
radiation lengths of lead (the pre--shower detector), 
and a transition--radiation detector (TRD).  
A \v{C}erenkov detector was used primarily
for pion identification.  The threshold detector was replaced by a
Ring--Imaging \v{C}erenkov (RICH) detector in 1998.  The RICH allowed
pions, kaons, and protons to be separated. Both \v{C}erenkov detectors
also helped in lepton identification.

\paragraph{\underline {The Calorimeter:}}
The calorimeter \cite{hermes:Calo} has the following functions:
suppress hadrons by a factor of 10 in the trigger and 100 offline;
measure the energy of electrons/positrons and also of photons from
other sources, e.g. $\pi^0$ and $\eta$ decays.  It consists of two
halves each containing 420 blocks (42 $\times$ 10) of radiation
resistant F101 lead--glass.  The blocks are $9 \times 9$~cm$^2$ by
50~cm deep (about 18 radiation lengths).  Each block is viewed from
the back by a photomultiplier tube.

The response of the calorimeter blocks was studied in a test beam with
a $3 \times 3$ array.  The response to electrons was found to be
linear within $1\,\%$ over the energy range $1$--$30\,\GeV$.  The
energy resolution was measured to be $\sigma(E)/E\, [\%] = (5.1 \pm
1.1)/\sqrt{E [\GeV]} + (1.5 \pm 0.5)$

\paragraph{\underline{The Pre--Shower Detector:}}
The calorimeter is preceded by a scintillator hodoscope (H2) that has
two radiation lengths of lead in front of it.  The hodoscope H2
therefore acts as a pre--shower detector and contributes to the lepton
identification.  This detector consists of 42 vertical scintillator
modules in each of two halves.  Each paddle is $1\,\cm$ thick and
$9.3\times 91\,\cm^2$ in area.

The lead preceding the hodoscope initiates showers for leptons but
with a much reduced probability for hadrons.  Pions deposit only about
$2\,\MeV$ of energy on average while electrons/positrons deposit
roughly $20$--$40\,\MeV$.  H2 suppresses hadrons by a factor of about
10 with $95\,\%$ efficiency for detection of electrons/positrons.

\paragraph{\underline{Transition Radiation Detector:}}
The transition radiation detector (TRD) rejects hadrons by a factor
exceeding 300 at $90\,\%$ electron/positron detection efficiency.
Each of the upper and lower halves of the spectrometer contains six
TRD modules with an active area of $325\times 75\,\cm^2$.  Each module
consists of a radiator and a proportional wire chamber to detect the
TR photons.  The radiators consist of a pseudo--random but
predominantly two--dimensional array of polyethylene fibers with
$17$--$20\,\mum$ diameter.  The proportional chambers have a wire
spacing of $1.27\,\cm$, use Xe:CH$_4$ (90:10) gas, and are $2.54\,\cm$
thick.

\paragraph{\underline{\v{C}erenkov Detector:}}
In 1995--97, a threshold \v{C}eren\-kov detector was operated, which
was located between the two sets of back tracking chambers.  During
the 1996 and 1997 data taking periods, a mixture of $70\,\%$ nitrogen
and $30\,\%$ C$_4$F$_{10}$ was used as the radiator, resulting in
momentum thresholds for pions, kaons, and protons of 3.8, 13.6, and
$25.8\,\GeV$ respectively.

As for the other components of the spectrometer, the \v{C}eren\-kov
detector consists of two identical units in the upper and lower half
of the spectrometer.  The numbers given in the following refer to one
detector half.  An array of 20 spherical mirrors (radius of curvature:
$156\,\cm$) mounted at the rear of the gas volume focused the
\v{C}erenkov photons onto phototubes of diameter $12.7\,\cm$.
Hinterberger--Winston light cones with an entrance diameter of
$21.7\,\cm$ helped maximize light collection.  The mean number of
photoelectrons for a $\beta \approx 1$ particle was measured to be
around five.

% +++++++ NOTE: This is the reference for the number of p.e. for the real
% +++++++       gas radiator mixture used during running:
%  
% Date: Fri, 7 Mar 2003 14:12:24 -0600 (CST)
% From: Harold Jackson <jackson@mep.phy.anl.gov>
% To: Marc.Beckmann@desy.de
% Subject: # p.e.
%
% Marc,
%
% Per your question, I check my files and found our calibration data for the
% 70/30 mixture. The yield fluctuated over the face of the detectors, but
% the yield for the mixture was on the average 5 photoelectrons for a fully
% relativistic particle.
%
% Regards, Hal
%
% ++++++++++++++++++++++++++++++++++++++++++++++++++++++++++++++++++++++++

\paragraph{\underline{RICH:}}
The threshold \v{C}erenkov detector was replaced in 1998 by a ring
imaging \v{C}erenkov detector (RICH) which allowed kaons and protons
to be identified as well as pions \cite{hermes:RICH}.  The RICH uses a
novel two--radiator design to achieve separation of pions, kaons, and
protons over the entire kinematic range of interest
($4$--$13.8\,\GeV$; see Fig.~\ref{fig:richangles}).  One of the
radiators is C$_4$F$_{10}$ gas with an index of refraction of
$n=1.0014$, while the second radiator consists of aerogel tiles with
index of refraction $n= 1.03$ mounted just behind the entrance window.
The aerogel tiles are $1.1\,\cm$ thick and they are stacked in five
layers for a total length of $5.5\,\cm$.
A mirror array with a radius of curvature of $220\,\cm$ focuses the
\v{C}erenkov photons onto $1934$ photomultiplier tubes of $1.92\, \cm$
diameter per detector half.  Details on the analysis of the RICH data are
given in section~\ref{sect:data}.

\begin{figure*}[htp]
  \centering \includegraphics[width=\textwidth]{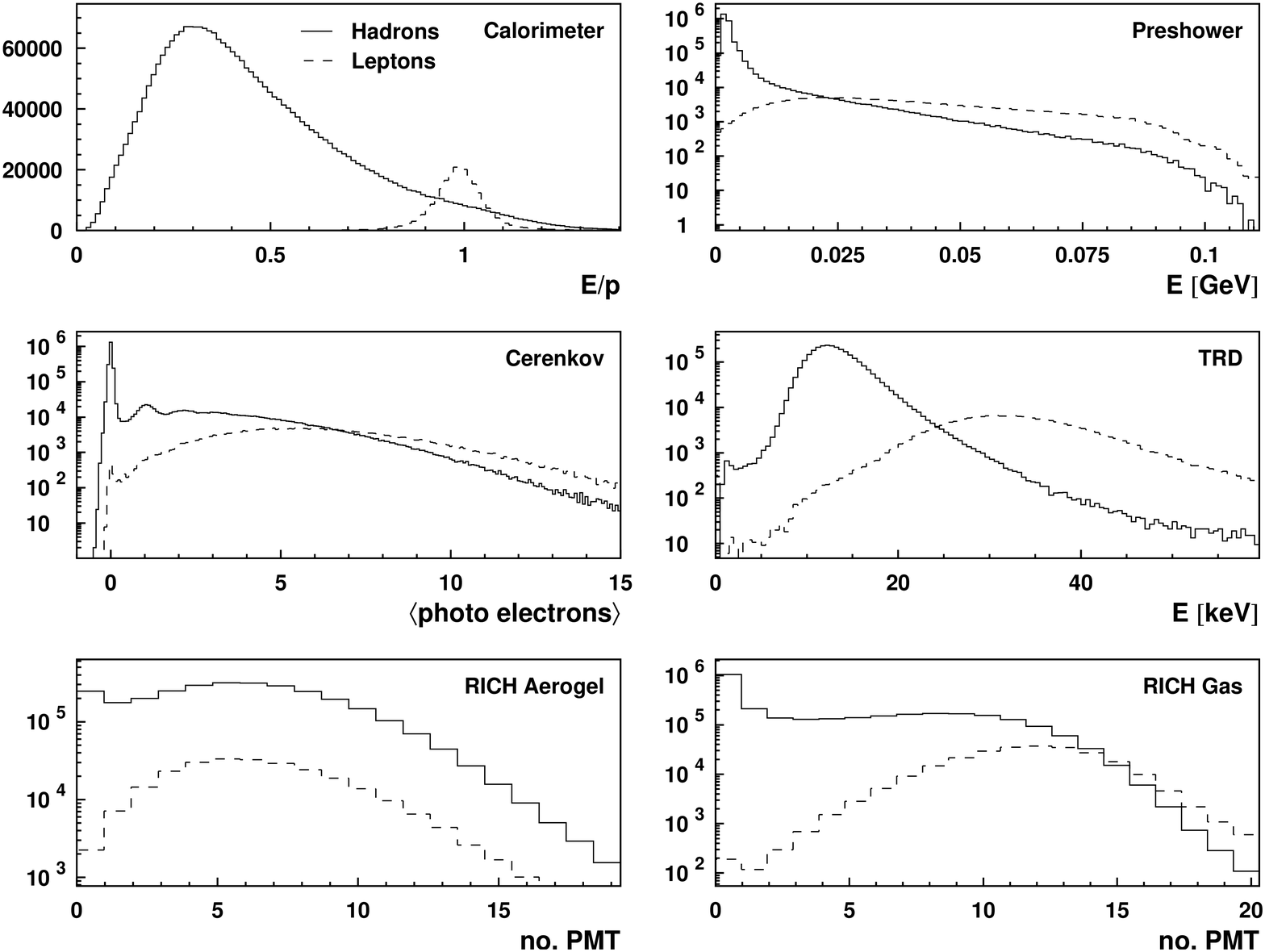}
  \caption{\label{fig:detresp}Typical PID detector responses.
    The distributions are based on a small set of the data collected
    in 2000, except for the threshold \v{C}erenkov response, which was
    computed from a data set of similar size collected in 1997.  The
    truncated mean is shown in the case of the TRD. The relative size
    of the lepton (dashed line) and hadron distributions (solid line)
    was scaled to the flux ratio in the respective data--taking
    periods to give a better idea of the level of contamination
    possible from each detector.  The flux ratio of electrons to
    hadrons is typically $\sim 10\,\%$ for these data.}
%%
%%  Plot macro: PID-DetResponses.kumac
%%
\end{figure*}

\paragraph{\underline{PID Detector Performance:}}
Plots of the responses of the PID detectors are shown in
Fig.~\ref{fig:detresp}.  A description of the PID analysis,
integrating all the detectors, is given in section~\ref{sect:data}.

\subsubsection{Event Trigger}
\label{sec:trigger}
Before discussing the trigger itself, two more detectors used
specifically for the trigger must be introduced: two scintillator
hodoscopes H0 and H1.  The hodoscope H1 is identical to H2 except that
it does not have any lead in front of it.  It is situated between BC4
and the TRD.  The scintillator H0 was added after the first year of
running to help discriminate against particles traveling backwards in
the spectrometer.  These particles originate in showers initiated by
the proton beam.  The H0 hodoscope is placed just in front of the
magnet and therefore has enough separation from H1 and H2 that it can
determine whether a particle is going forward or backwards in the
spectrometer.

The DIS trigger selects electron/positron events by requiring hits in
the three scintillator hodoscopes (H0, H1, and H2) together with
sufficient energy deposited in two adjacent columns of the
calorimeter, in coincidence with the accelerator bunch signal (HERA
clock).  The requirement of hits in H0 and H1 suppresses neutral
particle background.  The calorimeter has a high efficiency for
electromagnetic showers, but relatively low efficiency for hadronic
showers.  The calorimeter threshold was set at $1.4\,\GeV$
($3.5\,\GeV$ for the first period in 1996).

\subsubsection{Luminosity Monitor}
\label{sec:lumi}

The luminosity was measured using elastic scattering of beam particles
by the electrons in the target gas: Bhabha scattering and annihilation
for a positron beam, M{\o}ller scattering for an electron
beam \cite{hermes:lumi}.
The scattered particles exit the beam pipe $7.2\,\m$ downstream of the
target.  They are detected in coincidence by a pair of small
calorimeters with a horizontal acceptance of $4.6$--$8.9\,\mrad$.  The
calorimeters consist of \v{C}erenkov crystals of NaBi(WO$_4$)$_2$ that
are highly resistant to radiation damage.

\subsubsection{Data Acquisition System and Event Structure}
\label{sec:daq}

The backbone of the data acquisition system is constructed in Fastbus.
It consists of 10 front--end crates, an event collector crate, and an
event receiver crate, connected to the online workstation cluster via
two SCSI interfaces.  CERN Host Interfaces (CHI) act as Fastbus
masters, and their performance is enhanced by Struck Fastbus Readout
Engines (FRE) containing two Motorola 96002 DSPs.

The drift chambers were read out by LeCroy multi--hit, multi--event
16--bit 96 channel TDCs (model 1877).  Charge from the photomultipliers
and from the TRD was digitized by LeCroy multi--event 64 channel 1881M
multi block ADCs.  These ADCs and the TDCs are capable of sparsifying
the data, {\it i.e.}~online suppressing channels with pedestal levels from
the readout.  The magnet chamber readout was instrumented with the
LeCroy VME based PCOS4 system. The vertex chamber data arrived from
the detector as a 16 bit ECL STR330/ECL data stream and were processed
in one of the VC DSPs.
Double buffering was implemented in the dual DSPs of the Fastbus
masters.  Event collection on one DSP was done in conjunction with
readout from the second DSP to the DAQ computer.

In addition to the standard readout, a series of asynchronous
independent events from the luminosity monitor and from
monitoring equipment could be read out at rates exceeding
$5\,\mathrm{kHz}$.  One VME branch with 4 crates and three CAMAC
branches with 9 crates were used for these events.  The DAQ dead--time
was typically less than $10\,\%$ with a total trigger rate of about
$300\,\Hz$.

The data are arranged into the following time structure:
\begin{itemize}
\item \underline{{\em Burst:}} Events are grouped into {\em bursts},
  defined as the interval between two successive reads of the
  experiment scalers.  A burst is roughly $10\,\seconds$ long.  Data
  quality is checked on the burst level.
\item \underline{{\em Run:}} The size of the files stored on disk and
  tape is adjusted so that an integral number of {\em runs} can fit on
  a tape.  At high instantaneous luminosity, one run can be as short
  as $10\,\minutes$.  A run is the basic unit of data for analysis.
  Calibration constants are applied at the run level, although not all
  detectors are calibrated with this time granularity.
\item \underline{{\em Fill:}} Runs are grouped into {\em fills}, which
  are simply defined as data collected during a given fill of the
  electron storage ring.
\end{itemize}

%%
%% ====================== End of file 'sect-III.tex' ===================== %%
%%
%% ----------------------------------------------------------------------- %%
%%
%%   File        : sect-IV.tex
%%
%%   Description : Source file for section IV of the second delta-q paper
%%                 (DC number 42)
%%   
%%   Main Author : Juergen Wendland
%%
%%   Date        : 19-Aug-2002
%%
%%   Remarks     : 
%%
%%   Modified    : Apr. 7, 2003
%%                 June-20-2003: jw
%%                    implemented Andy's comments
%%
%% ----------------------------------------------------------------------- %%

\section{Data Analysis}
\label{sect:data}

%%
%% ----------------------------------------------------------------------- %%
%%

\subsection{Data Quality}

The data used to compute the asymmetries and multiplicities were
selected by a number of quality criteria applied at the burst level:
\begin{itemize}
%%\item {\it the H and D target polarizations were larger than $75\,\%$,} 
\item the beam polarization was between $30\,\%$ and $80\,\%$, and the
  beam current was between 5 and 50\,mA. (The upper bounds are beyond
  values observed during data--taking. They are imposed to reject
  faulty records.),
\item the trigger dead--time was less than $50\,\%$ and the data acquisition
  worked satisfactorily,
\item the PID system and the tracking detectors worked properly,
\item there were no high voltage trips in any of the detectors,
\item the experiment was in polarized running mode,
\end{itemize}
and the target system was required to be fully operational. This
requirement resulted in polarizations in excess of $75\,\%$ for
both the hydrogen and deuterium targets.
See Refs.~\cite{hermes:simani,hermes:juergen} for more details on the
selection of good quality data.

%%
%% ----------------------------------------------------------------------- %%
%%

\subsection{Tracking Algorithm}

Particle tracks were reconstructed using the pattern of hits in the
front and back tracking systems \cite{hermes:wander}. In the first
step of this procedure, the partial front and back tracks, which are
approximately straight lines, are reconstructed separately in each of
the $U$, $V$, and $X$ orientations. The algorithm is based on a fast
tree search.  For each orientation, the algorithm begins by
considering the entire plane and successively doubles the resolution
by discarding the halves without a hit. In each step the combined
patterns of all planes in a given orientation are compared to a data
base of physically possible tracks and only corresponding patterns are
kept. After about $11$ steps the search reaches a resolution that is
sufficient for track finding. The projections in the three planes are
then combined to form the partial tracks in the front and the back
respectively.

\begin{figure}[htb]
  \centering
  \includegraphics[width=\columnwidth]{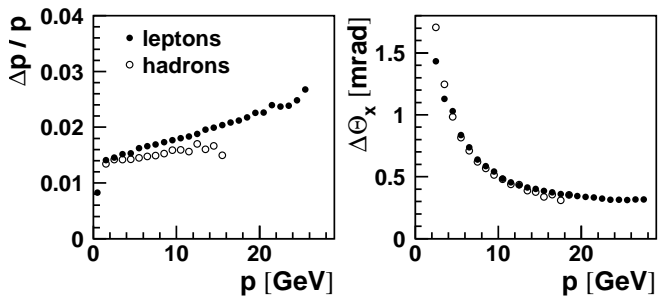}
  \caption{\label{fig:resolution}Tracking system resolution for
    lepton and hadron tracks for the detector configuration used since
    1998.  In the left panel the relative momentum resolution is
    displayed, and the right panel shows the resolution in the horizontal
    scattering angle $\theta_x$, both as a function of the track
    momentum $p$.}
%%
%%  Plot macro: Resolutions.kumac
%%
\end{figure}

The front and back tracks are associated by matching pairs that
intersect in the center of the magnet within a given tolerance.  For
each associated pair, the front track is forced to agree with the
magnet mid--point of the back track, and the front track is recomputed
accordingly.  This procedure improves the resolution of the front
tracking system, which relies on the FC
chambers since only they were installed and operational during the entire
data taking period from 1996 until 2000.  
Because the tracking information
from the other chambers was not available for this entire period,
they were not used in order to avoid possible biases for different data
taking periods.  The particle momentum is determined using another
data base of $520,000$ tracks which contains the momentum as a
function of the front and back track parameters. Multiple scattering in the
spectrometer material leads to reduced resolutions 
of less than $0.03$ for the
reconstructed track momenta and less than $1.5\,\mrad$ for the
reconstructed scattering angles.
Fig.~\ref{fig:resolution} shows the resolutions for the deuterium data
sample as obtained from a Monte Carlo simulation of the entire
spectrometer.  The momentum and angular resolution 
of the hydrogen
data are better, because of the shorter radiation length of the
\v{C}erenkov detector compared to the RICH.

%%
%% ----------------------------------------------------------------------- %%
%%

\subsection{Particle Identification Algorithm}

The PID system discriminates between electrons/{}po\-si\-trons
(referred to as leptons in the following), pions, kaons, and other
hadrons.  It provides a factor of about 10 in hadron suppression at the
trigger level to keep data acquisition rates reasonable.  The hadron
rate from photo production exceeds the DIS rate by a factor of up
to 400:1 in some kinematic regions. In offline analysis, the HERMES
PID system suppresses hadrons misidentified as leptons by as much as
$10^4$ with respect to the total number of hadrons, while identifying
leptons with efficiencies exceeding $98\,\%$.

%%
%%  Bayes
%%
The identification of hadrons and leptons is based on a Bayesian
algorithm that uses the conditional probability $P(A|B)$ defined as
the probability that $A$ is true, given that $B$ was observed. For each
track the conditional probability $P(H_{l(h)}|E,p,\theta)$ that the
track is a lepton (hadron) is calculated as
\begin{equation}
  P(H_{l(h)}|E,p,\theta)=\frac{P(H_{l(h)}|p,\theta)\; P(E|H_{l(h)},p)}
  {\sum_{i=l,h}P(H_i|p,\theta)\; P(E|H_i,p)} \, .
\end{equation}
Here $H_{l(h)}$ is the hypothesis that the track is a lepton
(hadron), $E$ the response of the considered detector, and $p$ and
$\theta$ are the track's momentum and polar angle. The parent
distributions $P(E|H_{l(h)},p)$ of each detector (i.e. the typical
detector responses) were extracted from data with stringent restrictions on
the other PID detectors to isolate a particular particle type.  See
Fig.~\ref{fig:detresp} for plots of the individual PID detector
responses.

%%
%%  PID without flux
%%
In a first approximation, uniform fluxes
$P(H_{l}|p,\theta)=P(H_{h}|p,\theta)$ are assumed so that the ratio
\begin{equation}
  \label{eqn:pidratio}
  \log_{10}\frac{P(H_{l}|E,p,\theta)}{P(H_{h}|E,p,\theta)}
\end{equation}
reduces to:
\begin{equation}
  \mathrm{PID}_{det}=\log_{10}\frac{P(E|H_{l},p)}{P(E|H_{h},p)} \, .
\end{equation}

\begin{figure}[htbp]
  \centering
  \includegraphics[width=\columnwidth]{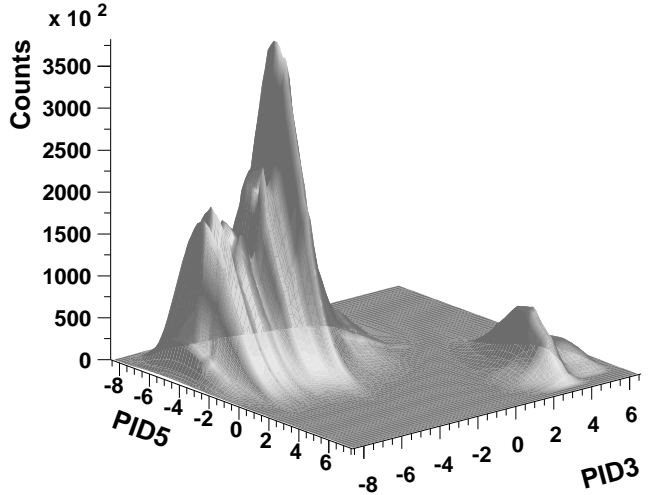}
  \caption{\label{fig:PID3vsPID5}Two--dimensional distribution of PID
    values for all particles in the acceptance.  The quantities
    $\mathrm{PID3}$ and $\mathrm{PID5}$ are defined as
    $\mathrm{PID3}\equiv\mathrm{PID}_{cal}+\mathrm{PID}_{pre}
    +\mathrm{PID}_{ric}$ and $\mathrm{PID5}\equiv\mathrm{PID}_{trd}$.
    }
%%
%%  Plot macro: PID3vs5.kumac
%%
\end{figure}

The quantity PID$_{det}$ is defined for the calorimeter ($cal$), the
pre--shower detector ($pre$), the \v{C}erenkov detector ($cer$) (the
RICH detector ($ric$) since 1998), and the TRD ($trd$).  In the case
of the RICH and the TRD this ratio is the sum over the $\mathrm{PID}$
values of the two radiators and the six TRD modules respectively. The
$\mathrm{PID}$ distribution of the TRD ($\mathrm{PID5}$) is shown in
Fig.~\ref{fig:PID3vsPID5} versus the sum of the $\mathrm{PID}$ values
of the calorimeter, the pre--shower, and the threshold
\v{C}erenkov/RICH ($\mathrm{PID3}$).  The leptons (small bump) are
seen to be clearly separable from the hadrons (large peak).

\begin{figure}[htb]
  \centering
  \includegraphics[width=\columnwidth]{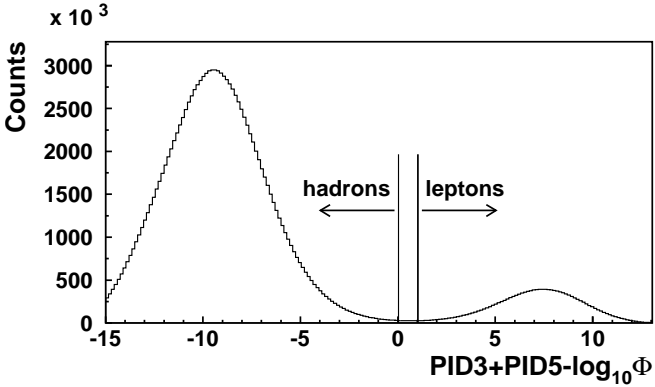}
  \caption{\label{fig:PID3+PID5}The distribution of the total
    $\mathrm{PID}$ value.  This logarithmic ratio of probabilities
    includes the particle fluxes and the responses of all PID
    detectors.  The left hand peak is the hadron peak, while the right
    hand peak originates from leptons.  The limits that were applied in
    the analysis are shown as vertical lines.}
%%
%%  Plot macro: PID35.kumac
%%
\end{figure}

\begin{figure}[htb]
  \centering
  \includegraphics[width=.45\textwidth]{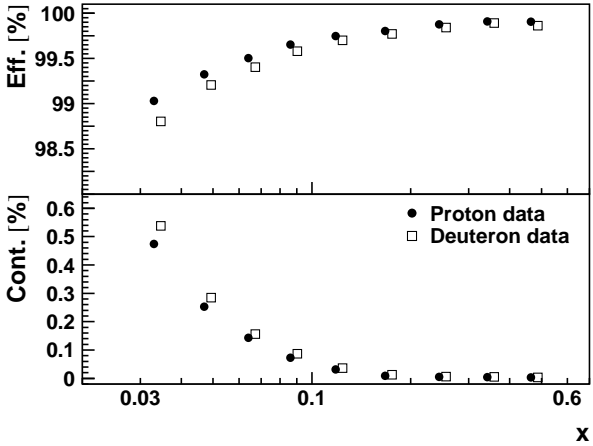}
  \caption{\label{fig:pidconeff}Identification efficiency and hadron
    contamination of the DIS lepton sample as a function of $x$.  
    Because correlations between the responses of the PID 
    detectors were neglected, the contaminations are 
    uncertain by a factor of two. The
    deuteron data have slightly worse efficiencies and contaminations
    because of the better hadron--lepton discrimination of the
    threshold \v{C}erenkov counter compared to the RICH.}
%
%  The threshold Cerenkov has a higher threshold (than the
%  aerogel). Therefore hadrons Cerenkov--radiate less and are better
%  distinguishable from leptons.
%
%%
%%  Plot macro: PID-ContEff.kumac
%%
\end{figure}

%%
%%  Fluxes
%%
The particle fluxes $P(H_{l(h)}|p,\theta)$ were computed in an
iterative procedure by comparing the calculated ratio
Eq.~\eqref{eqn:pidratio} to data and varying the fluxes. These fluxes
were then combined with $\mathrm{PID3+PID5}$ to form the
%%
%%  total PID
%%
total $\mathrm{PID}$ value
\begin{equation}
  \mathrm{PID}=\mathrm{PID3}+\mathrm{PID5}-\log_{10}\Phi \, ,
\end{equation}
where $\Phi=P(H_{h}|p,\theta)/P(H_{l}|p,\theta)$ is the ratio of
hadron and lepton fluxes.  A plot of the quantity PID which was used
to discriminate hadrons and leptons is shown in
Fig.~\ref{fig:PID3+PID5}, where the two peaks for hadrons and leptons
are seen to be well separated. Hadrons and leptons were identified
with limits requiring $\mathrm{PID}<0$ and $\mathrm{PID}>1$
respectively.
The lepton restriction provided excellent discrimination of DIS leptons from
the large hadron background, with efficiencies larger than $98\,\%$ and
contaminations below $1.0\,\%$ over the entire range in $x$ (see
Fig.~\ref{fig:pidconeff}).  Semi--inclusive hadrons were identified
with efficiencies larger than $99\,\%$ and lepton contaminations
smaller than $1.0\,\%$, which were determined~\cite{hermes:juergen} from 
data collected during the normal operation of the experiment.

\subsection{The \v{C}erenkov detectors and hadron identification}
\label{sect:hadID}

%%
%%  The threshold Cerenkov's pion identification
%%
The threshold \v{C}erenkov detector identified pions with momenta
between $4$ and $13.8\,\GeV$.  A hadron track was identified as a
pion, if the number of detected photo--electrons was above the noise
level.  The contamination of the pion sample by other hadrons as well
as leptons is negligible.

\begin{figure}[htbp]
  \centering
  \includegraphics[width=\columnwidth]{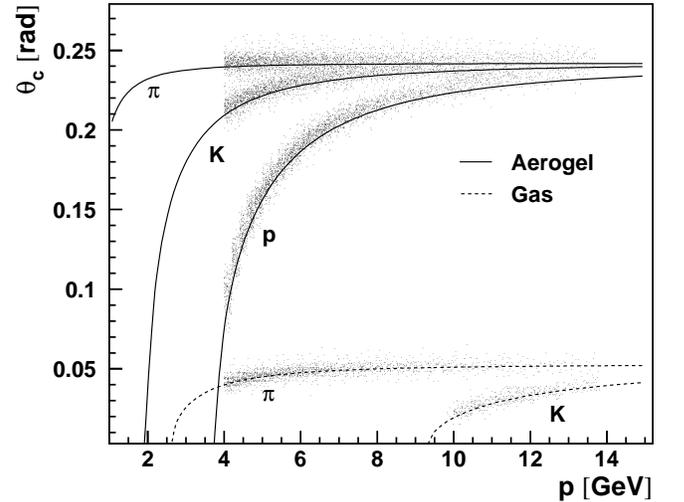}
  \caption{\label{fig:richangles}\v{C}erenkov angles associated with
    the three particle hypotheses as a function of the particle
    momentum. The characteristic angles of \v{C}erenkov light emitted
    in the aerogel ($n=1.03$) are given by the solid lines. The
    characteristic angles for emission in the gas ($n=1.0014$) are
    shown as the dashed lines.  The corresponding histogram entries
    are experimentally determined angles of a sample of SIDIS
    hadrons.}
%%
%%  Plot macro: RICH.kumac
%%
\end{figure}

%%
%%  The RICH detector id's Kaons and Pions
%%
The RICH detector identifies pions, kaons, and protons in the momentum
range $2\,\GeV<p<15\,\GeV$. In the semi--inclusive analysis reported
in this paper a momentum range of $4\,\GeV<p<13.8\,\GeV$ was used for
consistency with the threshold \v{C}erenkov detector.  The pattern of
\v{C}erenkov photons emitted by tracks passing through the aerogel or
the gas radiators on the photomultiplier matrix was associated with
tracks using inverse ray tracing. For each particle track, each hadron
hypothesis, and each hypothesis for the radiator emitting the photons,
aerogel or gas, the photon emission angle was computed.
%%For each particle track the emission
%%angles of photons were computed assuming that the photon was emitted
%%in the aerogel and the gas, respectively. 
The average \v{C}erenkov
angles $\langle \theta \rangle^{a,g}_{\pi,K,p}$ were calculated for
each radiator ($a,g$) and particle hypothesis ($\pi,K,p$) by including
only photons with emission angles within $2\,\sigma_\theta$ about the
theoretically expected emission angle
$\theta^{\mathrm{theo};a,g}_{\pi,K,p}$, where $\sigma_\theta\simeq
8\,\mrad$ is the single photon resolution. This procedure rejects
background photons, and photons due to other tracks or the other
radiator.  Fig.~\ref{fig:richangles} shows the distribution of angles
in the two radiators as a function of the particle momentum.
Based on the Gaussian likelihood,
\begin{equation}
  \mathcal{L}^{a,g}_{i}=
  \exp\left[-\left(\theta^{\mathrm{theo};a,g}_{i}
      -\langle\theta\rangle^{a,g}_{i}\right)^2
    \frac{1}{2\sigma^2_{\langle\theta\rangle^{a,g}_{i}}}\right]
\end{equation}
%%
%% some more possibilities to write this equation.
%%
% \begin{equation}
%   \mathcal{L}^{a,g}_{i}=
%   \exp\left[-
%     \left(\theta^{\mathrm{theo};a,g}_{i}
%       -\langle\theta\rangle^{a,g}_{i}\right)^2
%     /2\sigma^2_{\langle\theta\rangle^{a,g}_{i}}\right]\quad i=\pi,K,p
% \end{equation}
% \begin{equation}
%   \mathcal{L}^{r}_{i}=
%   \exp\left[-
%     \left(\theta^{\mathrm{theo};r}_{i}
%       -\langle\theta\rangle^{r}_{i}\right)^2
%     /2\sigma^2_{\langle\theta\rangle^{r}_{i}}\right]\quad r=a,g,\;i=\pi,K,p
% \end{equation}
                              %
a particle hypothesis $i=\pi,K,p$ with the largest total likelihood
$\mathcal{L}_i^{\mathrm{tot}}=\mathcal{L}_i^a\cdot\mathcal{L}_i^g$ is
assigned to each hadron track.

%%
%%  Contamination and efficiency of the RICH.
%%
Identification efficiencies and probabilities for contamination
of hadron populations from misidentification of other hadrons
were estimated with a
Monte Carlo simulation which had been calibrated 
with pion and kaon tracks from
experimentally reconstructed $\rho^0$, $\phi$, $K_s^0$ meson and
$\Lambda$ hyperon decays. In the analysis, each pion and kaon track
was assigned a weight $\omega_i^{K,\pi}$ accordingly. The number of
counts of pions and kaons,
\begin{equation}
  N_{K,\pi}=\sum_{i} \omega_i^{K,\pi}
  \label{eq:WpiK}
\end{equation}
were computed as the sums of these weights.

%%
%% ======================= End of file 'sect-IV.tex' ===================== %%
%%

%%% Local Variables: 
%%% mode: latex
%%% TeX-master: "dqpaper"
%%% End: 

%% ----------------------------------------------------------------------- %%
%%
%%   File        : sect-V.tex
%%
%%   Description : Source file for section V of the second delta-q paper
%%                 (DC number 42)
%%   
%%   Main Author : Juergen Wendland
%%
%%   Date        : 19-Aug-2002
%%
%%   Remarks     : 
%%
%%   Modified    : April 8, 2003
%%                 April 30, 2003 jw
%%                 June 20, 2003, jw:
%%                      Details of radcorr into appendix
%%                      Implemented Andy's comments
%%
%%                 July 24, 2003, jw: Finalized data and plots
%%
%% ----------------------------------------------------------------------- %%

\section{Asymmetries}
\label{sect:asyms}

%%
%% ----------------------------------------------------------------------- %%
%%

\subsection{Measured Asymmetries}
\label{sect:asyms:meas}

In the data sample that satisfied the data quality criteria described
in the previous section, events were selected for analysis if they
passed the DIS trigger (see section~\ref{sect:exp}).
Tracks with a minimum energy of $3.5\,\GeV$ in the calorimeter that
were identified as leptons by the PID system were selected as 
candidates for the scattered DIS particle by imposing additional 
requirements on the track kinematics.  A requirement of $Q^2>1\,\GeV^2$
selected hard scattering events.  Events from the nucleon
resonance region were eliminated by requiring $W^2>10\,\GeV^2$.  
The requirement
$y < 0.85$ reduced the number of events with large radiative
corrections.

Hadron tracks coincident with the DIS positron were identified as
semi--inclusive hadrons if the fractional energy $z$ of the hadron was
larger than $0.2$ and $x_F$ was larger than $0.1$.  
These limits suppress contributions from target fragmentation. Hadrons from
exclusive processes, such as diffractive vector meson production, were
suppressed by requiring that $z$ be smaller than $0.8$.

%%
%%  Equivalent count rates
%%
The final statistics of inclusive DIS events and SIDIS hadrons are
given in Tab.~\ref{tab:DISrates}. The numbers are presented
in terms of statistically equivalent numbers of events $N_{\mathrm{eq}}$. 
This quantity is the number of unweighted events with the same
relative error as the sum of weighted events $N$,
\begin{equation}
  \frac{\sigma_{N_{\mathrm{eq}}}}{N_{\mathrm{eq}}}
  \equiv\frac{\sqrt{N_{\mathrm{eq}}}}{N_{\mathrm{eq}}}
  =\frac{\sqrt{\sum_i(\omega_i)^2}}{\sum_{i'}\omega_{i'}}
  \equiv\frac{\sigma_N}{N} \, .
\end{equation}
The weights $\omega_i$ are defined in section~\ref{sect:hadID} for
hadrons identified by the RICH detector in semi--inclusive events.
For pions identified by the threshold \v{C}erenkov counter, for
undifferentiated hadrons, and for inclusive DIS events $\omega_i = 1$.
An additional weight factor of $\pm 1$ is applied according to the event
classification as signal or charge--symmetric background (see below).

\begin{table}[htbp]
  \caption{\label{tab:DISrates}Statistically equivalent number of counts of DIS
    events and SIDIS hadrons for the hydrogen and the deuterium data.}  
  \centering
  \renewcommand{\extrarowheight}{2pt}
  \renewcommand{\tabcolsep}{3pt}
  \begin{ruledtabular}
    \begin{tabular}{c|c|rrrr}
      & & \multicolumn{4}{c}{SIDIS events} \\
      Target & \multicolumn{1}{c|}{DIS evts.}
      & \multicolumn{1}{c}{$\pi^+$}
      & \multicolumn{1}{c}{$\pi^-$}
      & \multicolumn{1}{c}{$K^+$}
      & \multicolumn{1}{c}{$K^-$} \\ \hline
%% count rates 
%%     H& 1.8M & 120,000 & 84,000 \\
%%     D& 6.8M & 497,000 & 401,000 & 108,000 &  45,000 \\
%% equivalent count rates
%%      H& 1.7M & 117,000 & 82,000 \\
%%      D& 6.7M & 491,000 & 385,000 & 76,000 &  33,000 \\
%% equivalent count rates in exponential notation
      H & $1.7 \times 10^6$ & $117 \times 10^3$ &  $82 \times 10^3$ & & \\
      D & $6.7 \times 10^6$ & $491 \times 10^3$ & $385 \times 10^3$ & 
      $76 \times 10^3$ & $33 \times 10^3$ \\
    \end{tabular}
  \end{ruledtabular}
  \renewcommand{\extrarowheight}{0pt} % default
  \renewcommand{\tabcolsep}{2pt}      % default (at least in revtex)
\end{table}

The inclusive (semi--inclusive) data samples were used to calculate
the measured positron--nucleon asymmetry $A^{(h)}_{\|}$ in bins of $x$
(or $z$),
\begin{equation}
  A_{\|}^{(h)}=\frac{N_{(h)}^\sant L^\spar - N_{(h)}^\spar L^\sant}
  {N_{(h)}^\sant L_P^\spar + N_{(h)}^\spar L_P^\sant} \, .
  \label{eq:aparallel}
\end{equation}
Here $N^\spar$ ($N^\sant$) is the number of DIS events for target
spin orientation parallel (anti--parallel) to the beam spin orientation, and
$N_{h}^\spar$ ($N_{h}^\sant$) are the corresponding numbers of
semi--inclusive DIS hadrons.  The luminosity $L^\spar$ ($L^\sant$) for
the parallel (anti--parallel) spin state is corrected for dead time,
while $L_P^\spar$ ($L_P^\sant$) is the luminosity corrected for dead
time and weighted by the product of beam and target polarizations for
the parallel (anti--parallel) spin state.  Values for the beam and
target polarizations are given in Tabs.~\ref{tab:beampol} and
\ref{tab:tar}. The bins in $x$ used in the analysis are defined in
Tab.~\ref{tab:xbins}.

\begin{table}[htb]
  \caption{\label{tab:xbins}The bins in $x$ used in the analyses
    presented in this paper.}
  \centering
  \begin{ruledtabular}
    \begin{tabular}{l|lllllllll}
      bin & \multicolumn{1}{c}{1} & \multicolumn{1}{c}{2} &
      \multicolumn{1}{c}{3} & \multicolumn{1}{c}{4} &
      \multicolumn{1}{c}{5} & \multicolumn{1}{c}{6} &
      \multicolumn{1}{c}{7} & \multicolumn{1}{c}{8} & \multicolumn{1}{c}{9} \\
      \hline
      $x_{\mathrm{low}}$ \rule{0mm}{2.7ex}
      & 0.023 & 0.040 & 0.055 & 0.075 & 0.1  & 0.14 &
      0.2 & 0.3 & 0.4 \\
      $x_{\mathrm{up}}$  & 0.040 & 0.055 & 0.075 & 0.1   & 0.14 & 0.2  &
      0.3 & 0.4 & 0.6 \\
    \end{tabular}
  \end{ruledtabular}
\end{table}

%%
%%  Tensor polarization
%%
In the deuteron --- a spin--1 particle --- another polarized
structure function $b_1^d$ arises from binding effects associated with
the D-wave component of the ground state \cite{Hoodbhoy:1989am}. 
This structure function may contribute to the
cross section if the target is polarized with a population
of states with spin--projection $S_z=0$ that is not precisely 1/3, {\it i.e.} 
a substantial tensor polarization.
Because the maximum vector polarization can only be accomplished with a high
tensor polarization in a spin 1 target, measurements in HERMES, of necessity,
can include significant contributions from the tensor analyzing power of
the target. For inclusive scattering, the spin asymmetry is of the form
\begin{equation}
  A_{1}=\frac{\sigma_{1/2}-\sigma_{3/2}}
  {\sigma_{1/2}+\sigma_{3/2}}[1+\frac{1}{2}TA_{T}]
  \label{eq:tensor}
\end{equation} 
where $T$ is the tensor polarization, and $A_T$ is the tensor analyzing
power. The $b_1$ structure function is measured by $A_T$, 
{\it i.e.} $A_{T}\approx 2b_{1}/3F_1$. Studies by the HERMES
collaboration indicate that $b_1^d$ is small \cite{hermes:b1d}, and that
the tensor contribution to the the inclusive deuteron asymmetry is less than
$\approx 0.5-1.0\%$ of the measured asymmetry. 
For this reason, tensor contributions were assumed 
to be negligible for all the spin asymmetries presented here.
%%In the present analysis these polarization
%%states were excluded and any small dilution due to non--perfect
%%polarization cancels in the inclusive and semi--inclusive asymmetries
%%$A_{1,d}^{(h)}$. In addition preliminary measurements by the HERMES
%%collaboration indicate that $b_1^d$ is small \cite{hermes:b1d}.

%%
%% ----------------------------------------------------------------------- %%
%%

\subsection{Charge--Symmetric Background}
\label{sect:asyms:background}

%%
%%  Background count rate corrections.
%%
The particle count rates were corrected for charge symmetric
background processes (\emph{e.g.} $\gamma\rightarrow e^++e^-$).  The
rate for this background was estimated by considering lepton tracks with
a charge opposite to the beam charge that passed the DIS restrictions.  It
was assumed that these leptons stemmed from pair--production processes.
The rate for
the charge symmetric background process 
(where the particle is detected with the
same charge as the beam but originating from pair production) is the
same.  The number of events with an opposite sign lepton is therefore
an estimate of the number of charge symmetric events that masquerade
as DIS events.  They were subtracted from the inclusive DIS count
rate.
Hadrons that were coincident with the background DIS track and that
passed the SIDIS limits were also subtracted from the corresponding
SIDIS hadron sample. The DIS background rate was $\sim 6\,\%$ with
respect to the total DIS rate in the smallest $x$--bin, falling off
quickly with increasing $x$.  The overall background fraction from
this source was $1.4\,\%$.

%%
%% ----------------------------------------------------------------------- %%
%%

\subsection{Azimuthal Acceptance Correction}
\label{sect:asyms:acceptance}

The measured semi--inclusive asymmetries were corrected for acceptance
effects due to the nonisotropic azimuthal acceptance of the
spectrometer.  These acceptance effects arise because of an azimuthal
dependence of the polarized and unpolarized
semi--inclusive cross sections due, e.g. to non--zero 
intrinsic transverse parton
momenta
\cite{th:mulders-tangerman}. Taking into account the azimuthal
dependence, the measured semi--inclusive asymmetry given in
Eq.~\eqref{eq:aparallel} is modified \cite{Oganessyan:2002er},
\begin{equation}
  \frac{A_\|^h+\frac{C_1}{C_0}\langle\cos\phi\rangle_{LL}}
  {1+\frac{C_1}{C_0}\langle\cos\phi\rangle_{UU}} 
  = 
  \frac{N_{(h)}^\sant L^\spar - N_{(h)}^\spar L^\sant}
  {N_{(h)}^\sant L_P^\spar + N_{(h)}^\spar L_P^\sant} \, ,
\end{equation}
where $\langle\cos\phi\rangle_{LL}$ and $\langle\cos\phi\rangle_{UU}$
are the $\cos\phi$ moments of the semi--inclusive polarized and
unpolarized cross section, respectively and $C_0$ and $C_1$ are
the lowest order Fourier coefficients of the spectrometer's azimuthal
acceptance.

The semi--inclusive asymmetries $A_{\|}^h$ were corrected for the
unpolarized moment $\langle\cos\phi\rangle_{UU}$ using a
determination of the moments from HERMES data.  The polarized
moment $\langle\cos\phi\rangle_{LL}$ was found to be negligible as
expected in Ref.~\cite{Oganessyan:2002er}.  The correction for the
unpolarized moment to the asymmetries is 
$10\,\%$ for $x<0.1$.  In the measured $x$ range the absolute
correction of the semi--inclusive asymmetries is small, because of the
small size of the asymmetries at low $x$ (see below) and because the
correction is small for $x>0.1$.  The correction to the asymmetries as
function of $z$ is about $10\,\%$ at small $z$ and becomes smaller for
larger values of $z$.

%%
%% ----------------------------------------------------------------------- %%
%%

\subsection{Radiative and Detector Smearing Effects}
\label{sect:asyms:radcor}

%%
%% Detector and QED radiative corrections: Intro
%%
The asymmetries were corrected for detector smearing and QED radiative
effects to obtain the Born asymmetries which correspond to pure single 
photon exchange in the scattering process.  
The corrections were applied using an unfolding algorithm
that accounts for the kinematic migration of the events.
%%\cite{hermes:miller}.  
As opposed to iterative techniques described in
Ref.~\cite{Akushevich:1994dn} for example, this algorithm does not
require a fit of the data.  The final Born asymmetries 
shown in Figs.~\ref{fig:A1p} and \ref{fig:A1d} depend only on
the measured data, on the detector model, on the known
unpolarized cross sections, and on the models for the background
processes.  Another advantage is the unambiguous determination of the
statistical variances and covariances on the Born asymmetries based on
the simulated event migration.  A description of the unfolding 
algorithm and the input Monte
Carlo data is found in App.~\ref{sect:app:radcorr}.

\begin{figure}[htbp]
  \centering
  \includegraphics[width=\columnwidth]{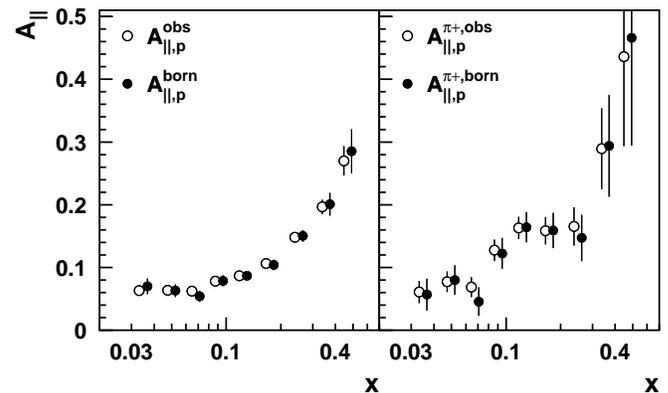} %% FINAL
  \caption{\label{fig:Aparap-born_meas}The observed asymmetries $A_\|$
    and $A_\|^{\pi^+}$ on the proton target compared with the
    corresponding Born asymmetries.  The Born asymmetries are offset
    horizontally for better presentation. See the text for details.}
%%
%%  Plot macro: Aparap-meas_born.kumac
%%
\end{figure}
\begin{figure}[htbp]
  \centering
  \includegraphics[width=\columnwidth]{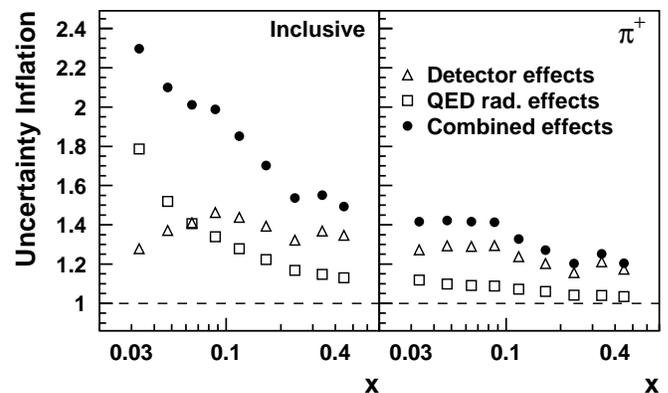} %% FINAL
  \caption{\label{fig:ErrAmp}Uncertainty inflation caused by detector
    smearing and QED radiative effects. The left hand panel shows the
    uncertainty inflation of the inclusive asymmetry for the proton and
    the right hand panel that of the positive pion asymmetry.  In both
    panels the open triangles present the uncertainty inflation caused
    by detector effects, the open squares present the inflation caused
    by QED radiative effects, and the filled circles show the total
    uncertainty inflation.}
%%
%%  Plot macro: ErrAmplification.kumac
%%
\end{figure}

\begin{figure*}[t]
  \centering
  \includegraphics[width=\textwidth]{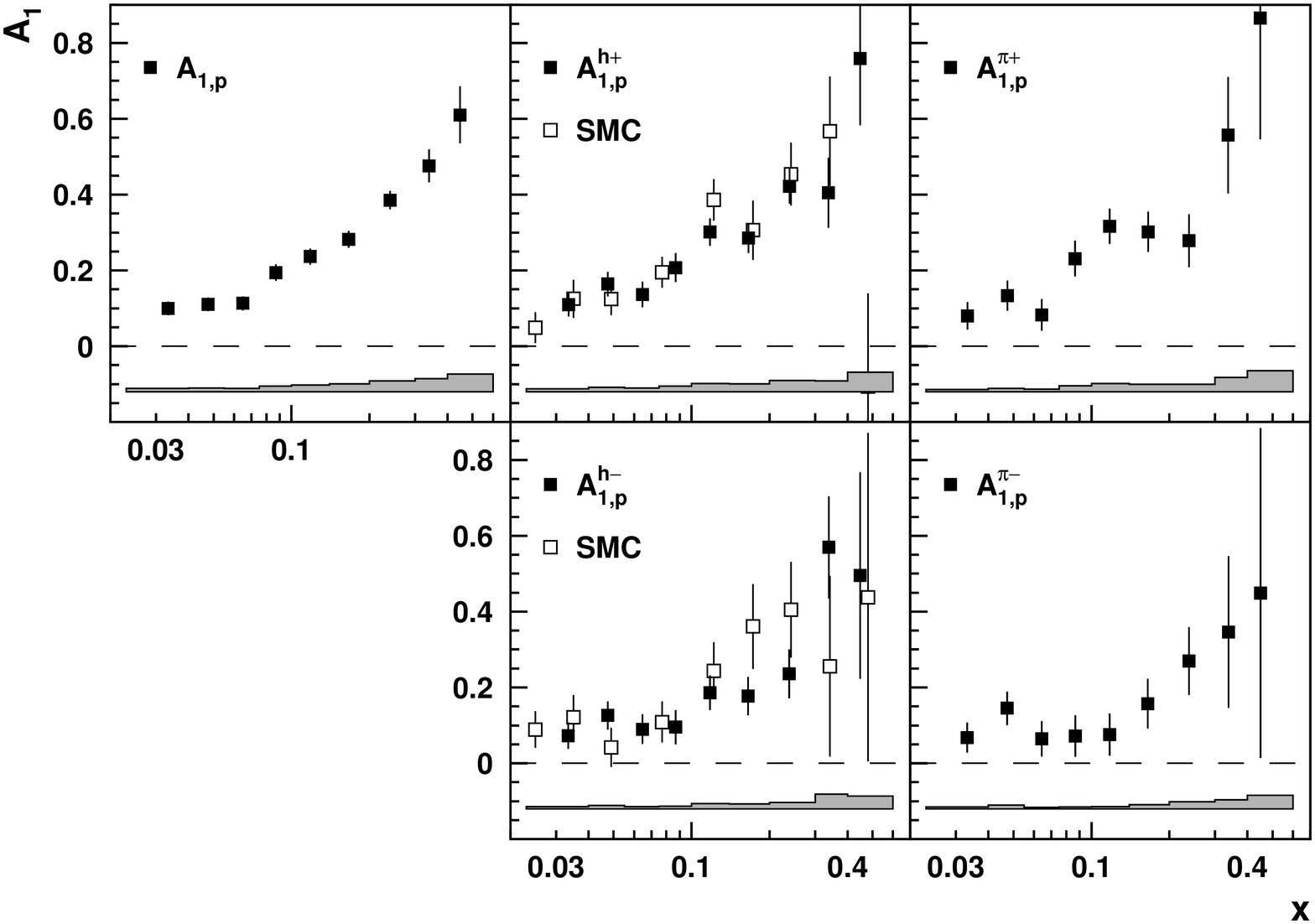} %% FINAL
  \caption{\label{fig:A1p}The inclusive and semi--inclusive Born level
    asymmetries on the proton, corrected for instrumental smearing and
    QED radiative effects.  The error bars give the statistical
    uncertainties, and the shaded bands indicate the systematic
    uncertainty.  The open squares show the positive and negative
    hadron asymmetries measured by the SMC collaboration, limited
    to the HERMES $x$--range \cite{smc:deltaq}.}
%%
%%  Plot macro: A1p-all.kumac
%%
\end{figure*}

%%
%%  Unfolding: Before and aft
%%
The impact of the unfolding on the asymmetries is illustrated with the
inclusive and the positive pion asymmetries on the proton in
Fig.~\ref{fig:Aparap-born_meas}.  The unfolding procedure shifts the
central asymmetry values only by a small amount. This is expected for
cross section asymmetries. Smearing results in a loss of information
about more rapid fluctuations that may be present in the data. Therefore 
correcting for this loss by effectively enhancing ``higher frequency''
components inevitably results in an inflation of the uncertainty of each
data point.  
The uncertainty inflation introduced by
the unfolding is shown in Fig.~\ref{fig:ErrAmp}.  The uncertainty at
low $x$ is significantly increased by the QED background in the case
of the inclusive asymmetry.  At large values of $x$ and in the case of
the semi--inclusive asymmetries, the inflation by QED radiation is
mainly due to inter--bin event migration.  Detector smearing effects,
which are largest at high $x$, increase the uncertainties through
inter--bin migration and a small number of events that migrate into
the acceptance. Uncertainty inflation due to interbin migration 
increases rapidly as the bin size is reduced to be comparable to the
instrumental resolution.

%%
%%  Comment: Unfolding wasn't done before
%%
It should be noted that this unfolding procedure is more rigorous than
the procedure applied in previous DIS experiments and previous
analyses of this experiment.  The size of the uncertainties is larger
in the current analysis due to the explicit inclusion of correlations
between $x$--bins and the model--independence of the unfolding
procedure.  This should be borne in mind when comparing the current
data to other results.

%%
%% ----------------------------------------------------------------------- %%
%%

\subsection{Results for the Asymmetries}
\label{sect:asyms:results}

\begin{figure*}[t]
  \centering \includegraphics[width=\textwidth]{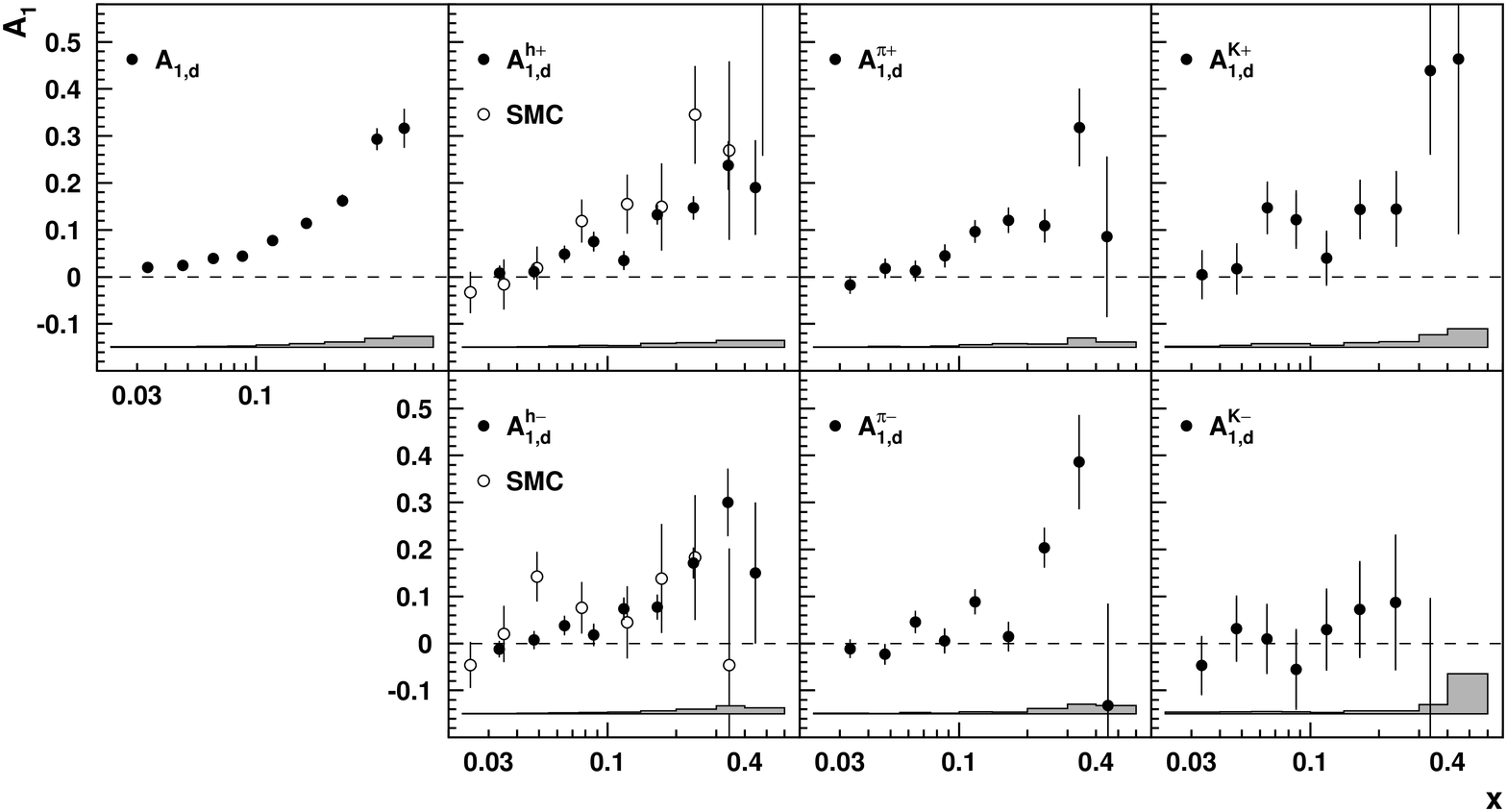} %% FINAL
  \caption{\protect\label{fig:A1d} The inclusive and semi--inclusive
    Born level asymmetries on the deuteron.  One data point at $x =
    0.45$ for the $K^-$ asymmetry including its large error bar is
    outside the displayed range; all data points are listed in
    Tab.~\ref{tab:A1d_results}.  See Fig.~\ref{fig:A1p} for details.}
%%
%%  Plot macro: A1d-all.kumac
%%
\end{figure*}

The asymmetries $A^{(h)}_\|$ are related to the inclusive and
semi--inclusive photon--nucleon asymmetries $A^{(h)}_1$ through the
kinematical factors $\eta$ and $\gamma$ and the depolarization factor
$D$ (see Eq.~\eqref{eqn:apar1}).  The Born level asymmetries
$A_1^{(h)}$ on the proton and on the deuteron targets are shown in
Figs.~\ref{fig:A1p} and \ref{fig:A1d} and are listed in
Tabs.~\ref{tab:A1p_results} and \ref{tab:A1d_results} in
App.~\ref{sect:app:results}, respectively.  
The present results on the proton target supersede earlier results
published in \cite{hermes:dq1999}.

%%
%%  Inclusive asymmetries
%%
The inclusive asymmetries on the proton and the deuteron are
determined with high precision.  On both targets, they are large and
positive.  A detailed discussion and a determination of the spin
structure functions $g_1$ from inclusive scattering data is given in
Ref.~\cite{hermes:g1p1998} for the proton and a forthcoming paper in
the case of the deuteron.

%%
%%  Semi-inclusive asymmetries
%%
The asymmetries for undifferentiated positive and negative hadrons on
both targets are compared with measurements performed by the SMC
collaboration \cite{smc:deltaq}. The statistical uncertainties of the
HERMES data are significantly better than those of the SMC data.  Pion
asymmetries on the proton and pion and kaon asymmetries on the
deuteron were measured for the first time.  The pion asymmetries are
determined with good precision, whereas the kaon asymmetries have
larger statistical uncertainties. Except for the $K^-$ asymmetry, all
asymmetries are seen to be mostly positive, which is attributed to the
dominance of scattering off the $u$--quark.  The fragmentation into
negative kaons ($\bar{u}s$--mesons) has in comparison to the other
hadrons an increased sensitivity to scattering off $\bar{u}$ and
$s$--quarks, which makes the $K^-$ asymmetry a useful tool to
determine the polarization of these flavors.

%%
%% ----------------------------------------------------------------------- %%
%%

\subsection{$z$--Dependence of the Asymmetries}
\label{sect:asyms:z-dep}

%%
%%  Asymmetries as function of z
%%
%%It has been suggested \cite{Gluck:2001vz} that spin--dependent
%%fragmentation functions $\Delta D_f^h$ may be non--zero and therefore
%%play a relevant role in polarized SIDIS.  In leading order these
%%functions generate terms like $q_f(x,Q^2)\,\Delta D_f^h(z,Q^2)$, which
%%could significantly affect the spin--dependent structure function
%%$g_1^h(x,Q^2)$.  This contribution would vary with the fractional
%%energy $z$ of the produced hadron.  However, for the analysis presented
%%here which involves (pseudo--)scalar mesons, it has been 
%%demonstrated in Ref.~\cite{th:mulders-tangerman} that the relevant 
%%spin-dependent fragmentation functions vanish in leading order.
%%In addition, a direct search for polarization dependent fragmentation was
%%done by extracting pion multiplicities
%%\begin{equation}
%%  \label{eq:multdef}
%%  n_{\pi^\pm} = \frac{N_{\pi^\pm}(x)}{N_{\mathrm{DIS}}(x)} \, ,
%%\end{equation}
%%separately for parallel and anti--parallel beam and target spin state
%%configurations.  In Eq.~\eqref{eq:multdef}, $n_{\pi^\pm}(x)$ is the
%%observed number of charged pions per observed number of DIS events,
%%$N_{\mathrm{DIS}}(x)$.  The extracted multiplicities were fully
%%consistent between both spin state configurations so that no evidence
%%for spin--dependent fragmentation functions was found.

Because the ratio of favored to unfavored fragmentation
functions is known to vary substantially with $z$, a $z$--dependence 
of the asymmetries could
be induced by the variation of the relative contributions 
of the various quark flavors to fragmentation.
The observation of a $z$--dependence of the asymmetries could also be
caused by hadrons in the semi--inclusive data sample that originate
from target fragmentation as opposed to current fragmentation, 
which is associated with the struck quark.
Furthermore, hadrons from non--partonic processes such as diffractive
interactions could play an important role in the semi--inclusive DIS
data sample \cite{th:frag-nonpartonic}. For example, at high
fractional energies $z$, it is possible that hadrons from exclusive
processes are misinterpreted as SIDIS hadrons. 

\begin{figure}[!htb]
  \centering
  \includegraphics[width=\columnwidth]{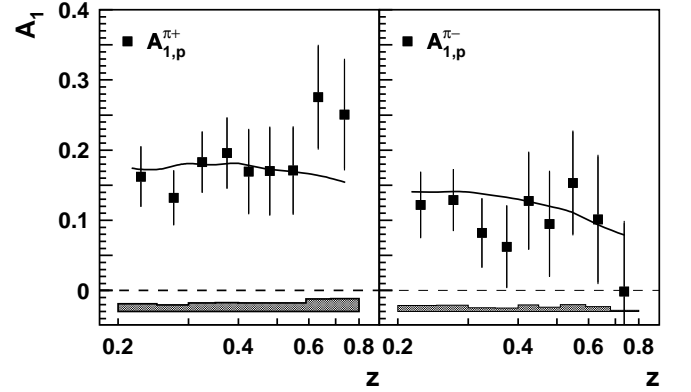} %% FINAL
  \caption{\label{fig:A1p-vsz}The semi--inclusive Born asymmetries for
    positive and negative pion production on the proton as a function
    of $z$.  The error bars indicate the statistical uncertainties and
    the error band represents the systematic uncertainties. The solid
    line is the z dependence from the Monte Carlo simulation of the 
    asymmetries.}
%%    The solid
%%    line is the result of a fit $A(z)=\mathrm{const}$. The reduced
%%    $\chi^2$--values of the fit are given in the figure.}
%%
%%  Plot macro: A1p-pi_vsz.kumac
%%
\end{figure}
To explore these possibilities, and to test the {\sc Jetset} fragmentation
model used here (see section~\ref{sec:quarkpol}) in the Monte Carlo
simulation of the scattering process, the semi--inclusive asymmetries
were extracted in bins of $z$. They were calculated with the same 
kinematical limits described above, except for the requirement on $x_F$, 
which is highly correlated with the limit on $z$ and was therefore discarded.  
Events were accepted over the range $0.023<x<0.6$.
The semi--inclusive pion asymmetries for the proton are shown in
Fig.~\ref{fig:A1p-vsz} together with a curve of the asymmetries
from the Monte Carlo simulation. The agreement between experimental 
and simulated data provide confirmation that the fragmentation process
is consistently modeled.
%%and for the deuteron in Fig.~\ref{fig:A1d-vsz}.A function
%%$A(z)=\mathrm{constant}$ was fitted to each asymmetry.  
%%Within their limited statistics, the data show
%%no indication of a significant systematic $z$--dependence of the
%%asymmetries.  While the reduced $\chi^2$--value of the fit to the
%%positive kaon asymmetry of $2.92$ seems high when viewed isolated from
%%the other asymmetries, there is no systematic trend visible in this
%%asymmetry data.  Furthermore, the global reduced $\chi^2$--value over
%%all six semi--inclusive asymmetries shown here is $1.0$.
                              %
%%\begin{figure}[htb]
%%  \centering
%%  \includegraphics[width=\columnwidth]{A1d-piK_born_vsz.eps} %% FINAL
%%  \caption{\label{fig:A1d-vsz}The semi--inclusive Born asymmetries for
%%    positive and negative pion and kaon production on the deuteron as
%%    a function of $z$.  See Fig.~\ref{fig:A1p-vsz} for details.}
%%
%%  Plot macro: A1d-piK_vsz.kumac
%%
%%\end{figure}

%%
%% ----------------------------------------------------------------------- %%
%%

\subsection{Systematic Uncertainties in $A_1$}
\label{sect:asyms:syst}

%%
%%  Systematic uncertainties
%%
Systematic uncertainties in the observed lepton--nucleon
asymmetries $A_\|^{(h)}$ arise from the systematic uncertainties in
the beam and target polarizations.  The unfolding of the observed
asymmetries also increases these uncertainties.  
A systematic uncertainty due to
the RICH hadron identification was estimated to be small as
the effect of neglecting the hadron misidentification (neglecting
the off--diagonal elements of $\omega$ appearing in Eq.~\eqref{eq:WpiK}) 
was found to be negligible. 
%%hadron misidentification is almost independent of the beam--target
%%spin configuration and therefore {\it as a multiplicative factor in both
%%numerator and denominator} cancels in the asymmetry.
Therefore, it was not
included in the semi--inclusive deuterium asymmetries.
\begin{table}[htbp]
  \caption{\label{tab:sysA1}The fractional systematic uncertainties on
    $A_1$ averaged over $x$.}
  \centering
  \renewcommand{\extrarowheight}{2pt}
  \begin{ruledtabular}
    \begin{tabular}{l|cc}
      Source              &Hydrogen data   &Deuterium data\\\hline
      Beam polarization      & $4.2\,\%$   & $2.3\,\%$\\
      Target polarization    & $5.1\,\%$   & $5.2\,\%$\\
      Azimuthal acc. (SIDIS) & $3.0\,\%$   & $3.1\,\%$\\
      QED rad.\ corr. (DIS)  & $2.0\,\%$   & $2.0\,\%$\\  %% FINAL
      QED rad.\ corr. (SIDIS)& $1.0\,\%$   & $1.0\,\%$\\
      Detector smearing      & $2.0\,\%$   & $2.0\,\%$\\
      $R$                    & $1.1\,\%$   & $1.1\,\%$ \\
      $g_2$                  & $0.6\,\%$   & $1.4\,\%$ \\
%% These numbers are the fractional uncertainties on the _Born_
%% asymmetries. They were calculated as the statistics weighted
%% average from the 96c2, 97c1, and the 98c1, 99b2, and 00c1
%% uncertainties. The uncertainties on beam and target pol are bigger
%% than those quoted in the exp section, because of the unfolding.
%% The script is ~/dqPaper/bin/getavgsyst-A.pl
%% data are
%% P:  dqPaper/Data/Asym/Prods/[9][76]*-xbins/A1.sys.born.txt 
%% D:  dqPaper/Data/Asym/Prods/[09][890]*-xbins/A1.sys.born.txt 
    \end{tabular}
  \end{ruledtabular}
  \renewcommand{\extrarowheight}{0pt} % default
\end{table}

Additional uncertainties arise due to the finite MC statistics, when
the corrections for detector smearing and QED radiation are applied.
They are included in the statistical error bars in the figures and are
listed in a separate column in the tables shown in the appendix.

In forming the photon--nucleon asymmetries $A_1^{(h)}$, systematic
uncertainties due to the parameterization of the ratio $R$ and the
neglect of the contribution from the second polarized structure
function $g_2$ were included \cite{hermes:marc,hermes:juergen}. The
relative systematic uncertainties are summarized in
Tab.~\ref{tab:sysA1}. The total systematic uncertainties on the
asymmetries are shown as the error bands in the figures.

The interpretation of the extracted asymmetries may be complicated
by contributions of pseudo--scalar mesons from the decay of exclusively 
produced vector mesons, mostly $\rho^0$'s producing charged pions. The 
geometric acceptance of the spectrometer is insufficient to identify and
separate these events, as typically only one of the decay mesons is detected. 
However, the fractional contributions of diffractive vector mesons to the
semi--inclusive yields  were estimated using a {\sc Pythia6} event 
generator~\cite{PYTHIA6} that has been tuned for the HERMES
kinematics~\cite{ourPYTHIA6}.
The results range from 2\%(3\%) at large $x$ to 10\%(6\%)
at small $x$ for pions(kaons) for both proton and deuteron targets. 
Although some data of limited precision for double--spin asymmetries in
$\rho ^0$ and $\phi$ production have been measured by 
HERMES~\cite{hermes:vectormesons}, no information is available on the effects
of target polarization on the angular distributions for the production
and decay of vector mesons. Therefore it was not possible at this time
to estimate the effect of the decay of exclusively produced vector mesons
on the semi--inclusive asymmetries. 
%%{\it no systematic uncertainty from contamination of semi-inclusive hadron
%%yields by decay products from exclusive production of vector mesons is
%%included in Tab.~\ref{tab:sysA1}. Some 
%%data of limited precision for double--spin asymmetries
%%in $\rho ^0$ and $\phi$ production have been measured by 
%%HERMES~\cite{hermes:vectormesons}. These data were used in a Monte Carlo
%%simulation tuned to measured exclusive meson production data to estimate
%%the range of possible contributions to the semi-inclusive hadron
%%asymmetries, that could arise from this contamination. Although corrections
%%for this effect could be comparable to those already discussed, the
%%uncertainty attendent to any such calculation precludes a reliable 
%%estimate of the magnitude of its contribution to the pion and kaon
%%asymmetries, until more detailed data becomes available on exclusive
%%vector meson double-spin asymmetries.}

%%
%%  Comments on acceptance
%%
The measurement of asymmetries as opposed to total cross sections has
the advantage that acceptance effects largely cancel.  Nevertheless,
the forward acceptance of the spectrometer restricts the topology of
the DIS electron and the SIDIS hadron in the final state.  It was
suggested \cite{th:bass} that a resulting cutoff in transverse hadron
momentum leads to a bias in the contributions of photon gluon fusion
(PGF) and QCD Compton (QCDC) processes to the total DIS cross section.
This bias could lead to an incorrect measurement of the polarizations
of the quarks using SIDIS asymmetries.  The momentum cut
($4\,\GeV<p<13.8\,\GeV$) on the coincident hadron tracks for particle
identification using the \v{C}erenkov/ RICH (cf.
section~\ref{sect:hadID}) could potentially introduce further bias.

\begin{figure}[htbp]
  \centering
  \includegraphics[width=\columnwidth,bb=1 128 569 567]{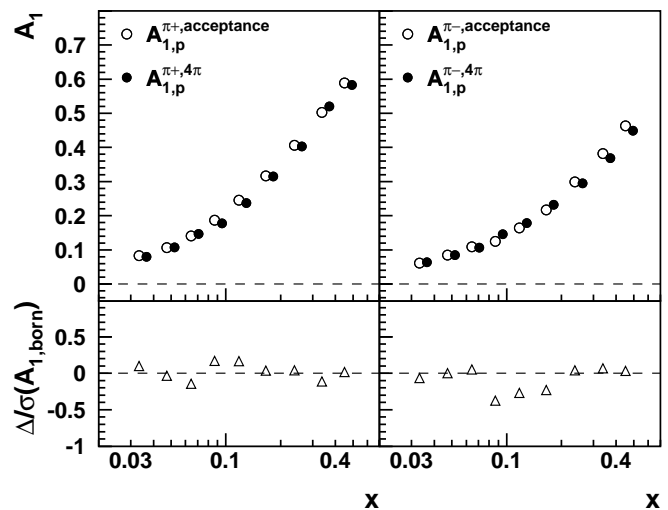}
  \caption{\label{fig:A1p-MC-4pi}Born level Monte Carlo asymmetries on
    the proton in the experimental acceptance and in $4\pi$.  The
    left hand plot compares the semi--inclusive asymmetries 
    $A_{1,p}^{\pi^+}$ and
    the right hand plot the semi--inclusive asymmetries
    $A_{1,p}^{\pi^-}$.  The asymmetries in the
    experimental acceptance also include the hadron momentum cut.
    For display purposes the full data points have been offset
    horizontally. The lower panels present the same data in the form
    of the difference in the asymmetries divided by the total 
    experimental uncertainty
    $\sigma_{born}$ in the corresponding measured Born asymmetry.}
%%
%%  Plot macro: A1-MC-4pi.kumac
%%
\end{figure}

Possible effects on the asymmetries due to the acceptance of the
HERMES spectrometer were studied with the HERMES Monte Carlo simulation.  
%%Born level data were generated and analyzed both in the
%%acceptance and in $4\pi$. In both data sets contributions to the cross
%%sections from the PGF and the QCDC process were found to be smaller
%%than $5\,\%$ and $15\,\%$, respectively.  The 
%%asymmetries were determined to be independent of the
%%experimental acceptance. 
%%The spectrometer samples the full ranges of $x$ and $z$,
%%although the geometrical acceptance limits the ranges of $\theta$ and
%%$\phi$.
Born level data were generated using a scenario in which
contributions to the cross sections from PGF and QCDC processes were
found to be smaller than $7\,\%$ and $18\,\%$ respectively. These
values were obtained in a scheme of cut offs against divergences in
the corresponding QCD matrix elements which require quark-antiquark
pairs to have masses $m_{q\overline{q}}>1\,\GeV$ and $>\, 0.005\, W^2$.
This is to be compared with the default values used in the purity
analysis of $m_{q\overline{q}}>2\,\GeV$ and $>\, 0.005\, W^2$. In this 
default case the contributions from PGF and QCDC processes were less
than $1.5\,\%$ and $3\,\%$ respectively. In the scenario 
employed for the acceptance study the
effect of the experimental acceptance was determined to be negligible
compared to the uncertainties in the data. The semi--inclusive $\pi^+$
and $\pi^-$ asymmetries in $4\pi$ and inside the acceptance are 
compared in Fig.~\ref{fig:A1p-MC-4pi}. Acceptance effects on the Born
level asymmetries are small and corrections are not necessary.

%%The simulated inclusive asymmetry on the deuteron and the
%%semi--inclusive $\pi^+$ asymmetry are shown in
%%The semi--inclusive $\pi^+$ 
%%and $\pi^-$ asymmetries for the proton are shown in
%%Fig.~\ref{fig:A1p-MC-4pi}.  The asymmetries in $4\pi$ and inside the
%%acceptance are in good agreement and no systematic trends are
%%observed. In conclusion, acceptance effects on the Born level
%%asymmetries are small and corrections are not necessary. 

%%
%% ====================== End of file 'sect-V.tex' ======================= %%
%%

%%% Local Variables: 
%%% mode: latex
%%% TeX-master: "dqpaper"
%%% End: 
 
%% ----------------------------------------------------------------------- %%
%%
%%   File        : sect-VI.tex
%%
%%   Description : Source file for section VI of the second delta-q paper
%%                 (DC number 42)
%%   
%%   Main Author : Juergen Wendland, Marc Beckmann
%%
%%   Date        : 19-Aug-2002
%%
%%   Remarks     : 
%%
%%   Modified    : 2003/05/01 jw
%%                 2003/07/25 jw: final data
%%
%% ----------------------------------------------------------------------- %%

\section{Quark Helicity Distributions}
\label{sect:deltaq}

%%
%% ----------------------------------------------------------------------- %%
%%

\subsection{Quark Polarizations and Quark Helicity Densities}
\label{sec:quarkpol}

A ``leading order'' analysis which included the PDF and QCDC processes
discussed in the previous section was used to compute
quark polarizations from the Born asymmetries.
The contribution of exclusively produced vector mesons is not 
distinguished in this extraction.
The analysis based on Eq.~\eqref{eq:dsigma} combines the Born
asymmetries in an over--constrained system of equations,
\begin{equation}
  \label{eq:A-ose}
  \vec{A}_1(x) = \left[\mathcal{N}(x)
    \mathcal{P}(x)\right] \vec{Q}(x) \, ,
\end{equation}
where the elements of the vector $\vec{A}_1(x)$ are the measured inclusive and
semi--inclusive Born asymmetries and the vector $\vec{Q}(x)$
contains the unknown quark polarizations.  The matrix $\mathcal{N}$ 
is the nuclear
mixing matrix that accounts for the probabilities for scattering off a
given nucleon in the deuteron nucleus and the nucleon's relative polarization.
The matrix $\mathcal{P}$ contains as elements the effective purities
for the proton and the neutron. These elements were obtained
by integrating Eq.~\eqref{eq:dsigma} over the range in $z$ and $Q^2$
giving
\begin{equation}
  \label{eq:a1siint}
  A_1^{h}(x)= 
  \sum_q {\cal P}_q^h(x) \, \frac{\Delta q(x)}{q(x)} \, ,
\end{equation}
where ${\cal P}_q^h(x)$ is now the effective spin--independent purity
\begin{equation}
  \label{eq:puridefint}
  {\cal P}_q^h(x) = \frac{e_q^2 \, q(x) \, \int_{0.2}^{0.8} D_q^h(z)
  \mathrm{d}z}
  {\sum_{q'} \, e_{q'}^2 \, q'(x) \, \int_{0.2}^{0.8} D_{q'}^h(z)
  \mathrm{d}z} \, .
\end{equation}
%%
%%  More details: Asymmetries
%%

The vector $\vec{A}_1(x)$ includes the inclusive and the
semi--inclusive pion asymmetries on the proton, and the inclusive
and the semi--inclusive pion and kaon asymmetries on the
deuteron:
\begin{equation}
  \vec{A}_1(x) = \left(A_{1,p}(x), \, A_{1,p}^{\pi^+}(x), \,
    \dots ,\, A_{1,d}^{K^-}(x)\right) \, .
\end{equation}
The semi--inclusive asymmetries of undifferentiated hadrons were not
included in the fit because they are largely redundant with the pion
and kaon asymmetries and thus do not improve the precision of the
results.

The nuclear mixing matrix $\mathcal{N}$ combines the proton and
neutron purities into effective proton and deuteron purities.  The
relation is trivial for the proton.  In the case of the deuteron
purities, the matrix takes into account the different probabilities
for scattering off the proton and the neutron as well as the effective
polarizations of the nucleons.  The probabilities were computed with
hadron multiplicities measured at HERMES and using the NMC
parameterization of $F_2$ \cite{Arneodo:1995cq}.  The D--state
admixture in the deuteron wave function of $(5 \pm 1)\,\%$
% \cite{ph:D-state:ericson,ph:D-state:lacombe,ph:D-state:desplanques,
%   ph:D-state:kotthoff,ph:D-state:machleidt,ph:D-state:tourreil,
%   ph:D-state:wiringa,ph:D-state:nagels1,ph:D-state:nagels2,
%   ph:D-state:zuilhof} 
\cite{ph:D-state:desplanques,ph:D-state:machleidt} 
leads to effective polarizations of the nucleons
in the deuteron of $p_{p,D}=p_{n,D}=(0.925 \pm 0.015)\,p_D$.

%%
%%  Purities
%%
The purities depend on the unpolarized quark densities and the
fragmentation functions.  The former have been measured with high precision
in a large number of unpolarized DIS experiments. The CTEQ5L parton
distributions \cite{pdf:cteq5} incorporating these data were used
in the purity determination.  
Much is known about fragmentation to mesons at collider energies.
However, the application of this information to fixed target energies
presents difficulties, especially regarding strange fragmentation,
which at lower energies no longer resembles that of lighter quarks. 
A recent treatment \cite{Kretzer:2002} using
extracted fragmentation functions in an analysis of inclusive and
semi--inclusive pion asymmetries for the proton demonstrates the
shortcomings of using the limited avaiable data base for fragmentation
functions. Hence the interpretaton of the present asymmetry data requires
a description of fragmentation that is constrained by meson multiplicities
measured at a similar energy. Such multiplicities within the HERMES 
acceptance are available, but those for kaons not yet 
available corrected to $4 \pi$ acceptance.
Hence the approach taken here was to tune the parameters of the LUND
string model implemented in the {\sc Jetset} 7.4 package
\cite{Sjostrand:1994yb} to fit HERMES multiplicities as observed in
the detector acceptance.  
%%The LUND model describes the fragmentation
%%from the current as well as the target fragmentation regions.  

In the LUND model, mesons are generated as the string connecting the diquark
remnant and the struck quark is stretched. 
Quark--antiquark pairs are generated at 
each breaking of the string. Even though the leading hadron
is often generated at one of the string breaks and not at the end,
the flavor composition of 
any hadron observed at substantial $z$ retains a
strong correlation  with the flavor of the struck quark. It is this
correlation which provides semi--inclusive flavor tagging.
Contrary to some speculations based on a misunderstanding of the
{\sc Jetset} code \cite{stringbreak}, this feature of the Lund
string model is independent of $W^2$ in lepton--nucleon scattering.
The quarks associated with either half of the string retain the 
information on the flavor of the struck quark. 
The LUND model has proved to
be a reliable widely accepted means of describing the fragmentation
process. 

The string breaking parameters of the LUND model were tuned to fit the
hadron multiplicities measured at HERMES in order to achieve a
description of the fragmentation process at HERMES energies
\cite{hermes:felix}.
\begin{figure}
  \centering
  \includegraphics[width=\columnwidth]{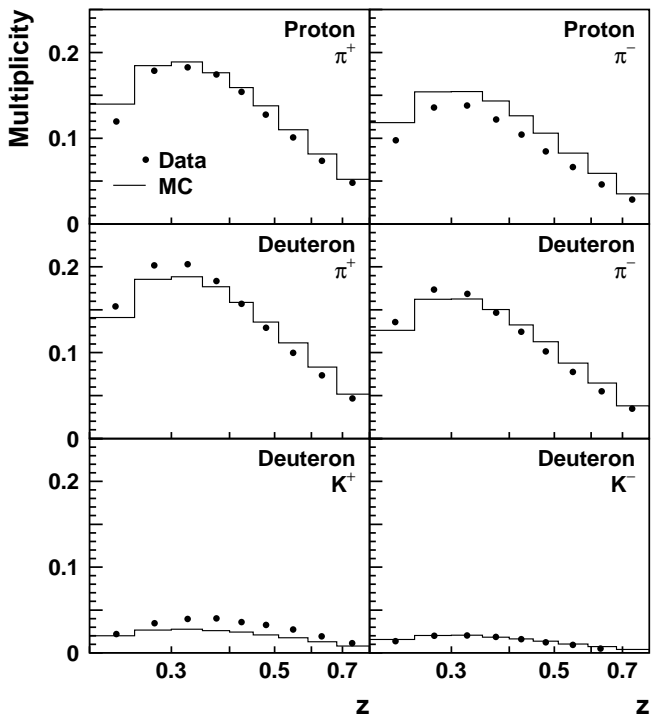}
  \caption{\label{fig:mult-mc-data}Multiplicities of charged pions and
    kaons in the HERMES acceptance compared with Monte Carlo data as
    function of $z$. Statistical uncertainties in the data are too small 
    to be visible. The multiplicities shown were not corrected for
    radiative and instrumental effects.}
%%
%%  Plot macro: Mult-MC-data.kumac
%%
\end{figure}
A comparison of the measured and the simulated hadron multiplicities
is shown in Fig.~\ref{fig:mult-mc-data}.  The tuned Monte Carlo
simulation reproduces the positive and negative pion multiplicities
and the negative kaon multiplicities while the simulated positive kaon
multiplicities are smaller than those measured.  A similar
disagreement is also reported by the EMC experiment
\cite{Arneodo:1985nf}.  
%%However, the results on the helicity densities
%%are fairly insensitive to the positive kaon asymmetry and hence this
%%disagreement has little impact in the fit.

\begin{figure*}[t]
  \centerline{
    \includegraphics[width=\textwidth]{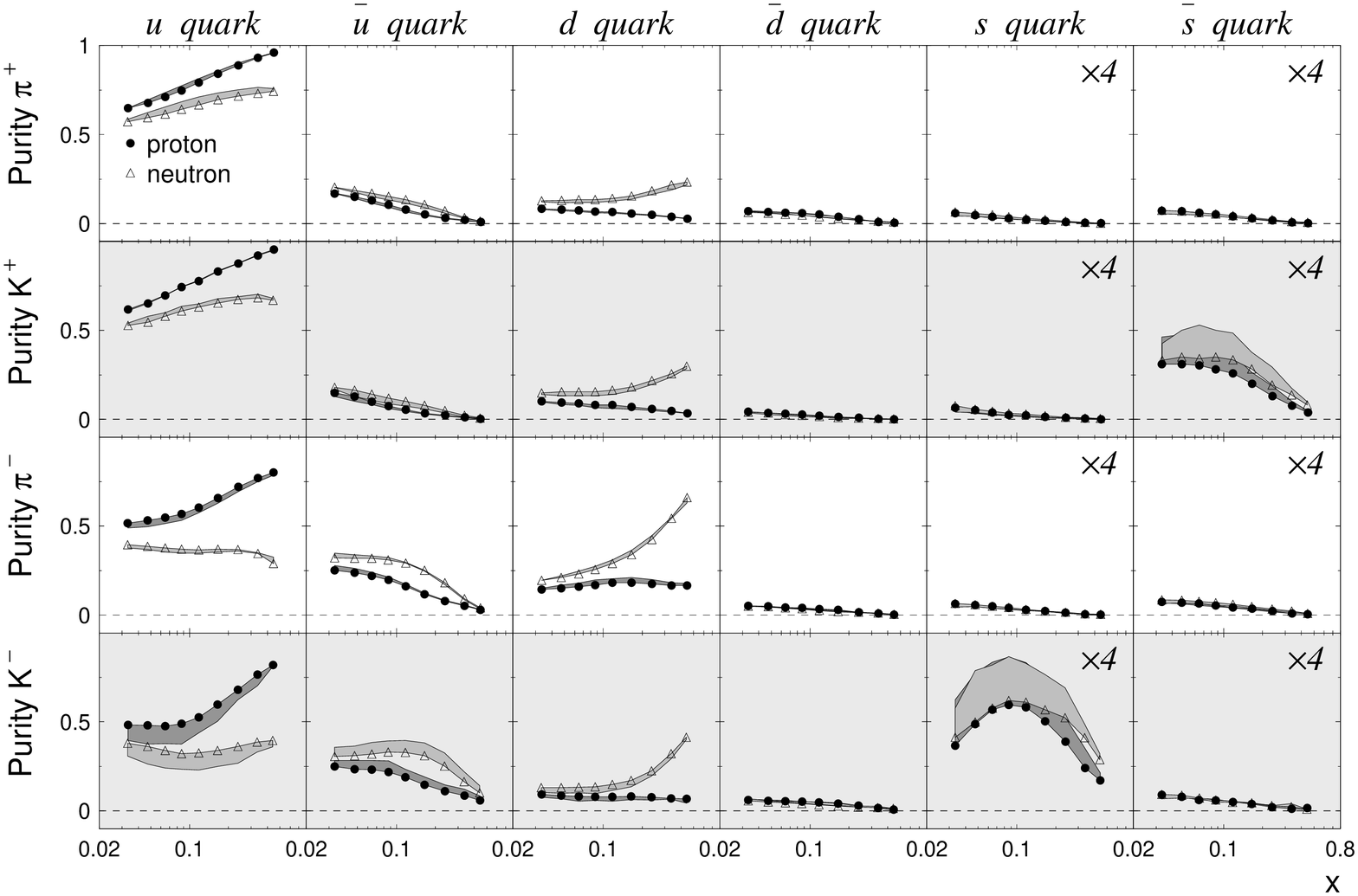}
    }
  \caption{\label{fig:puri_kaons}Purities for positive and negative
    pions and kaons on a proton and a free neutron target.  Each
    column corresponds to scattering off a certain quark flavor.  The
    shaded bands indicate the estimated systematic uncertainties due
    to the fragmentation parameters.  The estimate derives from a
    comparison of two parameter sets.  See text for details.}
%%
%%  Plot macro: Purities.kumac
%%
\end{figure*}

The purities were computed in each $x$--bin $i$ from the
described tuned Monte Carlo simulation of the entire scattering
process as
\begin{equation}
  \label{eq:puri-calc}
  P_q^h(x_i) = \frac{N_q^h(x_i)}{\sum_{q^\prime}N_{q^\prime}^h(x_i)} \, .
\end{equation}
In this expression, $N_q^h$ is the number of hadrons of type $h$
in bin $i$ passing all kinematic restrictions when a quark of flavor $q$ was
struck in the scattering process.  The purities include effects from
the acceptance of the spectrometer.  In
Fig.~\ref{fig:puri_kaons} the purities for the proton and the neutron
are shown.
It is evident from these plots that the $u$--quark dominates the
production of hadrons, due to its charge of $2/3$ and its large number
density $u(x)$ in the nucleon.  In particular the large contribution
by the $u$--quark to $\pi^+$ production from both proton and neutron
targets provides excellent sensitivity to the polarization of the
$u$--quark.  The $d$--quark becomes accessible through the production
of negative pions, which also separates the $\bar{u}$ and
$\bar{d}$ flavors.  More generally, contributions from the sea quarks
can be separated from the valence quarks through the charge of the
final state hadrons.  Finally the measurement of negative kaons is
sensitive to strange quarks and the anti--strange quark can be
accessed through positive kaons in the final state. However, large
uncertainties in the strange sea distributions are expected, because
the strange and anti--strange purities are small in comparison
to those for the other flavors. Some of the strange quark purities
vary rapidly with $Q^2$.

%% 
%%  That is it, here goes the second part of explaining the fit.
%%
The quark polarizations $[\Delta q/q](x)$ are obtained by solving
Eq.~\eqref{eq:A-ose}.  A combined fit was carried out for all $x$--bins
to account for the statistical correlations of the Born asymmetries
(cf.~Eq.~\eqref{eq:covABorn}).  Accordingly, the vectors $\vec{A}_1$
and $\mathcal{N}\,\mathcal{P}\vec{Q}$ are arranged such that they
contain consecutively their respective values in all $x$--bins,
\begin{equation}
  \vec{A}_1 = \left( \vec{A}_1(x_1), \,\vec{A}_1(x_2), \,
    \dots , \, \vec{A}_1(x_9)\right) \, ,
\end{equation}
and analogously for $\mathcal{N}\,\mathcal{P}\vec{Q}$. The
polarizations follow by minimizing
\begin{equation} \label{eq:dq_chi2min}
  \chi^2 = \left(\vec{A}_1-\mathcal{N}\mathcal{P}\,\vec{Q}\right)^T
  {\cal V}^{-1}_{A}
  \left(\vec{A}_1-\mathcal{N}\mathcal{P}\,\vec{Q} \right) \, .
\end{equation}
where ${\cal V}_{A}$ is the statistical covariance matrix
(Eq.~\eqref{eq:covABorn}) of the asymmetry vector $\vec{A}_1$. It
accounts for the correlations of the various inclusive and
semi--inclusive asymmetries as well as the inter--bin correlations.

The systematic uncertainties of the asymmetries were not included
in the calculation of $\chi^2$. The dominant contribution to
these uncertainties arises from the beam and target polarizations,
which affect the asymmetries in a nonlinear manner. It is natural
to linearly approximate these contributions as off-diagonal
inter-bin correlations in the systematic covariance matrix of the set
of asymmetries for all bins. However, when such a matrix  is 
included  in the fit based on linear recursion, the inaccuracies in
this linearization were found to introduce a significant bias in the
fit. Hence, the systematic uncertainties were excluded from the fit,
but were included in the propagation of all of the uncertainties in
the asymmetries into those on the results of the fit.

%%
%%  The constraint Delta sbar=0
%%
                                %
%%The statistical precision of the asymmetries provides little
%%constraint on the polarizations of the anti--strange quarks when it is
%%fitted together with all other quark flavor polarizations.  Therefore
%%the assumption
%%\begin{equation}
%%  \label{eq:constraint}
%%  \Delta \bar{s}(x) \equiv 0
%%\end{equation}
%%was imposed on the fit.  This constraint was chosen because some models
%%suggest that the helicity density of the strange sea is much larger
%%than that of the anti--strange sea
%%\cite{Wakamatsu:2002vf-1,Wakamatsu:2002vf-2,Cao:2003}.  
It was found that the data do not significantly  constrain 
$\Delta\bar{s}(x)$. The results presented here were extracted with the
constraint $\Delta \bar{s}(x) \equiv 0$. A comparison
of the fit using this constraint with a fit without assumptions on the
polarizations of the quark flavors showed that the constraint had
negligible impact on the final results for the unconstrained flavors
and their uncertainties.
In addition the resulting polarizations were found to be in good
agreement with the results of a fit under the assumption of a
symmetrically polarized strange sea $[\Delta{s}/s](x) =
[\Delta{\bar{s}}/\bar{s}](x)$.

\begin{figure}[ht]
  \centering
  \includegraphics[width=\columnwidth]{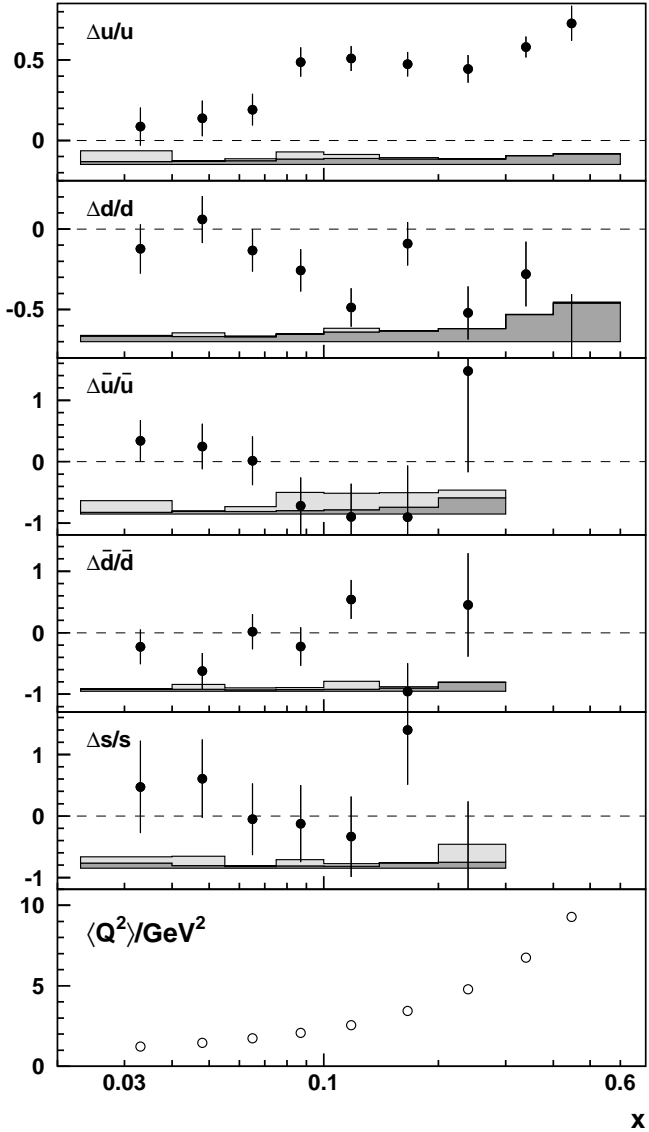} %% FINAL
  \caption{\label{fig:5Pfit}The quark polarizations in the
    5 parameter $\times$ 9 $x$--bins fit. 
%%    assuming zero anti--strange
%%    polarization, cf.~Eq.~\eqref{eq:constraint}. 
    The polarizations,
    shown as a function of $x$, were computed from the HERMES
    inclusive and semi--inclusive asymmetries. The error bars are the
    statistical uncertainties. The band represents the total
    systematic uncertainty, where the light gray area is the
    systematic error due to the uncertainties on the fragmentation
    model, and the dark gray area is from the contribution of the Born
    asymmetries.}
%%
%%  Plot macro: 6par-sbareq0-Pol.kumac
%%
\end{figure}

Assuming an unpolarized anti--strange sea the vector of polarizations
$\vec{Q}(x)$ in each $x$--bin is given by
\begin{equation} \label{eq:Qdef-std}
  \vec{Q}(x) = \left(\frac{\Delta u}{u}(x), \,
    \frac{\Delta d}{d}(x), \,
    \frac{\Delta \bar{u}}{\bar{u}}(x), \,
    \frac{\Delta \bar{d}}{\bar{d}}(x), \,
    \frac{\Delta s}{s}(x)\right) \, .
\end{equation}
As a further constraint the polarizations of the $\bar{u}$, $\bar{d}$,
and $s$--quarks were fixed at zero for values of $x>0.3$.  The effects
of this and of fixing the $\bar{s}$ polarization at zero were included
in the systematic error. The constraints reduced the number of free
parameters by fifteen, leaving $39$ parameters in the fit. 
The solution obtained by applying linear regression is
\begin{equation} \label{eq:Qconstrd.}
\vec{Q}=\left( {\cal P}^{T}_{ef}({\cal V}_{A})^{-1}{\cal P}_{ef}\right)^{-1}
{\cal P}^{T}_{ef}({\cal V}_{A})^{-1}\vec{A'_{1}},
\end{equation} 
where $\vec{A'_{1}}\equiv\vec{A}_{1}-{\cal N}{\cal P}
\vec{Q}_{fix}$, ${\cal P}_{ef}\equiv {\cal N}{\cal P}$, and
$\vec{Q}_{fix}$ is the set of constrained polarizations. The covariance 
matrix of the quark polarizations propagated from the Born asymmetries is
\begin{equation} \label{eq:Qcovar-constrd}
\begin{aligned}
{\cal V}(\vec{Q})=&\left[ \left( {\cal P}^{T}_{ef}({\cal V}_{A})^{-1}
{\cal P}_{ef}\right)^{-1}{\cal P}^{T}_{ef}({\cal V}_{A})^{-1} 
\right]{\cal V}_{A}^{tot} \\
&\times\left[ ({\cal V}_{A})^{-1}{\cal P}_{ef}
\left( {\cal P}^{T}_{ef}({\cal V}_{A})^{-1}
{\cal P}_{ef}\right)^{-1}\right],
\end{aligned}
\end{equation}
where the covariance matrix ${\cal V}_{A}^{tot}$ includes the statistical
and the systematic covariances, ${\cal V}_{A}^{tot}=
{\cal V}_{A}+{\cal V}_{A}^{sy}$. 
The resulting solution is shown in
Fig.~\ref{fig:5Pfit}.  The value of the $\chi^2/ndf$ of the fit is
$0.91$. The reasonable $\chi^2$ 
value confirms the consistency of the data set with the quark parton 
model formalism of section II.C. Removing the inclusive asymmetries
from the fit has only a small effect on the quark polarizations and 
their uncertainties.

The polarization of the $u$--quarks is positive in the measured range
of $x$ with the largest polarizations at high $x$ where the valence
quarks dominate.  The polarization of the $d$--quark is negative and
also reaches the largest (negative) polarizations in the range where
the valence quarks dominate.  The polarization of the light sea
flavors $\bar{u}$ and $\bar{d}$, and the polarization of the strange
sea are consistent with zero.  The values of $\chi^2/ndf$ for the zero
hypotheses are $7.4/7$, $11.2/7$, and $4.3/7$ for the $\bar{u}$, the
$\bar{d}$, and the $s$--quark, respectively.

%%
%%  Scaling, quark densities
%%
The quark polarizations in Fig.~\ref{fig:5Pfit} are presented at the
measured $Q^2$--values in each bin of $x$.  The $Q^2$--dependence is
predicted by QCD to be weak and the inclusive and semi--inclusive
asymmetries measured by HERMES (cf.~Figs.~\ref{fig:A1p},~\ref{fig:A1d}
and Ref.~\cite{hermes:dq1999}) and SMC \cite{smc:deltaq} at very
different average $Q^2$ show no significant $Q^2$--dependence when
compared to each other.  The quark polarizations $[\Delta q/q](x)$ are
thus assumed to be $Q^2$--independent.

\begin{figure}[ht]
  \centering
  \includegraphics[width=\columnwidth]{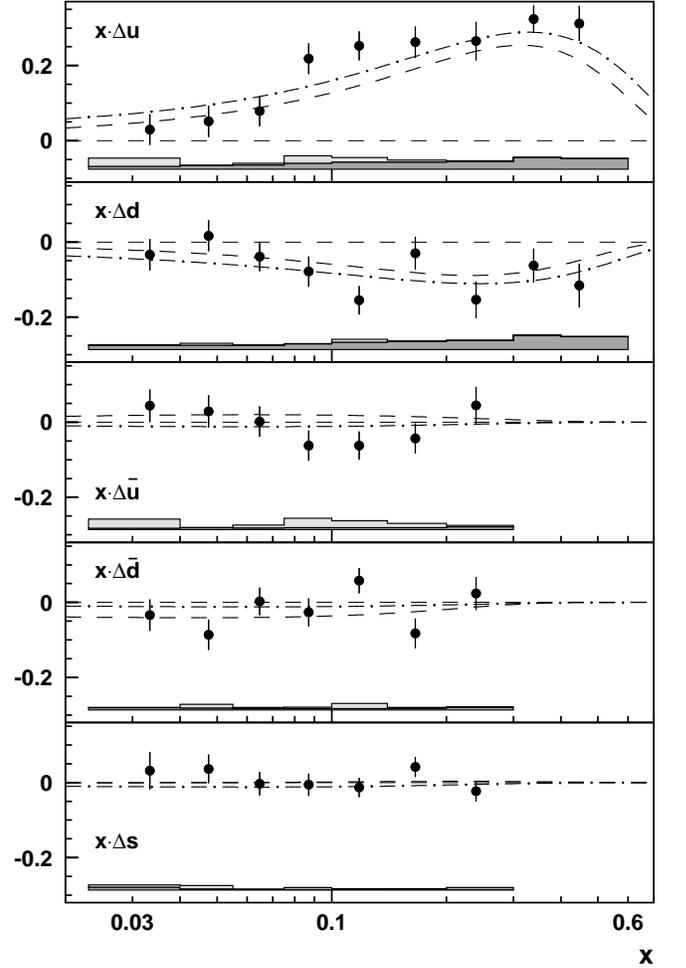} %% FINAL
  \caption{\label{fig:5Pfit-xdens} The quark helicity distributions
    $x\,\Delta q(x,Q_0^2)$ evaluated at a common value of
    $Q^2_0=2.5\,\GeV^2$ as a function of $x$.  The dashed line is the
    GRSV2000 parameterization (LO, valence scenario)
    \cite{pdf:grsv2000} scaled with $1/(1+R)$ and the dashed--dotted
    line is the Bl\"umlein--B\"ottcher (BB) parameterization (LO,
    scenario 1) \cite{Blumlein:2002be}.  See Fig.~\ref{fig:5Pfit} for
    explanations of the uncertainties shown.}
%%
%%  Plot macro: 6par-sbareq0-xDens.kumac
%%
\end{figure}

The quark helicity densities
$\Delta q(x,Q_0^2)$ are evaluated at a common $Q_0^2=2.5\,\GeV^2$
using the CTEQ5L unpolarized parton distributions. Because the CTEQ5L
compilation is based on fits to experimental data for $F_{2}(x)$, the 
relationship between $F_{2}(x)$ and $F_{1}(x)$ as given 
by Eq.~\eqref{eq:FL-contrib} is here taken into account. The factor 
$C_R \equiv (1+R)/(1+\gamma^2)$ connects CTEQ5L tabulations with
the parton distributions $q(x)$ required here. In the present analysis the
parameterization for $R(x,Q^2)$ given in Ref.~\cite{R1998} was used.
The results are presented in Fig.~\ref{fig:5Pfit-xdens}.
The data are compared with two parton helicity distributions
\cite{pdf:grsv2000,Blumlein:2002be} derived from LO fits to inclusive
data.  The GRSV2000 parameterization, which was fitted using the
assumption $R=0$, is shown with the scaling factor $1/(1+R)$ to match
the present analysis.  While in the Bl\"umlein--B\"ottcher (BB) 
analysis equal helicity densities for all sea flavors are assumed, in
the GRSV2000 ``valence fit'' a different assumption is used, which leads to
a breaking of flavor symmetry for the sea quark helicity 
densities.  In Tab.~\ref{tab:chi2models} the $\chi^2$--values of the
comparison of the measured densities with these parameterizations and
the zero hypothesis are given.
\begin{table}[htb]
  \caption{\label{tab:chi2models}Comparison of the measured quark
    helicity densities with the parameterizations obtained from LO QCD
    fits to inclusive data and with the zero hypothesis.  Listed are the
    reduced values $\chi^2/ndf$ for each hypothesis.}
  \centering
  \renewcommand{\extrarowheight}{2pt}
  \renewcommand{\tabcolsep}{3pt}
  \begin{ruledtabular}
    \begin{tabular}{l|ccccc} %% FINAL
      &$x\Delta u(x)$& 
      $x\Delta d(x)$&  
      $x\Delta \bar{u}(x)$&
      $x\Delta \bar{d}(x)$&
      $x\Delta s(x)$\\
      \hline
%      GRSV2000 std.&
%      $1.41$ & $1.12$ & $0.97$ & $1.34$ & $0.91$\\
      GRSV2000 val.&
      $1.45$ & $0.93$ & $1.54$ & $1.44$ & $0.60$\\
      BB (scenario 1)&
      $1.02$ & $ 1.06$ & $ 0.97$ & $ 1.32$ & $ 0.95$\\
      $x\Delta q(x)\equiv0$ 
      & $13.19$ & $2.50$  &$1.06$  & $1.60$  &$0.61$\\
      \hline
      $ndf$ & $9$ &$9$ &$7$ &$7$ &$7$\\
    \end{tabular}
  \end{ruledtabular}
  \renewcommand{\extrarowheight}{0pt} % default
  \renewcommand{\tabcolsep}{2pt}      % default (at least in revtex)
\end{table}
The measured densities are in good agreement with the
parameterizations.  The data slightly favor the BB parameterization of
the $u$-- and $\bar{u}$--flavors, while for the other flavors the
agreement with both parameterizations is equally good.  Within its
uncertainties the measured strange density is in agreement with the
very small non--zero values of the parameterizations as well as with
the zero hypothesis.

The total systematic uncertainties in the quark polarizations and the
quark helicity densities include contributions from the input
asymmetries and systematic uncertainties on the purities, which may
arise from the unpolarized parton distributions and the
fragmentation model.  Since the applied CTEQ5L PDFs \cite{pdf:cteq5}
are provided without uncertainties, no systematic uncertainty from this
source was assigned to the purities.

The uncertainties of the fragmentation model would be ideally calculated by
surveying the (unknown) $\chi^2$--surface of the space of {\sc Jetset}
parameters that were used to tune the Monte Carlo simulation.  At the
time of publication such a computationally intensive scan was not
available.  Instead the uncertainties were estimated by comparing the
purities obtained using the best tune of {\sc Jetset} parameters described
above to a parameter set which was derived earlier
\cite{hermes:dq1999}.  This earlier parameter set was also obtained
from a similar procedure of optimizing the agreement between simulated
and measured hadron multiplicities.  However, because of the lack of
hadron discrimation in a wide momentum range before the availability
of the RICH detector and limited available computer power, this earlier
parameter tune optimized only three {\sc Jetset}
parameters, while in the current tune \cite{hermes:felix} eleven
parameters were optimized from their default settings.  The
differences in the resulting purities from using these two different
tunes of {\sc Jetset}  parameters are shown as the shaded bands in
Fig.~\ref{fig:puri_kaons}.

The contributions from this systematic uncertainty estimate on the
purities to the total systematic uncertainties of the resulting
helicity densities and quark polarizations are shown as the light
shaded bands in Figs.~\ref{fig:5Pfit}, \ref{fig:5Pfit-xdens}, and
Fig.~\ref{fig:ub-db}.  In the case of the $u$ and $d$ quark, the
resulting uncertainty contributions due to the fragmentation model are
small compared with those related to the systematic uncertainties on
the asymmetries.  They are of equal or larger size in the case of the
sea quarks and they dominate in case of the light sea quark helicity
difference discussed in the following subsection.

%%
%% ------------------------ Results: dubar-ddbar ------------------------- %%
%%

\subsection{Isospin Asymmetry in Helicity Densities of Light Sea Quarks}

In the unpolarized sector, the breaking of flavor symmetry for the
light sea quarks $(\bar{d}(x) - \bar{u}(x) > 0)$ and consequently the
violation of the Gottfried sum rule is experimentally well established
\cite{ub-db:NA51, ub-db:HERMES, ub-db:E866} and is described by various
non--perturbative models.  Two comprehensive reviews of these models
can be found in \cite{Steffens-Thomas:1997,Kumano:1998}.
                                %
%%
%%  Antje:
%%
Such models also predict a flavor asymmetry in the light sea
helicity densities.  
Sizeable asymmetries of similar magnitude are predicted by
the chiral quark soliton model ($\chi$QSM)
\cite{Christov:1996vm,Dressler:1999zg, Diakonov:1996sr,
  Diakonov:1997vc,Wakamatsu:2002vf-1,Wakamatsu:2002vf-2}, which is an
effective theory where baryons appear as soliton solutions of the
chiral Lagrangian, a statistical model \cite{Bourrely:2001du} that
describes the nucleon as a gas of massless partons, models based on
the meson cloud picture \cite{Fries:2002um,Cao:2003} as well as the
chiral chromodielectric model, which is a bag--like confinement model
\cite{Barone:2003}. 
The meson cloud model published in \cite{Cao:2003} deviates most from
all other mentioned models as it predicts an asymmetry which is
smaller in magnitude but has the opposite sign $\Delta\bar{u}(x) -
\Delta\bar{d}(x) < 0$ to the other models, which all predict a positive
value of this quantity.

\begin{figure}[ht]
  \centering
    \includegraphics[width=\columnwidth]{%
%%      x_Dubar-Ddbar.CQSM.eps}%% Just the Bochum CQSM
      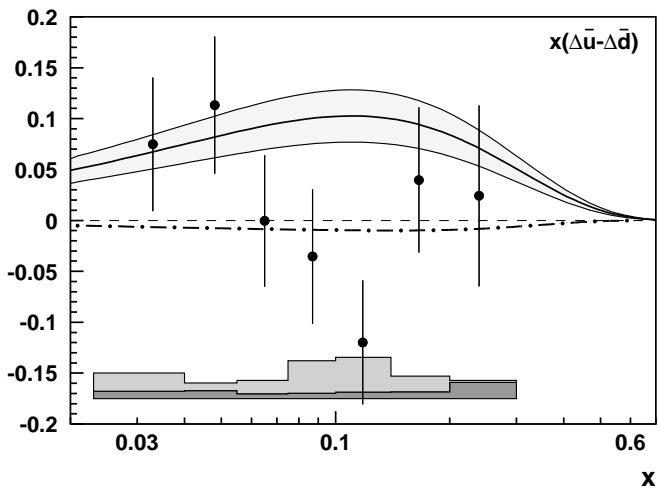}%% The Bochum CQSM and the meson cloud model
%%      x_Dubar-Ddbar.CQSM.Bourelly.eps}%% + the Bourrely stat. model
  \caption{\label{fig:ub-db} The flavor asymmetry 
    in the helicity densities of the light
    sea evaluated at $Q_0^2=2.5\,\GeV^2$.  The data are
    compared with predictions in the $\chi$QSM \cite{Dressler:1999zg}
    and a meson cloud model \cite{Cao:2003}.  The solid line with the
    surrounding shaded band show the $\chi$QSM prediction together
    with its $\pm 1\,\sigma$ uncertainties while the dash--dotted line
    shows the prediction in the meson cloud model.  The uncertainties
    in the data are presented as in Fig.~\ref{fig:5Pfit}.}
%%
%%  Plot macro: x_dubar-ddbar.kumac
%%
\end{figure}

The measured (semi--)inclusive asymmetries discussed above were used
in a modified fit to compute this flavor asymmetry.  In this fit the
parameter $[\Delta \bar{u}/\bar{u}](x)$ was replaced by $[(\Delta
\bar{u}-\Delta\bar{d})/(\bar{u}-\bar{d})](x)$ and the system of linear
equations \eqref{eq:A-ose} was solved for the following vector
$\vec{Q}^\prime(x)$ of quark polarizations:
\begin{multline} \label{eq:Qdef-cqsm}
  \vec{Q}^\prime(x) = \\
  \left(\frac{\Delta u}{u}(x), \,
    \frac{\Delta d}{d}(x), \,
    \frac{\Delta \bar{u}-\Delta \bar{d}}{\bar{u}-\bar{d}}(x), \,
    \frac{\Delta \bar{d}}{\bar{d}}(x), \,
    \frac{\Delta s}{s}(x)\right) \, .
\end{multline}
The flavor asymmetry $x\,[\Delta \bar{u}(x)-\Delta\bar{d}(x)]$ is
presented in Fig.~\ref{fig:ub-db}.  For comparison the same quantities
calculated in the $\chi$QSM \cite{Dressler:1999zg} and in a meson
cloud model \cite{Cao:2003} are shown.  These models were chosen for
presentation as they indicate the most positive and most negative
predictions.  All other cited model predictions yield similar
$x$--dependences and positive values $\Delta\bar{u}(x) -
\Delta\bar{d}(x) > 0$.  Because of the close similarity of these model
curves when plotted at the scale of Fig.~\ref{fig:ub-db} they
are not displayed in this figure.

The value of the $\chi^2/ndf$ for the symmetry hypothesis
$\Delta\bar{u}(x) = \Delta\bar{d}(x)$ is $7.7/7$.  The 
$\chi^2/ndf$ values of the comparisons with the predictions shown for the
$\chi$QSM and for the meson cloud model are $17.6/7$ and $8.1/7$,
respectively, where the value for the $\chi$QSM takes into account its
uncertainty.  This analysis of the HERMES data therefore favors a 
symmetric polarized
light flavor sea and exclude the prediction of the $\chi$QSM at the 
97\% confidence level.
%% within the context of the LO analysis used here.
%%
%% chi2 values are FINAL
%%
                                %

%%
%% ====================== End of file 'sect-VI.tex' ====================== %%
%%

%%% Local Variables: 
%%% mode: latex
%%% TeX-master: "dqpaper"
%%% End: 
%% ----------------------------------------------------------------------- %%
%%
%%   File        : sect-VII.tex
%%
%%   Description : Source file for section VII of the second delta-q paper
%%                 (DC number 42)
%%   
%%   Main Author : Marc Beckmann
%%
%%   Date        : 19-Aug-2002
%%
%%   Remarks     : 
%%
%%   Modified    : ... a lot
%%                 18-Jun-2003 - Change plot and text to present
%%                               x*delta_s instead of the strange quark
%%                               polarization //MB
%%
%% ----------------------------------------------------------------------- %%

\section{$\Delta s$ from the Isoscalar Extraction Method}
\label{sect:ds}

%%
%%  Introduction / motivation
%%
The strange quark polarization in the nucleon is one of the most
interesting quantities that can be determined from SIDIS
data.  Kaon asymmetries provide the largest sensitivity to the strange
quark polarization because the kaon contains a valence strange
(anti--)quark.  Unfortunately, experimental data on separate
favored and disfavored fragmentation of strange quarks into kaons are
scarce at best in the kinematic region of HERMES, thereby
introducing large systematic uncertainties on the strange quark
purities (cf.~Fig.~\ref{fig:puri_kaons}).  This section presents an
alternative approach for the extraction of the strange quark
polarization, which uses only data on 
the asymmetry of the total ($K^+
+ K^-$) charged kaon flux for a deuterium target, and 
which does not rely on a Monte Carlo
model of the fragmentation process.

%%
%%  Formalism
%%
The total strange quark helicity density $\Delta S(x) \equiv \Delta
s(x) + \Delta\bar{s}(x)$ carries no isospin.  It can hence be
extracted from scattering data off a deuterium target alone, which is
isoscalar.  Furthermore, the fragmentation functions $D_{q=u,d,s}^{K^+
  + K^-}$ for the total kaon flux were measured at $e^+ e^-$ collider
experiments with satisfactory precision.

The analysis is performed as a two component extraction of the
non--strange and total strange quark polarizations, $[\Delta Q/Q](x)$
and $[\Delta S/S](x)$.  Here, $Q(x) \equiv u(x) + \bar{u}(x) + d(x) +
\bar{d}(x)$.  Only two measured asymmetries, the inclusive $A_{1,d}$
and the semi--inclusive $A_{1,d}^{K^+ + K^-}$ asymmetries on the
deuterium target, are used for the extraction. The same purity
approach detailed in sections~\ref{sect:dis} and \ref{sect:deltaq} is
used, except in the present case the purity matrix ${\cal P}(x)$
consists of a $2 \times 2$ matrix only:
\begin{equation}
  \label{eq:puri_ds}
  {\cal P}(x) = \left(
  \begin{array}{cc}
    P_{Q}(x)              &  P_{S}(x)             \\
    P_{Q}^{K^+ + K^-}(x)  &  P_{S}^{K^+ + K^-}(x)
  \end{array}
  \right) \, .
\end{equation}
Due to row--wise unitarity, there are only two 
independent elements in this matrix.

Based on the wealth of $e^+ e^-$ collider data, parameterizations of
the total charged kaon fragmentation functions $D_{q+\bar{q}}^{K^+ +
  K^-}\!(z,Q^2)$ are available as a function of $z$ and $Q^2$
\cite{frag:bkk:1995,frag:kkp:2000}.  
%%The universality of these parameterizations 
%%was shown in \cite{frag:kkp:2001} by the good agreement with hadron
%%multiplicities measured in $p \bar{p}$ or $e p$ collisions.  
The light quarks $u, d, s$ are assumed to be massless in the analyses
of the $e^+e^-$ collider data \cite{frag:bkk:1995,frag:kkp:2000}.
This is justified in view of the high center--of--mass energies
available in these experiments.  However, this approximation is not
valid for the lower energies available at fixed target experiments.
Here, fragmentation of non--strange quarks into kaons is suppressed by
an additional factor $\lambda_s < 1$, in order to account for the
lower probability to generate an $s\bar{s}$ quark pair rather than a
lighter $u\bar{u}$ pair.  A value of $\lambda_s =
0.2$ was assumed for the HERMES kinematic domain, based on results
from deep--inelastic muon and neutrino scattering experiments at
similar center--of--mass energies \cite{BEBC:1983,EMC:D_s,E665:1994}.
This approach allows one to compute $P_{Q}^{K^+ + K^-}(x)$ 
directly according to
Eq.~\eqref{eq:puridef} instead of using a Monte Carlo model of the
fragmentation process.
In this analysis, the elements of the $2 \times 2$ purity matrix were
calculated at the central kinematic values $\langle x \rangle$ and
$\langle Q^2 \rangle$ in each $x$--bin, using the LO parameterizations
of the kaon fragmentation functions from Ref.~\cite{frag:bkk:1995} and
the CTEQ5L unpolarized parton densities from Ref.~\cite{pdf:cteq5}.

By computing the purities from parameterizations of the fragmentation
functions instead of from a Monte Carlo model, one does not take into
account acceptance effects on the measured asymmetries.  However, as
explained in section~\ref{sect:asyms}, the effects of the limited polar
acceptance of the HERMES spectrometer on selected semi-inclusive 
asymmetries are negligible. The
asymmetries were corrected for azimuthal acceptance effects.  The
precision of the method presented in this section is limited by the
statistical uncertainties of the measured kaon asymmetries, and the
knowledge of the strangeness suppression factor $\lambda_s$.

%%
%%  Results
%%
\begin{figure}[ht]
  \centerline{
    \includegraphics[width=\columnwidth]{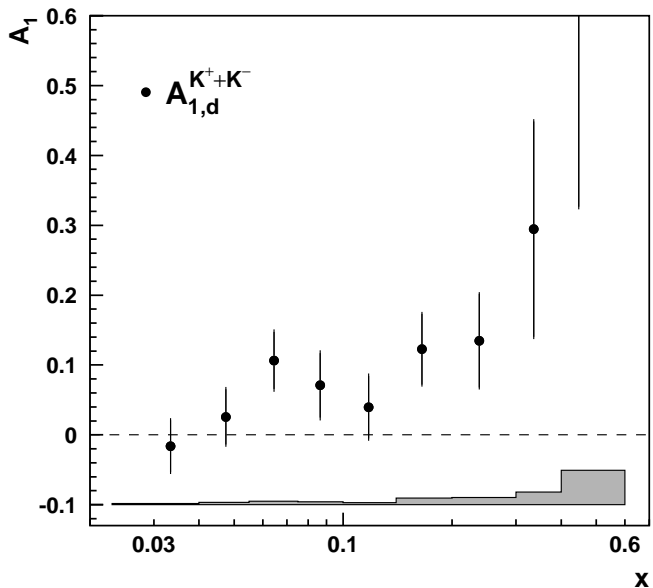}
    }
  \caption{\label{fig:a1dkchsum} The semi--inclusive Born level asymmetry
    $A_{1,d}^{K^+ + K^-}$ for the total charged kaon flux on a
    deuterium target, corrected for instrumental smearing and QED
    radiative effects.  The uncertainties are presented analogously to
    Fig.~\ref{fig:A1p}.}
%%
%%  Plot macro: A1d-Kchsum.kumac
%%
\end{figure}

\begin{figure}[htbp]
  \centerline{ 
    \includegraphics[width=\columnwidth]{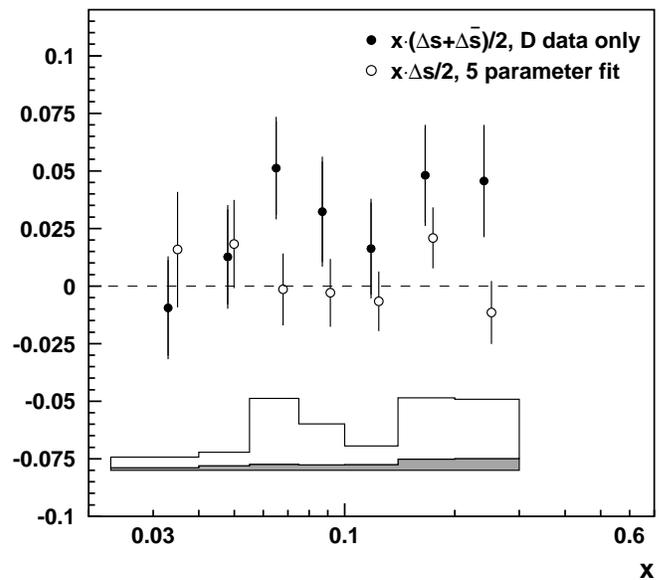}
    }
  \caption{\label{fig:ds} The average strange quark helicity density
    $x\cdot[\Delta s(x)+\Delta\bar{s}(x)]/2$ from the isoscalar
    extraction method (full points).  For comparison the open symbols
    denote the results from a five parameter fit (see text), which are
    offset horizontally for presentation.  The band in the bottom part
    gives the total systematic uncertainty on the results from the
    isoscalar extraction.  The dark shaded area corresponds to the
    uncertainty from the input asymmetries, and the open part relates
    to the uncertainty in $\lambda_s$.}
\end{figure}

Fig.~\ref{fig:a1dkchsum} shows the semi--inclusive $K^+ + K^-$
asymmetry on a deuterium target, corrected for effects of 
QED radiation and detector
smearing, and azimuthal detector acceptance.  The same kinematic limits
and extraction method as described in section~\ref{sect:asyms} were
applied.  The numerical values can be found in
Tab.~\ref{tab:A1d_results} in App. B.  Combined with the
inclusive asymmetry $A_{1,d}$ shown in Fig.~\ref{fig:A1d}, the total
strange quark polarization $[\Delta S/S](x)$ was extracted from this
data set.  The average strange quark helicity density
\begin{equation}
  \label{eq:dsavg}
  \begin{split}
    \frac{1}{2} \left[\Delta s(x)+\Delta\bar{s}(x)\right] &=
    \frac{\Delta S}{S}(x) \cdot \frac{s(x)+\bar{s}(x)}{2}\\
%    &\stackrel{s(x)=\bar{s}(x)}{=}
    &=\frac{\Delta S}{S}(x) \cdot s(x)
  \end{split}
\end{equation}
is obtained by multiplication with the unpolarized strange quark
density $s(x) = \bar{s}(x)$.  The second equality in
Eq.~\eqref{eq:dsavg} holds because the unpolarized strange quark
density is symmetric by construction for the quark and anti--quark in
the parameterization employed \cite{pdf:cteq5}.  The resulting average
strange quark helicity density is shown in Fig.~\ref{fig:ds}.
In analogy to the determination of the polarizations of the sea
flavors in the previous section, this extraction was restricted to
the range $0.023 \leq x \leq 0.3$.

The uncertainty in the strangeness suppression factor
contributes the dominant part of the systematic uncertainty on the
extracted values of $x\,\Delta S(x)$.  It was estimated by varying
$\lambda_s$ in the range $0.15 \leq \lambda_s \leq 0.3$
covered by the
experimental results \cite{BEBC:1983,EMC:D_s,E665:1994}.  This
contribution to the total systematic uncertainty is shown as the unshaded
band in Fig.~\ref{fig:ds}.  The remaining contribution shown as the
shaded band arises from the systematic uncertainties on the Born level
asymmetries treated in the previous section.

The results from the five flavor fit to
the full data set on both the proton and deuteron targets are also
shown in Fig.~\ref{fig:ds}.  
%%The modified fit was obtained by
%%substituting Eq.~\eqref{eq:constraint} by the alternative constraint
%%\begin{equation}
%%  \label{eq:s-eq-sbar}
%%  \Delta s(x) \equiv \Delta \bar{s}(x)
%%\end{equation}
%%to be consistent with Eq.~\eqref{eq:dsavg}.  The resulting impact on
%%the strange quark polarizations in the five parameter fit is
%%negligible and is well contained within the systematic uncertainties.
%%
%%  Plot macro: isods-xDs.kumac
%%
A comparision of the first moments of the helicity densities in the
measured $x$ region provides a good measure of the agreement between the
two methods. It should be noted that these two analyses use the same 
input data. For the five flavor fit, from Tab.~\ref{tab:moments},
$\Delta s(5par)=0.028\pm 0.033\pm 0.009$. This is to be compared
with $\Delta S(iso)=0.129\pm 0.042\pm 0.129$ where the systematic 
uncertainty is dominated by that in $\lambda_s$ cited above. 
Considering statistical and systematic uncertainties, both
results are consistent with zero. (After 
this analysis was completed it was learned that the charged kaon 
multiplicities were  consistent with a value somewhat larger than
this range.)
The helicity densities from the isoscalar method and from the full five flavor
separation agree within their uncertainties.  The isoscalar
method yields results that are consistent with a vanishing total
strange quark helicity density $[\Delta s(x) + \Delta\bar{s}(x)]=0$.
In analogy to the previous section, the $\chi^2/ndf$ value for
this hypothesis is $5.3/7$ for the results from the
isoscalar extraction method. Neither the isoscalar nor the five flavor 
measurement provides any indication of a negative polarization 
of the strange sea.
%%with the constraint Eq.~\eqref{eq:s-eq-sbar}, respectively. \\
%%The reduced statistical uncertainties for the isoscalar method reflect
%%the smaller number of degrees of freedom in the extraction.  In
%%contrast to the results from section~\ref{sect:deltaq}, the isoscalar
%%extraction method is independent of a Monte Carlo modeling of the
%%fragmentation process. 
%%The consistency with the results from the
%%five parameter fit thereby offers support for the JETSET fragmentation
%%model as tuned and applied at HERMES.

%%
%% ====================== End of file 'sect-VII.tex' ===================== %%
%%

%%% Local Variables: 
%%% mode: latex
%%% TeX-master: "dqpaper"
%%% End: 
%% ----------------------------------------------------------------------- %%
%%
%%   File        : sect-VIII.tex
%%
%%   Description : Source file for section VIII of the second delta-q paper
%%                 (DC number 42)
%%   
%%   Main Author : Marc Beckmann
%%
%%   Date        : 19-Aug-2002
%%
%%   Remarks     : 
%%
%%   Modified    : 18-Jan-2003
%%                 May 8, 2003 (jw)
%%
%% ----------------------------------------------------------------------- %%

\section{Moments of Helicity Distributions}
\label{sect:moments}

%%
%% ----------------------------------------------------------------------- %%
%%

\subsection{Determination of the Moments}
\label{sect:moments:measured}

%%
%%  The moments in the measured range
%%
In the measured region, the $n$--th moment of the helicity distributions
is given by
\begin{equation} \label{eq:momdef}
  \Delta^{(n)}q(Q_0^2) = \sum_i \left[ \frac{\Delta q}{q}(x_i)
  \int_{\xi_i}^{\xi_{i+1}} \!\!\!dx\: x^{n-1} \: q(x,Q_0^2) \right] \, ,
\end{equation}
where the quark polarization $[\Delta q/q](x_i)$ is assumed to be
constant in each $x$--bin $i$ $[\xi_i, \xi_{i+1}]$, and $q(x,Q_0^2)$
is obtained from the CTEQ5L LO parameterization of the unpolarized
quark densities from Ref.~\cite{pdf:cteq5} at a fixed scale of $Q_0^2
= 2.5\:\text{GeV}^2$.  For simplicity, the first moment $\Delta^{(1)}
q(Q_0^2)$ is denoted $\Delta q(Q_0^2)$ hereafter.

The simultaneous fit of all quark flavors in all $x$--bins yields the
correlations between different $x$--bins as well as between different
quark flavors, which are taken into account for the computation of the
uncertainties on the moments.  Specifically, from the solution of
Eq.~\eqref{eq:dq_chi2min} one obtains the statistical and systematic
covariance matrices ${\cal V}_Q^{\mathrm{stat}}$ and ${\cal
  V}_Q^{\mathrm{sys}}$ for the quark polarizations.  The statistical
and systematic uncertainties on the moments,
$\delta^{\,\mathrm{stat}}_{\Delta q}$ and
$\delta^{\,\mathrm{sys}}_{\Delta q}$, are obtained from these
covariance matrices as
\begin{multline}
  \label{eq:momerr}
  \left(\delta^{\,\mathrm{stat/sys}}_{\Delta q}\right)^2 =
  \sum_{i,j} 
  {\cal V}_Q^{\mathrm{stat/sys}}\! \left(
  {\textstyle\frac{\Delta q}{q}(x_i),\frac{\Delta q}{q}(x_j)}
  \right) 
  \\
  \times
  \int_{\xi_i}^{\xi_{i+1}} \!\!\!dx\: x^{n-1} \: q(x,Q_0^2) 
  \int_{\xi_j}^{\xi_{j+1}} \!\!\!dx\: x^{n-1} \: q(x,Q_0^2)\, ,
\end{multline}
where $i$ and $j$ run over all $x$--bins.  By the inclusion of the
correlations between different $x$--bins the total uncertainties on
the moments are reduced by 22 -- 39\,\%, depending on the quark
flavor.  This can be understood as a partial compensation of the
uncertainty enlargement on the asymmetries due to inter--bin event
migration in the unfolding procedure
(cf.~section~\ref{sect:asyms:radcor} and App.~\ref{sect:app:radcorr}):
The net migration of events from $x$--bin $i$ into $x$--bin $j$ causes
these bins to be anticorrelated and hence reduces the resulting
uncertainty in Eq.~\eqref{eq:momerr}.

%%
%%  The singlet and non-singlet flavor combinations
%%
Theoretical predictions are often made in terms of the first 
moments of the isospin
singlet, the isovector, and the octet combinations,
\begin{align}
  \Delta \Sigma &= \Delta u + \Delta \bar{u} +
  \Delta d + \Delta \bar{d} + \Delta s + \Delta \bar{s} \, , \\
  \Delta q_3 &= \Delta u + \Delta \bar{u} -
  (\Delta d + \Delta \bar{d}) \, , \\
  \Delta q_8 &= \Delta u + \Delta \bar{u} + \Delta d + \Delta \bar{d}
  - 2( \Delta s + \Delta \bar{s}) \, .
\end{align}
The resulting values in the measured range
$0.023 \leq x \leq 0.6$ are tabulated together with the first and
second moments of the individual helicity densities in
Tab.~\ref{tab:moments}.  For the sea flavors, the measured range is
$0.023 \leq x \leq 0.3$.  The contributions from $0.3 < x \leq 0.6$ to
these moments have been fixed at zero.  The valence distributions are
obtained as $\Delta u_{\mathrm{v}} \equiv \Delta u - \Delta\bar{u}$ and
$\Delta d_{\mathrm{v}} \equiv \Delta d - \Delta\bar{d}$.  In the
computation of the uncertainties of these flavor combinations the
correlations between the individual helicity densities are taken into
account.  The corresponding statistical and systematic correlations
for the first and second moments are listed in
Tabs.~\ref{tab:first_mom_correlations} and
\ref{tab:second_mom_correlations} in the Appendix.

\begin{table}
  \renewcommand{\arraystretch}{1.1}
  \caption{\label{tab:moments} First and second moments of various helicity
    distributions in the measured range at a scale of
    \mbox{$Q_0^2 = 2.5\:\text{GeV}^2$.}}
%%
%%  Include external file, output by the program ``moments'':
%%
  \include{moments}
\end{table}

%%
%% ----------------------------------------------------------------------- %%
%%

\subsection{Comments on Extrapolations outside the Measured Range}
\label{sect:moments:extrapol}

%%
%%  The extrapolations...
%%
In order to compute the full moments for the complete $x$--range $0
\leq x \leq 1$, the contributions outside the measured region would
have to
be estimated.  For this, one has to rely on models in the kinematic
regions where no data exist.  In particular, the extrapolation towards
$x = 0$ is very problematic with many competing models making
contradictory predictions
\cite{th:close-roberts:2,Bass:small-x,Ball:small-x}.  This may be
compared to the situation regarding the behavior of the unpolarized
structure function $F_2(x,Q^2)$ towards low values of $x$ before the
HERA data became available (see for instance \cite{hera:low-x}).

While in our earlier publication \cite{hermes:dq1999} we chose one
particular model (based on Regge theory), we do not pursue this
approach any longer.  The reason for this is the very strong model
dependence which cannot be quantified reliably in terms of an
associated systematic uncertainty, as some of the models yield
diverging integrals in the low--$x$ range \cite{e154:g1n}.

To illustrate this more clearly, we performed a study of the
robustness of a global QCD fit similar to \cite{Blumlein:2002be} (LO,
scenario 1), which does include preliminary HERMES inclusive data taken on a
polarized deuterium target \cite{hermes:g1d-prelim}.  In this study,
the fit was repeated with artificially offset values for the
parameters $\eta^\prime_{u_{\mathrm{v}}} = 0.730$ and
$\eta^\prime_{d_{\mathrm{v}}} = -0.270$ as compared to their default
settings $\eta_{u_{\mathrm{v}}} = 0.926$ and $\eta_{d_{\mathrm{v}}} =
-0.341$, obtained from the constants $F$ and $D$ measured in weak
neutron and hyperon $\beta$--decays.  The parameterizations for the
helicity densities in the QCD fit are constructed such that the
parameters $\eta_{u_{\mathrm{v}}}$ and $\eta_{d_{\mathrm{v}}}$ give
the first moments of the helicity valence quark distributions,
$\eta_{u_{\mathrm{v}}} \equiv \int_0^1 dx\; \Delta
u_{\mathrm{v}}(x,Q^2)$ and $\eta_{d_{\mathrm{v}}} \equiv \int_0^1 dx\;
\Delta d_{\mathrm{v}}(x,Q^2)$.

This modification implies a $21\,\%$ change of the value of the
Bjorken Sum Rule (BJSR) \cite{th:bjsr,th:bjsr2}
\begin{equation}
  \label{eq:bjsr}
  I_3 \equiv \Delta q_3 = \int_0^1 \!\! dx \; \Delta q_3(x) =
  \left| \frac{g_A}{g_V}\right|
\end{equation}
from its physical value $I_3 = |g_A/g_V| = 1.267$ to $I^\prime_3 =
1.0$.  The unmodified fit agrees well with the obtained helicity
distributions in the measured range as illustrated in
Fig.~\ref{fig:5Pfit-xdens} and Tab.~\ref{tab:chi2models} in
section~\ref{sec:quarkpol}.  This consistency is also reflected in the
good agreement of the first moments in the measured
range of the valence quark helicity
densities $\Delta u_{\mathrm{v}}$ and $\Delta d_{\mathrm{v}}$ as well
as the isotriplet flavor combination $\Delta q_3$ from the BB LO
parameterization with the present results from semi-inclusive 
data, as listed in
Tab.~\ref{tab:bb-study:meas}.  It is, however, remarkable that the
moments of the same quantities computed from the modified fit yield
almost as good consistency with the measured results, despite the
significant changes in the boundary conditions $\eta_{u_{\mathrm{v}}}$
and $\eta_{d_{\mathrm{v}}}$.

\begin{table}[ht]
%%
%% Helmut's study on the stability of the BB fit
%%
  \renewcommand{\arraystretch}{1.1}
  \caption{\label{tab:bb-study:meas}
    Comparison of first moments of the LO QCD fit \cite{Blumlein:2002be}
    and a modified version of it (see text) with the measured results, in
    the range $0.023 \leq x \leq 0.6$.  The values in square brackets 
    give the absolute deviation of the moments of the QCD fits from the
    measured SIDIS values in units of their combined statistical and systematic
    uncertainties.}
  \begin{ruledtabular}
    \begin{tabular}{l|rrrrr} % \hline
      & \multicolumn{2}{c}{Original fit}
      & \multicolumn{2}{c}{Modified fit}
      & \multicolumn{1}{c}{Measured SIDIS} \\ \hline
      $\Delta u_{\mathrm{v}}$ \rule{0mm}{3ex} & 
      $0.601$ & $[0.02\,\sigma]$ &   % original fit
      $0.569$ & $[0.40\,\sigma]$ &   % modified fit
      $0.603 \pm 0.081$ \\           % HERMES measured
      $\Delta d_{\mathrm{v}}$ & 
      $-0.215$ & $[0.50\,\sigma]$ &  % original fit
      $-0.220$ & $[0.56\,\sigma]$ &  % modified fit
      $-0.172 \pm 0.082$ \\          % HERMES measured
      $\Delta q_3$ & 
      $0.816$  & $[0.52\,\sigma]$ &  % original fit
      $0.789$  & $[0.74\,\sigma]$ &  % modified fit
      $0.880 \pm 0.116$ \\           % HERMES measured
    \end{tabular}
  \end{ruledtabular}
\end{table}

While the QCD fit appears to be stable in the kinematic region where
it is rather well constrained by data (the data sets entering the fit
\cite{Blumlein:2002be} cover a range $0.005 \leq x \leq 0.75$) it
seems to exhibit enough freedom in the unmeasured regions
to allow for a drastic reduction of the total integral $I_3$ from
$1.267$ to $1.0$.  Table \ref{tab:bb-study:lowx} lists the fractional
contributions to the moments from the original and the modified QCD
fit in the range $0 < x < 0.023$.  In order to match the smaller value
of $I_3^\prime = 1.0$, the low--$x$ contributions to all moments are
significantly reduced in the modified fit.  In particular, the
low--$x$ contribution from the original fit $\int_0^{0.023} \!\!  dx
\; \Delta q_3(x) = 0.411$ reduces to $\int_0^{0.023} \!\!  dx \;
\Delta q_3^\prime(x) = 0.178$ in the modified fit.  For comparison,
the high--$x$ extrapolation $\int_{0.6}^1 dx \; \Delta q_3(x) = 0.041$
vs.~$\int_{0.6}^1 dx \; \Delta q_3^\prime(x) = 0.033$ is only slightly
reduced and despite the large covered $x$--range its contribution to
the total moment is small.

\begin{table}[ht]
%%
%% Helmut's study on the stability of the BB fit
%%
  \renewcommand{\arraystretch}{1.1}
  \caption{\label{tab:bb-study:lowx}
    Fractional contributions from the low--$x$ range $(0 < x < 0.023)$
    to the first moments obtained from the original and modified
    global QCD fit (see text).}
  \begin{minipage}{0.6\columnwidth}
    \begin{ruledtabular}
      \begin{tabular}{l|cc} % \hline
        & \multicolumn{1}{c}{Original fit}
        & \multicolumn{1}{c}{Modified fit} \\ \hline
        $\Delta u_{\mathrm{v}}$ \rule{0mm}{3ex} & 
        $31\,\%$ &              % original fit
        $18\,\%$ \\             % modified fit
        $\Delta d_{\mathrm{v}}$ & 
        $36\,\%$ &              % original fit
        $18\,\%$ \\             % modified fit
        $\Delta q_3$ & 
        $32\,\%$  &             % original fit
        $18\,\%$  \\            % modified fit
      \end{tabular}
    \end{ruledtabular}
  \end{minipage}
\end{table}

This study provides information about the ``flexibility'' of
the chosen parameterizations rather than covering the full range of competing
theoretical models for the extrapolations into the unmeasured
kinematic regions.  Yet, it demonstrates the arbitrariness of these
extrapolations, in particular to the small $x$ range.  In the
measured region, our experimental result on $\Delta q_3$ agrees within
$0.52\,\sigma$ with the integral 
of this quantity from the original QCD fit over the same
$x$--range (see Tab.~\ref{tab:bb-study:meas}).  We nevertheless
refrain from interpreting this as an experimental confirmation of the
BJSR Eq.~\eqref{eq:bjsr}, as the above example has shown that one can
obtain a similar level of agreement with a $21\,\%$ reduction of
the predicted value, thereby rendering a sound estimate of the
associated systematic uncertainty impossible.  For the same reason we
restrict all results for the first and second moments to the
experimentally covered $x$--region, albeit the weight of the low--$x$
extrapolation becomes smaller for the higher moments.

We finally remark that in \cite{ph:knauf2002} a reverse approach to
this problem is presented, where the authors assume the validity of
the BJSR Eq.~\eqref{eq:bjsr} and use experimental data at $x > 0.005$
to restrict the exponent $\lambda = -0.40 \pm 0.31$ in an assumed
power--like behaviour of the polarized structure function
$g_1(x) \sim x^\lambda$, based on Regge theory.

%%
%% ----------------------------------------------------------------------- %%
%%

\subsection{Comparison with other Results}
\label{sect:moments:comparison}

SMC is the only other experiment that has published results from SIDIS
on the quark helicity distributions in the nucleon \cite{smc:deltaq}.
Due to limited statistics and the lack of discrimination
between different hadron types, SMC could extract quark helicity
distributions only under the assumption of $SU(3)$ flavor symmetry for the
sea quark flavors, {\it i.e.} $\Delta\bar{u}(x) \equiv \Delta\bar{d}(x)
\equiv \Delta s(x) \equiv \Delta\bar{s}(x)$. It is only because of this
assumption that their precision in sea quark polarization is comparable
to that of the present work. 
%%Therefore, when
%%comparing to the results obtained by SMC, one does not necessarily
%%expect agreement for the sea quark flavors, in particular for flavors
%%other than $u_{\mathrm{sea}}(x)$ and $\bar{u}(x)$.  
%%Because of the
%%weighting by the square of the charge in the DIS cross section
%%(Eqs.~\eqref{eqn:wmunu}--\eqref{eqn:g1}),
%%up--flavor quarks and antiquarks are favored in scattering by sea quarks.

The first and second moments for the valence and $\bar{u}$ sea quark
helicity densities in the measured range of the HERMES data are
compared in Table \ref{tab:compare_smc} to the results from the SMC
experiment, which were integrated over the same kinematic region.  It
should be noted that SMC evaluated their moments according to a
somewhat simpler expression \cite{smc:pretz}
\begin{equation}
  \label{eq:smc-moments}
  \Delta^{(n)} q(Q_0^2) = \sum_i (\xi_{i+1}-\xi_i) \: x_i^{n-1} \:
  \Delta q(x_i,Q_0^2) \, ,
\end{equation}
where the notations have been adapted to match the definition in
Eq.~\eqref{eq:momdef}.  The integration procedure assumes constant
helicity densities $\Delta q(x)$ over the entire width of an $x$--bin.
In particular towards the upper experimental limit on $x$, where the
$x$--bins are widest, the values determined for the helicity densities
$\Delta u_{\mathrm{v}}(x)$ and $\Delta d_{\mathrm{v}}(x)$ in some
cases violate the positivity limits given by the unpolarized parton
densities at the upper $x$--bin limits (see Fig.~2 in
\cite{smc:deltaq}).  This causes larger values for the moments when
computed according to Eq.~\eqref{eq:smc-moments} as compared to
Eq.~\eqref{eq:momdef}.  Nevertheless the results from the two experiments
are in good agreement within their combined uncertainties, while the
present data set has an improved precision.  When comparing the
precision of the results from both experiments, note that the HERMES
results are free of symmetry assumptions regarding the sea quark
helicity densities and that they are based on a new
unfolding technique to account for the proper propagation of
uncertainties while handling QED radiative and instrumental smearing
effects.
\begin{table}[t]
%%
%% Comparison with SMC
%%
  \renewcommand{\arraystretch}{1.1}
  \caption{\label{tab:compare_smc}
    Comparison of first and second moments in the measured range from this
    analysis with results from the SMC experiment.
    The SMC values were extrapolated to the same value of $Q_0^2 =
    2.5\:\text{GeV}^2$ and integrated over the HERMES $x$--range.}
  \begin{ruledtabular}
    \begin{tabular}{lrr} % \hline
      & \multicolumn{1}{c}{HERMES} & \multicolumn{1}{c}{SMC} \\ \hline
      $\Delta u_{\mathrm{v}}$ \rule{0mm}{3ex} & 
      $0.603 \pm 0.071 \pm 0.040$ &   % HERMES
      $0.614 \pm 0.082 \pm 0.068$ \\  % SMC
      $\Delta d_{\mathrm{v}}$ & 
      $-0.172 \pm 0.068 \pm 0.045$ &  % HERMES
      $-0.334 \pm 0.112 \pm 0.089$ \\ % SMC
      $\Delta \bar{u}$ & 
      $-0.002 \pm 0.036 \pm 0.023$ &  % HERMES
      $ 0.015 \pm 0.034 \pm 0.024$ \\ % SMC
      $\Delta^{(2)} u_{\mathrm{v}}$ & 
      $0.144 \pm 0.013 \pm 0.011$ &   % HERMES
      $0.152 \pm 0.016 \pm 0.013$ \\  % SMC
      $\Delta^{(2)} d_{\mathrm{v}}$ &
      $-0.047 \pm 0.012 \pm 0.012$ &  % HERMES
      $-0.056 \pm 0.026 \pm 0.015$ \\ % SMC
    \end{tabular}
  \end{ruledtabular}
\end{table}

%%
%% ===================== End of file 'sect-VIII.tex' ===================== %%
%%
% \include{sect-IX}
%% ----------------------------------------------------------------------- %%
%%
%%   File        : sect-X.tex
%%
%%   Description : Source file for section XI of the second delta-q paper
%%                 (DC number 42)
%%   
%%   Main Author : Mike Vetterli
%%
%%   Date        : 19-Aug-2002
%%
%%   Remarks     : 
%%
%%   Modified    : 02-Dec-2003 - This was sect-XI.tex until today... //mb
%%
%% ----------------------------------------------------------------------- %%

\section{Summary}
\label{sect:conclusions}

This paper describes the most precise available measurements of semi--inclusive
asymmetries in polarized deep inelastic electron/positron scattering.
Results from both inclusive and semi--inclusive measurements are
presented.  The unique semi--inclusive data are particularly important
for the unbiased extraction of sea quark helicity densities in the
nucleon.

Data were collected on longitudinally polarized atomic hydrogen and deuterium
gas targets.  Good particle identification in the HERMES spectrometer
allows hadrons that are coincident with the scattered lepton to be
separated into pion and kaon samples.  This gives sensitivity to the
individual quark \textit{and} antiquark helicity distribution
functions for light and strange quark flavors.

An unfolding technique --- new to this type of measurement --- was applied for
instrumental smearing and QED radiative effects. It takes into
account event migrations between bins.  This algorithm
provides more rigorous, yet larger, estimates of the inflation of the
uncertainties on the Born asymmetries with respect to the measured
asymmetries than correction methods previously applied.  The unfolding
procedure also yields previously unavailable estimates of the
statistical correlations between different kinematic bins.
All asymmetries were found to be positive and increase with $x$, with
the exception of the $K^-$ asymmetry which is consistent with zero
over the entire measured range.  The present data on undiscriminated
hadrons agree well with earlier measurements by SMC at higher $Q^2$,
albeit with much improved precision.

A ``leading order'' QCD analysis that relies on the technique of flavor
tagging was used to extract helicity distribution functions.  The
results presented here allow for the first time the independent
determination of five out of six quark polarizations in the nucleon.
Quark polarizations $[\Delta q/q](x)$ were obtained using purities 
that were calculated in a Monte Carlo based on
parameterizations of unpolarized parton densities and a modeling of
hadron multiplicities measured at HERMES.   
The polarization $[\Delta u / u](x)$ was
found to be positive and rising over the entire range in $x$, while
$[\Delta d/d](x)$ is negative.  These first results on the individual sea
quark polarizations $[\Delta \bar{u} / \bar{u}](x)$, $[\Delta \bar{d}
/ \bar{d}](x)$, and $[\Delta s/s](x)$ are consistent with zero.

Furthermore, the ``LO'' approach presented here results in helicity quark
distributions that are fully consistent with global QCD fits to world
data on inclusive DIS asymmetries.  While the BB fit favors a slightly
negative strange quark helicity density $\Delta s(x)$, the data yield
a small positive result. 
%%(cf.~Fig.~\ref{fig:5Pfit-xdens}).  
However, it should be noted that those particular fits invoke $SU(3)$ symmetry
to relate the triplet $a_3$ and octet $a_8$ axial couplings to the 
weak decay constants $F$ and $D$, assume a $SU(3)$ symmetric
sea, and require a model for the gluon
%%one should note that the fits to inclusive scattering data require
%%symmetry assumptions between the quark helicity distributions (they
%%usually assume an $SU(3)$ flavor symmetric sea), a model for the gluon
%%helicity density $\Delta g(x)$, and they constrain integrals to
%%combinations of the weak decay constants $F$ and $D$.  
helicity density $\Delta g(x)$. Because of
these assumptions and taking into account the combined statistical and
systematic uncertainties on the present experimental result, no
significant discrepancy remains.  However, the results from the
present semi--inclusive scattering data might reveal the possible
biases of global fits for certain quark flavors.  Within the
experimental uncertainty all obtained quark helicity densities $\Delta
q(x)$ are in good agreement with the most recent parameterizations.

To confirm the results for the helicity distribution of the strange
sea, a different isoscalar analysis leading to $\Delta
s(x) + \Delta\bar{s}(x)$ was performed.  This technique does not
depend on a Monte Carlo modeling of the strange fragmentation
functions.  Instead,
parameterizations of strange and non--strange fragmentation functions
measured at $e^+e^-$ colliders at higher energies were utilized.  The
results for the two analyses are in reasonable agreement.

The first direct experimental extraction of the helicity density 
asymmetry $\Delta\bar{u}(x) - \Delta \bar{d}(x)$ in the light quark sea,
which is predicted to
be non zero by many models in analogy to the unpolarized sector
($\bar{u}(x) \neq \bar{d}(x)$), does not establish broken
$SU(2)$ flavor symmetry.  The data disfavor the substantial positive asymmetry
predicted by the $\chi$QSM model and are consistent 
with the small negative asymmetry characteristic of the meson cloud model.

Moments were computed in the measured kinematic range for the quark
helicity distributions and singlet and non--singlet flavor
combinations.  The moments of the valence and sea quark
helicity densities $\Delta u_\mathrm{v}$, 
$\Delta d_\mathrm{v}$, and $\Delta \bar{u}$
agree with previous results from the SMC experiment.  In the measured
range, the non--singlet flavor combination $\Delta q_3$ is in good
agreement with the same quantity computed from global QCD fits, which
in turn by construction fulfill the Bjorken sum rule.
However, because extrapolations into the unmeasured kinematic regions
are not sufficiently constrained by the data reported here, an 
experimental confirmation of this fundamental sum rule was not possible.  
%%However, an
%%apparent arbitrariness in the extrapolations into the unmeasured
%%kinematic regions does not allow an experimental confirmation of this
%%fundamental sum rule.  
For the same reason we do not quote a total
value for the singlet quantity $\Delta \Sigma$, which gives the total
contribution from quark spins to the spin of the nucleon.  In the
measured range, a value of $\Delta\Sigma = 0.347 \pm 0.024 \pm 0.066$
was obtained.  
%%While the good consistency of the present results with
%%those from earlier experiments might justify a confirmation that
%%gluons and/or orbital angular momentum must contribute to the nucleon
%%spin we note that this interpretation is subject to explicit
%%assumptions on the behavior of the spin--dependent parton densities in
%%the unmeasured kinematic regions.

In conclusion, HERMES has made detailed semi--inclu\-sive measurements
of polarization asymmetries in deep inelastic lepton scattering.
These measurements allow quark helicity distribution functions to be
extracted with fewer and different model assumptions than in previous
inclusive measurements. 

\vspace*{0.5cm}
\centerline{\textbf{Acknowledgements}}
\vspace{0.5cm}

We gratefully acknowledge the DESY management for its support, the
staffs at DESY and the collaborating institutions for their
significant effort.  This work was supported by the FWO--Flanders,
Belgium; the Natural Sciences and Engineering Research Council of
Canada; the ESOP, INTAS and TMR network contributions from the
European Union; the German Bundesministerium f\"ur Bildung und
Forschung; the Italian Istituto Nazionale di Fisica Nucleare (INFN);
Monbusho International Scientific Research Program, JSPS and Toray
Science Foundation of Japan; the Dutch Foundation for Fundamenteel
Onderzoek der Materie (FOM); the U.K. Particle Physics and Astronomy
Research Council; and the U.S. Department of Energy and National
Science Foundation.

%%
%% ====================== End of file 'sect-XI.tex' ====================== %%
%%

% --------------------------------------------------------------------------
%   The appendices
% --------------------------------------------------------------------------

%
% The radiative corrections algorithm
%
\appendix

%% ----------------------------------------------------------------------- %%
%%
%%   File        : sect-XI.tex
%%
%%   Description : Source file for section XI of the second delta-q paper
%%                 (DC number 42)
%%   
%%   Main Author : J. Wendland
%%
%%   Date        : 19-Jun-2003
%%
%%   Remarks     : This is an appendix
%%
%%   Modified    : 02-Dec-2003 - This was sect-XII.tex until today... //mb
%%
%% ----------------------------------------------------------------------- %%

\section{Unfolding of Radiative and Detector Smearing Effects}
\label{sect:app:radcorr}

Corrections to the asymmetries for higher order QED and detector
smearing effects were carried out using an unfolding algorithm as
indicated in section~\ref{sect:asyms:radcor}.
%%
%%  MC data
%%
However, the algorithm is complex,
because the known absolute cross sections of the spin--dependent
background processes must be normalized to the data by comparing
simulated and measured unpolarized yields based on world data for
$F_2$.
The procedure uses two sets of Monte Carlo data.  Born data were
generated with the GRSV2000 (LO, standard scenario) spin--dependent
parton distributions \cite{pdf:grsv2000} and the LUND fragmentation
model implemented in {\sc Jetset} 7.4 \cite{Sjostrand:1994yb}. Within the
acceptance of the HERMES spectrometer
($|\theta_x|<170\,\mathrm{mrad}$,
$40\,\mathrm{mrad}<|\theta_y|<140\,\mathrm{mrad}$) an equivalent
number of $11.5\,\mathrm{M}$ DIS events were generated on hydrogen and
deuterium targets. The second Monte Carlo data set included, in
addition, internal radiative effects and a simulation of the HERMES
spectrometer. Radiative effects were simulated with RADGEN
\cite{Akushevich:1998ft}, and the spectrometer simulation was done
using GEANT \cite{Brun:1978fy}. An equivalent number of
$5.4\,\mathrm{M}$ DIS events was available for analysis for each
target. The total MC statistics are sufficiently large that the
uncertainty on the Born asymmetries from this source is small (cf.
Tabs.~\ref{tab:A1p_results}, \ref{tab:A1d_results} and
\ref{tab:Dq_results}).

%%
%%  The algorithm: n(i,j) and n^B
%%
Based on the latter Monte Carlo data, matrices $n_{\pap}(i,j)$ with
dimensions $n_X\times(n_B+1)$ were constructed for the parallel
($\spar$) and anti--parallel ($\sant$) spin--states, which describe the
count rates that fall in both bin $j$ of \underline{B}orn kinematics
and bin $i$ of e\underline{X}perimental kinematics. The experimental
kinematics reflect the radiative and instrumental effects. The bins
$j=1,\dots ,\, n_B$ describe the migration of DIS events, where both the
experimental and the Born kinematics are within the acceptance. Here,
acceptance refers to both the geometrical acceptance of the
spectrometer and the phase space defined by the DIS and SIDIS cuts.
The additional bin $j=0$ contains background rates that feed into the
experimental bins through QED radiative and detector smearing effects.
For example, in the case of the inclusive data sample on the proton,
the background stems from elastic and inelastic scattering events that
are radiated (through QED effects) or smeared (by interactions in the
detector) into the acceptance. The experimental rates in the parallel
and anti--parallel spin states in bin $i$ are given by the sum
$n_{\pap}^X(i)=\sum_{j=0}^{n_B} \,n_{\pap}(i,j)$.  The corresponding
Born rates, $n_{\pap}^B(j)$, in each spin state and bin $j$ were
calculated from the Born Monte Carlo data.

These data, normalized with respect to each other provide access to
the migration matrices ${\cal S}_{\pap}(i,j)$, which are given by
\begin{multline}
  {\cal S}_{\pap}(i,j)\equiv\frac{\partial \sigma^X_{\pap}(i)}
  {\partial \sigma^B_{\pap}(j)} \, , \\
  \quad i=1,\dots, \, n_X, \; j=1,\dots, \, n_B \, .
\end{multline}
In terms of $n_{\pap}(i,j)$ and $n_{\pap}^B(j)$, these matrices
can be written as
\begin{equation}
  {\cal S}_{\pap}(i,j)=\frac{n_{\pap}(i,j)}{n_{\pap}^B(j)} \, ,
\end{equation}
provided the first derivatives are constant.  The matrices ${\cal
  S}_{\pap}$ describe the kinematical migration inside of the
acceptance.  Backgrounds are taken into account through the
normalization with the Born rates $n_{\pap}^B$ and as will be seen
below in the $j=0$ row of $n_{\pap}(i,j)$. The ${\cal S}_{\pap}$
matrices are insensitive to the Monte Carlo model of the Born
distributions, because both the numerator and the denominator scale
with the relative number of Born events generated in bin number $j$.

The unknown Born rates $B_\pap(j)$ are related to the measured
asymmetries (Eq.~\eqref{eq:aparallel} corrected for the azimuthal
acceptance effects) through the migration matrices, the unpolarized
experimental rates $n_u^X(i)\equiv n_\sant^X(i)+n_\spar^X(i)$ and the
spin--dependent background $n_p(i,0)\equiv n_\sant(i,0)-n_\spar(i,0)$:
                              %
%\begin{equation}
%  \label{unfold1}
%  \begin{split}
%    \sum_{j=1}^{n_B} \left[{\cal S}_\sant(i,j)B_\sant(j) 
%      - {\cal S}_\spar(i,j) B_\spar(j) \right]=&\\
%    A_\|^X(i)\,n_u^X(i)-n_p(i,0), \quad i&=1,\dots, \, n_X \, .\\
%  \end{split}
%\end{equation}
\begin{multline} \label{unfold1}
  \sum_{j=1}^{n_B} \left[{\cal S}_\sant(i,j)B_\sant(j) 
    - {\cal S}_\spar(i,j) B_\spar(j) \right]= \\
  A_\|^X(i)\,n_u^X(i)-n_p(i,0), \quad i=1,\dots, \, n_X \, .
\end{multline}
Here the unpolarized Born rate $B_\sant(j) + B_\spar(j) \equiv B_u(j)$
is known from previous experiments and incorporated in the Monte Carlo
simulation: $B_u(j)= n_u^B(j)\equiv n_\sant^B+n_\spar^B$. The
\begin{figure}[b]
  \centering
  \includegraphics[width=\columnwidth]{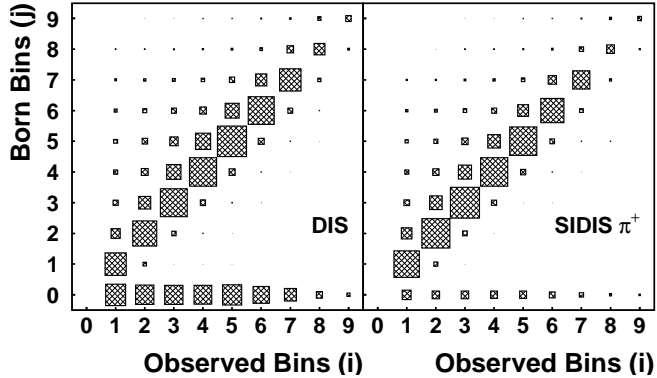}
  \caption{\label{fig:nijmatrices}Matrices
    $n_u(i,j)=n_\spar(i,j)+n_\sant(i,j)$ for DIS events and SIDIS
    $\pi^+$ events on the proton.  The binning shown corresponds to
    the 9 bins in $x$ that were used in the asymmetry and $\Delta q$
    analysis (Tab.~\ref{tab:xbins}).  See the text for details.}
\end{figure}
background term in the sum was moved to the right hand side of the
equation $n_p(i,0) = {\cal S}_\sant(i,0) B_\sant(0) - {\cal
  S}_\spar(i,0) B_\spar(0)$.  The rate $B_\spar(j)$ may now be
eliminated in favor of $B_\sant(j)$ and $B_u(j)$:
\begin{equation}
  \begin{split}
    \sum_{j=1}^{n_B} \left[{\cal S}_\sant(i,j) + {\cal S}_\spar(i,j) \right]
    B_\sant(j) = A_\|^X(i) \, n_u^X(i) &\\
    - n_p(i,0) + \sum_{j=1}^{n_B} {\cal S}_\spar(i,j) \, n_u^B(j),
    \quad i=1,\dots, \, \,n_X &\, .
  \end{split}
  \label{unfoldSSB}
\end{equation}
The Born asymmetry is found by solving Eq.~\eqref{unfoldSSB} for
$B_\sant(j)$ and substituting into 
\begin{equation}
  A_\|^B(j)=\frac{2B_\sant(j)-B_u(j)}{B_u(j)} \, .
\end{equation}
The final expression for the unfolded asymmetry is then:
\begin{equation}
  \label{eq:unfold2}
  \begin{aligned}
    A&^B_{\|}(j)
    = -1 + \frac{2}{n^B_u(j)} \sum_{i=1}^{n_X} \, [{\cal S}']^{-1}(j,i)
    \:\times\\
    &\left[ A^X_\|(i) \, n_u^X(i) \,
      - n_p(i,0) + \sum_{k=1}^{n_B} \, {\cal S}_\spar(i,k) \, n_u^B(k)
    \right] \, ,
  \end{aligned}
\end{equation}
for $j=1,\dots, \, n_B$, where ${\cal S}'(i,j)$ is the square
submatrix without the column $j=0$ of ${\cal S}(i,j)\equiv {\cal
  S}_\sant(i,j) + {\cal S}_\spar(i,j)$.  Generally the inverse of the
migration matrix may be ill--defined \cite{Blobel:2002pu}. However, in
this case the migration due to QED and detector effects is
sufficiently small that the matrix inversion yields satisfactory
results.

Fig.~\ref{fig:nijmatrices} presents the matrices $n_u(i,j)$ for DIS
events and SIDIS $\pi^+$ events on the proton calculated in the
$x$--Bjorken bins of the analysis presented in this paper.  The
background that migrates into the acceptance ($j=0$) is seen to be
large in case of the DIS events. Only little background is present in
the case of the SIDIS events, where the hadron--tag rejects elastic
events that are present in the inclusive case. The inter--bin
migration is of similar size in both data--samples. QED radiative
effects, specifically initial and final--state bremsstrahlung, cause
migration to smaller values of observed $x$ only. Multiple scattering
and finite resolution effects in the detector also increase but may in
some cases decrease the observed $x$ with respect to the Born $x$.

The statistical covariance matrix of the Born asymmetry follows from
Eq.~\eqref{eq:unfold2}
\begin{equation}
  \label{eq:covABorn}
  \begin{aligned}
  {\cal V}&(A^B_\|(j),A^B_\|(k)) = \\ 
  &\sum_{i_{1}=1}^{n_X}\sum_{i_{2}=1}^{n_X} 
  {\cal D}(j,i) {\cal D}(k,i) \, \sigma(A^X_\|(i_{1})) \,
  \sigma(A^X_\|(i_{2})) ,
  \end{aligned}
\end{equation}
where ${\cal D}(j,i)$ is the dilution matrix,
\begin{equation}
  {\cal D}(j,i) \equiv \frac{\partial A^B_\|(j)}{\partial A^X_\|(i)}
  = \frac{2[{\cal S}']^{-1}(j,i)\,n_u^X(i)}{n_u^B(j)} \, .
\end{equation}
These expressions unambiguously determine the statistical
uncertainties on the Born asymmetries, unlike previous methods that
compute the corrections bin--by--bin. In a similar manner, systematic
uncertainties due to beam and target polarization, as well as for the
azimuthal acceptance correction were determined for the measured aysmetries.
The systematic covariances from these sources follow via the dilution
matrix. For the beam and target polarization measurements the uncertainty
in the Born asymmetries is
\begin{equation}
  \label{eq:syscovABorn}
  \begin{aligned}
  &{\cal V}_{P_{T(B)}}(A^B_\|(j),A^B_\|(k)) = \\ 
  &\sum_{i_{1}=1}^{n_X}\sum_{i_{2}=1}^{n_X} 
  {\cal D}(j,i) {\cal D}(k,i) \, \sigma_{P_{T(B)}}(A^X_\|(i_{1})) \,
  \sigma_{P_{T(B)}}(A^X_\|(i_{2})) .
  \end{aligned}
\end{equation}
The double sum over the experimental uncertainties is a consequence of
the complete correlation of the beam and target polarization measurements
among the $x$ bins. In the case of the azimuthal acceptance correction, the
systematic uncertainty was assumed to be uncorrelated between the bins.

%%
%% ====================== End of file 'sect-XII.tex' ===================== %%
%%

%%% Local Variables: 
%%% mode: latex
%%% TeX-master: "dqpaper"
%%% End: 
% \clearpage

% 
% Tables of results
% 

\section{Tables of Results}
\label{sect:app:results}

\newcolumntype{q}[1]{D{.}{.}{#1}}

%%\onecolumngrid
%%\section{Tables of Results}
%%\label{sect:app:results}
%%\twocolumngrid

\begin{table*}
  \renewcommand{\arraystretch}{1.1} 
  \caption{\label{tab:A1p_results} Inclusive and semi--inclusive Born level 
    asymmetries on the proton target.}
\begin{ruledtabular}
  \begin{tabular}{q{4}q{3}q{4}|q{5}q{5}q{5}q{5}|q{5}q{5}q{5}q{5}}
    \multicolumn{1}{c}{$\langle x \rangle$\rule[-1.5ex]{0mm}{4.0ex}} &
    \multicolumn{1}{c}{\hspace*{-2.5ex}$\langle Q^2 \rangle
    \,/\:\mathrm{GeV}^2$\hspace*{-2.5ex}} &
    \multicolumn{1}{c}{\phantom{$\langle z \rangle$}} &
    \multicolumn{1}{c}{$A_{1,p}$} &
    \multicolumn{1}{c}{$\pm \mathrm{stat.}$} &
    \multicolumn{1}{c}{$\pm \mathrm{syst.}$} &
    \multicolumn{1}{c}{$\pm \mathrm{MC}$} & & & & \\ \hline
      0.033 &  1.22 & &  0.0996 &  0.0176 &  0.0091 &  0.0037
       & & & & \\
      0.048 &  1.45 & &  0.1102 &  0.0175 &  0.0093 &  0.0043
       & & & & \\
      0.065 &  1.68 & &  0.1131 &  0.0181 &  0.0088 &  0.0050
       & & & & \\
      0.087 &  1.93 & &  0.1941 &  0.0208 &  0.0143 &  0.0063
       & & & & \\
      0.118 &  2.34 & &  0.2366 &  0.0205 &  0.0178 &  0.0061
       & & & & \\
      0.166 &  3.16 & &  0.2819 &  0.0216 &  0.0208 &  0.0054
       & & & & \\
      0.240 &  4.54 & &  0.3854 &  0.0239 &  0.0283 &  0.0045
       & & & & \\
      0.340 &  6.56 & &  0.4760 &  0.0430 &  0.0344 &  0.0054
       & & & & \\
      0.447 &  9.18 & &  0.6102 &  0.0750 &  0.0467 &  0.0059
       & & & & \\
    \hline
    \multicolumn{1}{c}{$\langle x \rangle$\rule[-1.5ex]{0mm}{4.5ex}} &
    \multicolumn{1}{c}{\hspace*{-2.5ex}$\langle Q^2 \rangle
    \,/\:\mathrm{GeV}^2$\hspace*{-2.5ex}} &
    \multicolumn{1}{c}{$\langle z \rangle$} &
    \multicolumn{1}{c}{$A_{1,p}^{h^+}$} &
    \multicolumn{1}{c}{$\pm \mathrm{stat.}$} &
    \multicolumn{1}{c}{$\pm \mathrm{syst.}$} &
    \multicolumn{1}{c}{$\pm \mathrm{MC}$} &
    \multicolumn{1}{c}{$A_{1,p}^{h^-}$} &
    \multicolumn{1}{c}{$\pm \mathrm{stat.}$} &
    \multicolumn{1}{c}{$\pm \mathrm{syst.}$} &
    \multicolumn{1}{c}{$\pm \mathrm{MC}$} \\ \hline
      0.034 &  1.21 & 0.356 &  0.1097 &  0.0299 &  0.0077 &  0.0067
        &  0.0724 &  0.0341 &  0.0054 &  0.0073 \\
      0.048 &  1.44 & 0.375 &  0.1640 &  0.0316 &  0.0113 &  0.0075
        &  0.1262 &  0.0365 &  0.0087 &  0.0085 \\
      0.065 &  1.72 & 0.386 &  0.1361 &  0.0329 &  0.0096 &  0.0083
        &  0.0901 &  0.0382 &  0.0057 &  0.0094 \\
      0.087 &  2.06 & 0.393 &  0.2075 &  0.0370 &  0.0147 &  0.0095
        &  0.0953 &  0.0442 &  0.0066 &  0.0111 \\
      0.118 &  2.58 & 0.395 &  0.3011 &  0.0356 &  0.0212 &  0.0088
        &  0.1860 &  0.0445 &  0.0135 &  0.0109 \\
      0.166 &  3.52 & 0.390 &  0.2851 &  0.0384 &  0.0209 &  0.0082
        &  0.1770 &  0.0498 &  0.0127 &  0.0104 \\
      0.239 &  5.03 & 0.388 &  0.4223 &  0.0456 &  0.0292 &  0.0075
        &  0.2360 &  0.0639 &  0.0165 &  0.0106 \\
      0.338 &  7.09 & 0.377 &  0.4046 &  0.0914 &  0.0291 &  0.0101
        &  0.5696 &  0.1339 &  0.0387 &  0.0157 \\
      0.448 &  9.76 & 0.364 &  0.7586 &  0.1756 &  0.0519 &  0.0122
        &  0.4957 &  0.2715 &  0.0341 &  0.0231 \\
    \hline
    \multicolumn{1}{c}{$\langle x \rangle$\rule[-1.5ex]{0mm}{4.5ex}} &
    \multicolumn{1}{c}{\hspace*{-2.5ex}$\langle Q^2 \rangle
    \,/\:\mathrm{GeV}^2$\hspace*{-2.5ex}} &
    \multicolumn{1}{c}{$\langle z \rangle$} &
    \multicolumn{1}{c}{$A_{1,p}^{\pi^+}$} &
    \multicolumn{1}{c}{$\pm \mathrm{stat.}$} &
    \multicolumn{1}{c}{$\pm \mathrm{syst.}$} &
    \multicolumn{1}{c}{$\pm \mathrm{MC}$} &
    \multicolumn{1}{c}{$A_{1,p}^{\pi^-}$} &
    \multicolumn{1}{c}{$\pm \mathrm{stat.}$} &
    \multicolumn{1}{c}{$\pm \mathrm{syst.}$} &
    \multicolumn{1}{c}{$\pm \mathrm{MC}$} \\ \hline
      0.033 &  1.22 & 0.364 &  0.0800 &  0.0353 &  0.0058 &  0.0077
        &  0.0675 &  0.0388 &  0.0053 &  0.0081 \\
      0.047 &  1.50 & 0.416 &  0.1336 &  0.0387 &  0.0091 &  0.0088
        &  0.1450 &  0.0431 &  0.0095 &  0.0092 \\
      0.064 &  1.87 & 0.449 &  0.0829 &  0.0408 &  0.0067 &  0.0093
        &  0.0649 &  0.0461 &  0.0039 &  0.0100 \\
      0.087 &  2.38 & 0.471 &  0.2312 &  0.0459 &  0.0157 &  0.0102
        &  0.0714 &  0.0536 &  0.0047 &  0.0109 \\
      0.118 &  3.08 & 0.487 &  0.3163 &  0.0458 &  0.0212 &  0.0093
        &  0.0754 &  0.0547 &  0.0057 &  0.0101 \\
      0.166 &  4.22 & 0.490 &  0.3017 &  0.0525 &  0.0201 &  0.0090
        &  0.1572 &  0.0645 &  0.0105 &  0.0102 \\
      0.238 &  5.83 & 0.504 &  0.2784 &  0.0695 &  0.0197 &  0.0092
        &  0.2696 &  0.0889 &  0.0187 &  0.0108 \\
      0.337 &  7.97 & 0.506 &  0.5566 &  0.1530 &  0.0373 &  0.0138
        &  0.3461 &  0.1995 &  0.0233 &  0.0177 \\
      0.449 & 10.49 & 0.496 &  0.8651 &  0.3185 &  0.0558 &  0.0175
        &  0.4490 &  0.4343 &  0.0352 &  0.0270 \\
  \end{tabular}
\end{ruledtabular}
\end{table*}

\begin{table*}
  \renewcommand{\arraystretch}{1.1}
  \caption{\label{tab:A1d_results} Inclusive and semi--inclusive Born level 
    asymmetries on the deuterium target. }
\begin{ruledtabular}
  \begin{tabular}{q{4}q{3}q{4}|q{5}q{5}q{5}q{5}|q{5}q{5}q{5}q{5}}
    \multicolumn{1}{c}{$\langle x \rangle$\rule[-1.5ex]{0mm}{4.0ex}} &
    \multicolumn{1}{c}{\hspace*{-2.5ex}$\langle Q^2 \rangle
    \,/\:\mathrm{GeV}^2$\hspace*{-2.5ex}} &
    \multicolumn{1}{c}{\phantom{$\langle z \rangle$}} &
    \multicolumn{1}{c}{$A_{1,d}$} &
    \multicolumn{1}{c}{$\pm \mathrm{stat.}$} &
    \multicolumn{1}{c}{$\pm \mathrm{syst.}$} &
    \multicolumn{1}{c}{$\pm \mathrm{MC}$} & & & & \\ \hline
      0.033 &  1.22 & &  0.0203 &  0.0078 &  0.0015 &  0.0034
       & & & & \\
      0.048 &  1.45 & &  0.0248 &  0.0080 &  0.0017 &  0.0040
       & & & & \\
      0.065 &  1.69 & &  0.0396 &  0.0085 &  0.0023 &  0.0048
       & & & & \\
      0.087 &  1.95 & &  0.0440 &  0.0100 &  0.0031 &  0.0061
       & & & & \\
      0.118 &  2.35 & &  0.0777 &  0.0099 &  0.0054 &  0.0060
       & & & & \\
      0.166 &  3.18 & &  0.1137 &  0.0107 &  0.0081 &  0.0054
       & & & & \\
      0.240 &  4.55 & &  0.1621 &  0.0121 &  0.0114 &  0.0046
       & & & & \\
      0.339 &  6.58 & &  0.2932 &  0.0228 &  0.0195 &  0.0057
       & & & & \\
      0.446 &  9.16 & &  0.3161 &  0.0412 &  0.0236 &  0.0065
       & & & & \\
    \hline
    \multicolumn{1}{c}{$\langle x \rangle$\rule[-1.5ex]{0mm}{4.5ex}} &
    \multicolumn{1}{c}{\hspace*{-2.5ex}$\langle Q^2 \rangle
    \,/\:\mathrm{GeV}^2$\hspace*{-2.5ex}} &
    \multicolumn{1}{c}{$\langle z \rangle$} &
    \multicolumn{1}{c}{$A_{1,d}^{h^+}$} &
    \multicolumn{1}{c}{$\pm \mathrm{stat.}$} &
    \multicolumn{1}{c}{$\pm \mathrm{syst.}$} &
    \multicolumn{1}{c}{$\pm \mathrm{MC}$} &
    \multicolumn{1}{c}{$A_{1,d}^{h^-}$} &
    \multicolumn{1}{c}{$\pm \mathrm{stat.}$} &
    \multicolumn{1}{c}{$\pm \mathrm{syst.}$} &
    \multicolumn{1}{c}{$\pm \mathrm{MC}$} \\ \hline
      0.033 &  1.21 & 0.355 &  0.0080 &  0.0146 &  0.0007 &  0.0068
        & -0.0125 &  0.0162 &  0.0013 &  0.0071 \\
      0.048 &  1.44 & 0.374 &  0.0112 &  0.0156 &  0.0018 &  0.0077
        &  0.0074 &  0.0174 &  0.0014 &  0.0084 \\
      0.065 &  1.73 & 0.384 &  0.0484 &  0.0162 &  0.0028 &  0.0085
        &  0.0380 &  0.0185 &  0.0022 &  0.0095 \\
      0.087 &  2.07 & 0.391 &  0.0754 &  0.0185 &  0.0043 &  0.0101
        &  0.0179 &  0.0212 &  0.0033 &  0.0113 \\
      0.118 &  2.60 & 0.394 &  0.0350 &  0.0179 &  0.0038 &  0.0094
        &  0.0739 &  0.0213 &  0.0040 &  0.0108 \\
      0.166 &  3.56 & 0.392 &  0.1326 &  0.0194 &  0.0087 &  0.0087
        &  0.0775 &  0.0245 &  0.0065 &  0.0106 \\
      0.238 &  5.04 & 0.388 &  0.1469 &  0.0237 &  0.0104 &  0.0081
        &  0.1712 &  0.0315 &  0.0103 &  0.0107 \\
      0.338 &  7.12 & 0.382 &  0.2372 &  0.0504 &  0.0151 &  0.0115
        &  0.3001 &  0.0700 &  0.0175 &  0.0168 \\
      0.446 &  9.61 & 0.380 &  0.1901 &  0.0995 &  0.0149 &  0.0147
        &  0.1499 &  0.1481 &  0.0128 &  0.0242 \\
    \hline
    \multicolumn{1}{c}{$\langle x \rangle$\rule[-1.5ex]{0mm}{4.5ex}} &
    \multicolumn{1}{c}{\hspace*{-2.5ex}$\langle Q^2 \rangle
    \,/\:\mathrm{GeV}^2$\hspace*{-2.5ex}} &
    \multicolumn{1}{c}{$\langle z \rangle$} &
    \multicolumn{1}{c}{$A_{1,d}^{\pi^+}$} &
    \multicolumn{1}{c}{$\pm \mathrm{stat.}$} &
    \multicolumn{1}{c}{$\pm \mathrm{syst.}$} &
    \multicolumn{1}{c}{$\pm \mathrm{MC}$} &
    \multicolumn{1}{c}{$A_{1,d}^{\pi^-}$} &
    \multicolumn{1}{c}{$\pm \mathrm{stat.}$} &
    \multicolumn{1}{c}{$\pm \mathrm{syst.}$} &
    \multicolumn{1}{c}{$\pm \mathrm{MC}$} \\ \hline
      0.033 &  1.22 & 0.353 & -0.0172 &  0.0175 &  0.0011 &  0.0076
        & -0.0113 &  0.0183 &  0.0014 &  0.0078 \\
      0.047 &  1.50 & 0.405 &  0.0180 &  0.0192 &  0.0022 &  0.0089
        & -0.0231 &  0.0203 &  0.0012 &  0.0093 \\
      0.064 &  1.87 & 0.437 &  0.0130 &  0.0201 &  0.0016 &  0.0095
        &  0.0457 &  0.0218 &  0.0028 &  0.0100 \\
      0.087 &  2.36 & 0.458 &  0.0449 &  0.0226 &  0.0029 &  0.0106
        &  0.0056 &  0.0245 &  0.0020 &  0.0110 \\
      0.118 &  3.07 & 0.472 &  0.0966 &  0.0223 &  0.0061 &  0.0096
        &  0.0884 &  0.0249 &  0.0045 &  0.0103 \\
      0.165 &  4.18 & 0.479 &  0.1207 &  0.0257 &  0.0079 &  0.0093
        &  0.0144 &  0.0298 &  0.0035 &  0.0105 \\
      0.238 &  5.80 & 0.488 &  0.1089 &  0.0343 &  0.0073 &  0.0097
        &  0.2039 &  0.0413 &  0.0116 &  0.0111 \\
      0.338 &  7.93 & 0.494 &  0.3179 &  0.0815 &  0.0202 &  0.0153
        &  0.3860 &  0.0988 &  0.0209 &  0.0186 \\
      0.446 & 10.24 & 0.503 &  0.0856 &  0.1695 &  0.0115 &  0.0215
        & -0.1323 &  0.2159 &  0.0182 &  0.0282 \\
    \hline
    \multicolumn{1}{c}{$\langle x \rangle$\rule[-1.5ex]{0mm}{4.5ex}} &
    \multicolumn{1}{c}{\hspace*{-2.5ex}$\langle Q^2 \rangle
    \,/\:\mathrm{GeV}^2$\hspace*{-2.5ex}} &
    \multicolumn{1}{c}{$\langle z \rangle$} &
    \multicolumn{1}{c}{$A_{1,d}^{K^+}$} &
    \multicolumn{1}{c}{$\pm \mathrm{stat.}$} &
    \multicolumn{1}{c}{$\pm \mathrm{syst.}$} &
    \multicolumn{1}{c}{$\pm \mathrm{MC}$} &
    \multicolumn{1}{c}{$A_{1,d}^{K^-}$} &
    \multicolumn{1}{c}{$\pm \mathrm{stat.}$} &
    \multicolumn{1}{c}{$\pm \mathrm{syst.}$} &
    \multicolumn{1}{c}{$\pm \mathrm{MC}$} \\ \hline
      0.033 &  1.22 & 0.383 &  0.0048 &  0.0479 &  0.0022 &  0.0202
        & -0.0471 &  0.0597 &  0.0039 &  0.0212 \\
      0.048 &  1.50 & 0.424 &  0.0171 &  0.0496 &  0.0043 &  0.0232
        &  0.0312 &  0.0661 &  0.0041 &  0.0246 \\
      0.065 &  1.86 & 0.457 &  0.1469 &  0.0504 &  0.0083 &  0.0245
        &  0.0097 &  0.0701 &  0.0051 &  0.0262 \\
      0.086 &  2.33 & 0.484 &  0.1220 &  0.0561 &  0.0079 &  0.0270
        & -0.0554 &  0.0811 &  0.0046 &  0.0290 \\
      0.118 &  3.08 & 0.489 &  0.0399 &  0.0534 &  0.0046 &  0.0239
        &  0.0292 &  0.0830 &  0.0029 &  0.0274 \\
      0.165 &  4.23 & 0.493 &  0.1436 &  0.0593 &  0.0104 &  0.0229
        &  0.0722 &  0.0993 &  0.0069 &  0.0279 \\
      0.238 &  5.81 & 0.503 &  0.1445 &  0.0773 &  0.0120 &  0.0235
        &  0.0871 &  0.1411 &  0.0067 &  0.0313 \\
      0.336 &  7.76 & 0.516 &  0.4389 &  0.1747 &  0.0270 &  0.0373
        & -0.2504 &  0.3422 &  0.0202 &  0.0621 \\
      0.448 & 10.20 & 0.510 &  0.4641 &  0.3692 &  0.0400 &  0.0536
        &  1.4585 &  0.7001 &  0.0859 &  0.1010 \\
    \hline
    \multicolumn{1}{c}{$\langle x \rangle$\rule[-1.5ex]{0mm}{4.5ex}} &
    \multicolumn{1}{c}{\hspace*{-2.5ex}$\langle Q^2 \rangle
    \,/\:\mathrm{GeV}^2$\hspace*{-2.5ex}} &
    \multicolumn{1}{c}{$\langle z \rangle$} &
    \multicolumn{1}{c}{$A_{1,d}^{K^+ + K^-}$} &
    \multicolumn{1}{c}{$\pm \mathrm{stat.}$} &
    \multicolumn{1}{c}{$\pm \mathrm{syst.}$} &
    \multicolumn{1}{c}{$\pm \mathrm{MC}$} & & & & \\ \hline
      0.033 &  1.22 & 0.372 & -0.0161 &  0.0372 &  0.0016 &  0.0148
       & & & & \\
      0.047 &  1.50 & 0.415 &  0.0257 &  0.0395 &  0.0036 &  0.0168
       & & & & \\
      0.064 &  1.87 & 0.446 &  0.1064 &  0.0408 &  0.0049 &  0.0182
       & & & & \\
      0.086 &  2.36 & 0.470 &  0.0709 &  0.0460 &  0.0042 &  0.0201
       & & & & \\
      0.118 &  3.11 & 0.475 &  0.0395 &  0.0447 &  0.0030 &  0.0180
       & & & & \\
      0.165 &  4.26 & 0.481 &  0.1226 &  0.0506 &  0.0095 &  0.0177
       & & & & \\
      0.238 &  5.87 & 0.489 &  0.1347 &  0.0671 &  0.0103 &  0.0189
       & & & & \\
      0.336 &  7.81 & 0.504 &  0.2945 &  0.1540 &  0.0183 &  0.0316
       & & & & \\
      0.447 & 10.26 & 0.499 &  0.6517 &  0.3254 &  0.0491 &  0.0467
       & & & & \\
  \end{tabular}
\end{ruledtabular}
\end{table*}

\begin{table*}
  \renewcommand{\arraystretch}{1.1}
  \caption{\label{tab:Dq_results} The quark polarizations
    $[\Delta q/q](x)$, and the quark helicity densities
    $x\cdot\Delta q(x,Q_0^2)$ evolved to $Q_0^2=2.5\,\GeV^2$.
    The systematic uncertainty due to the purities
%%    and parton distribution function
    ($[\pm \mathit{Pur}]$) is included in the total systematic uncertainty
    ($\pm\mathrm{sys}$).}
\begin{ruledtabular}
  \begin{tabular}{q{4}|q{5}q{5}q{5}q{5}q{5}|q{5}q{5}q{5}q{5}q{5}}
    \multicolumn{1}{c}{$\langle x \rangle$ \rule[-1.5ex]{0mm}{4.4ex}} &
    \multicolumn{1}{c}{$ \Delta{u}/u $} &
    \multicolumn{1}{c}{$\pm \mathrm{stat}$} &
    \multicolumn{1}{c}{$\pm \mathrm{sys}$} &
    \multicolumn{1}{c}{$[\pm \mathit{Pur}]$} &
    \multicolumn{1}{c}{$\pm \mathrm{MC}$} &
    \multicolumn{1}{c}{$ x\cdot\Delta{u} $} &
    \multicolumn{1}{c}{$\pm \mathrm{stat}$} &
    \multicolumn{1}{c}{$\pm \mathrm{sys}$} &
    \multicolumn{1}{c}{$[\pm \mathit{Pur}]$} &
    \multicolumn{1}{c}{$\pm \mathrm{MC}$} \\ \hline
    0.033 &  0.0855 &  0.1180 &  0.0856 &  0.0837 &  0.0129 
          &  0.0296 &  0.0408 &  0.0300 &  0.0290 &  0.0045 \\ 
    0.048 &  0.1368 &  0.1107 &  0.0217 &  0.0037 &  0.0077 
          &  0.0515 &  0.0417 &  0.0101 &  0.0014 &  0.0029 \\ 
    0.065 &  0.1913 &  0.0978 &  0.0356 &  0.0264 &  0.0051 
          &  0.0785 &  0.0401 &  0.0160 &  0.0108 &  0.0021 \\ 
    0.087 &  0.4864 &  0.0909 &  0.0784 &  0.0703 &  0.0057 
          &  0.2185 &  0.0408 &  0.0357 &  0.0316 &  0.0026 \\ 
    0.118 &  0.5086 &  0.0774 &  0.0633 &  0.0501 &  0.0057 
          &  0.2525 &  0.0385 &  0.0317 &  0.0249 &  0.0029 \\ 
    0.166 &  0.4731 &  0.0757 &  0.0436 &  0.0267 &  0.0045 
          &  0.2623 &  0.0420 &  0.0244 &  0.0148 &  0.0025 \\ 
    0.239 &  0.4445 &  0.0855 &  0.0364 &  0.0109 &  0.0028 
          &  0.2652 &  0.0510 &  0.0217 &  0.0065 &  0.0017 \\ 
    0.339 &  0.5805 &  0.0650 &  0.0558 &  0.0050 &  0.0091 
          &  0.3241 &  0.0363 &  0.0313 &  0.0028 &  0.0051 \\ 
    0.447 &  0.7272 &  0.1087 &  0.0684 &  0.0139 &  0.0084 
          &  0.3121 &  0.0467 &  0.0294 &  0.0060 &  0.0036 \\ 
    \hline
    \multicolumn{1}{c}{$\langle x \rangle$ \rule[-1.5ex]{0mm}{4.4ex}} &
    \multicolumn{1}{c}{$ \Delta{d}/d $} &
    \multicolumn{1}{c}{$\pm \mathrm{stat}$} &
    \multicolumn{1}{c}{$\pm \mathrm{sys}$} &
    \multicolumn{1}{c}{$[\pm \mathit{Pur}]$} &
    \multicolumn{1}{c}{$\pm \mathrm{MC}$} &
    \multicolumn{1}{c}{$ x\cdot\Delta{d} $} &
    \multicolumn{1}{c}{$\pm \mathrm{stat}$} &
    \multicolumn{1}{c}{$\pm \mathrm{sys}$} &
    \multicolumn{1}{c}{$[\pm \mathit{Pur}]$} &
    \multicolumn{1}{c}{$\pm \mathrm{MC}$} \\ \hline
    0.033 & -0.1236 &  0.1529 &  0.0390 &  0.0199 &  0.0153 
          & -0.0337 &  0.0416 &  0.0128 &  0.0054 &  0.0042 \\ 
    0.048 &  0.0588 &  0.1452 &  0.0543 &  0.0435 &  0.0115 
          &  0.0167 &  0.0411 &  0.0166 &  0.0123 &  0.0032 \\ 
    0.065 & -0.1336 &  0.1307 &  0.0362 &  0.0189 &  0.0122 
          & -0.0394 &  0.0385 &  0.0120 &  0.0056 &  0.0036 \\ 
    0.087 & -0.2572 &  0.1303 &  0.0495 &  0.0165 &  0.0190 
          & -0.0789 &  0.0400 &  0.0159 &  0.0051 &  0.0058 \\ 
    0.118 & -0.4876 &  0.1185 &  0.0841 &  0.0592 &  0.0210 
          & -0.1552 &  0.0377 &  0.0271 &  0.0188 &  0.0067 \\ 
    0.166 & -0.0918 &  0.1337 &  0.0675 &  0.0091 &  0.0169 
          & -0.0296 &  0.0431 &  0.0219 &  0.0029 &  0.0055 \\ 
    0.239 & -0.5218 &  0.1646 &  0.0822 &  0.0015 &  0.0125 
          & -0.1536 &  0.0484 &  0.0242 &  0.0004 &  0.0037 \\ 
    0.339 & -0.2799 &  0.1988 &  0.1694 &  0.0166 &  0.0400 
          & -0.0628 &  0.0446 &  0.0382 &  0.0037 &  0.0090 \\ 
    0.447 & -0.8133 &  0.4074 &  0.2454 &  0.0564 &  0.0404 
          & -0.1158 &  0.0580 &  0.0349 &  0.0080 &  0.0057 \\ 
    \hline
    \multicolumn{1}{c}{$\langle x \rangle$ \rule[-1.5ex]{0mm}{4.4ex}} &
    \multicolumn{1}{c}{$ \Delta{\bar{u}}/\bar{u} $} &
    \multicolumn{1}{c}{$\pm \mathrm{stat}$} &
    \multicolumn{1}{c}{$\pm \mathrm{sys}$} &
    \multicolumn{1}{c}{$[\pm \mathit{Pur}]$} &
    \multicolumn{1}{c}{$\pm \mathrm{MC}$} &
    \multicolumn{1}{c}{$ x\cdot\Delta{\bar{u}} $} &
    \multicolumn{1}{c}{$\pm \mathrm{stat}$} &
    \multicolumn{1}{c}{$\pm \mathrm{sys}$} &
    \multicolumn{1}{c}{$[\pm \mathit{Pur}]$} &
    \multicolumn{1}{c}{$\pm \mathrm{MC}$} \\ \hline
    0.033 &  0.3382 &  0.3342 &  0.2189 &  0.2169 &  0.0520 
          &  0.0437 &  0.0432 &  0.0283 &  0.0280 &  0.0067 \\ 
    0.048 &  0.2484 &  0.3677 &  0.0471 &  0.0074 &  0.0448 
          &  0.0288 &  0.0426 &  0.0058 &  0.0009 &  0.0052 \\ 
    0.065 &  0.0166 &  0.3938 &  0.1200 &  0.1132 &  0.0512 
          &  0.0017 &  0.0403 &  0.0126 &  0.0116 &  0.0052 \\ 
    0.087 & -0.7151 &  0.4585 &  0.3508 &  0.3468 &  0.0546 
          & -0.0624 &  0.0400 &  0.0307 &  0.0302 &  0.0048 \\ 
    0.118 & -0.8989 &  0.5391 &  0.3395 &  0.3324 &  0.0626 
          & -0.0621 &  0.0372 &  0.0235 &  0.0229 &  0.0043 \\ 
    0.166 & -0.9022 &  0.8403 &  0.3491 &  0.3307 &  0.0815 
          & -0.0432 &  0.0402 &  0.0168 &  0.0158 &  0.0039 \\ 
    0.239 &  1.4742 &  1.6410 &  0.3868 &  0.2844 &  0.1302 
          &  0.0446 &  0.0496 &  0.0117 &  0.0086 &  0.0039 \\ 
    \hline
    \multicolumn{1}{c}{$\langle x \rangle$ \rule[-1.5ex]{0mm}{4.4ex}} &
    \multicolumn{1}{c}{$ \Delta{\bar{d}}/\bar{d} $} &
    \multicolumn{1}{c}{$\pm \mathrm{stat}$} &
    \multicolumn{1}{c}{$\pm \mathrm{sys}$} &
    \multicolumn{1}{c}{$[\pm \mathit{Pur}]$} &
    \multicolumn{1}{c}{$\pm \mathrm{MC}$} &
    \multicolumn{1}{c}{$ x\cdot\Delta{\bar{d}} $} &
    \multicolumn{1}{c}{$\pm \mathrm{stat}$} &
    \multicolumn{1}{c}{$\pm \mathrm{sys}$} &
    \multicolumn{1}{c}{$[\pm \mathit{Pur}]$} &
    \multicolumn{1}{c}{$\pm \mathrm{MC}$} \\ \hline
    0.033 & -0.2281 &  0.2819 &  0.0380 &  0.0198 &  0.0200 
          & -0.0338 &  0.0417 &  0.0067 &  0.0029 &  0.0030 \\ 
    0.048 & -0.6238 &  0.2921 &  0.1076 &  0.1033 &  0.0387 
          & -0.0862 &  0.0404 &  0.0151 &  0.0143 &  0.0053 \\ 
    0.065 &  0.0174 &  0.2847 &  0.0513 &  0.0457 &  0.0446 
          &  0.0022 &  0.0367 &  0.0068 &  0.0059 &  0.0058 \\ 
    0.087 & -0.2239 &  0.3103 &  0.0605 &  0.0550 &  0.0482 
          & -0.0267 &  0.0370 &  0.0073 &  0.0065 &  0.0057 \\ 
    0.118 &  0.5412 &  0.3144 &  0.1621 &  0.1588 &  0.0449 
          &  0.0577 &  0.0335 &  0.0173 &  0.0169 &  0.0048 \\ 
    0.166 & -0.9546 &  0.4561 &  0.0734 &  0.0545 &  0.0590 
          & -0.0828 &  0.0396 &  0.0064 &  0.0047 &  0.0051 \\ 
    0.239 &  0.4523 &  0.8380 &  0.1470 &  0.0323 &  0.1075 
          &  0.0237 &  0.0439 &  0.0077 &  0.0017 &  0.0056 \\ 
    \hline
    \multicolumn{1}{c}{$\langle x \rangle$ \rule[-1.5ex]{0mm}{4.4ex}} &
    \multicolumn{1}{c}{$ \Delta{s}/s $} &
    \multicolumn{1}{c}{$\pm \mathrm{stat}$} &
    \multicolumn{1}{c}{$\pm \mathrm{sys}$} &
    \multicolumn{1}{c}{$[\pm \mathit{Pur}]$} &
    \multicolumn{1}{c}{$\pm \mathrm{MC}$} &
    \multicolumn{1}{c}{$ x\cdot\Delta{s} $} &
    \multicolumn{1}{c}{$\pm \mathrm{stat}$} &
    \multicolumn{1}{c}{$\pm \mathrm{sys}$} &
    \multicolumn{1}{c}{$[\pm \mathit{Pur}]$} &
    \multicolumn{1}{c}{$\pm \mathrm{MC}$} \\ \hline
    0.033 &  0.4734 &  0.7492 &  0.1871 &  0.1659 &  0.0596 
          &  0.0317 &  0.0502 &  0.0134 &  0.0111 &  0.0040 \\ 
    0.048 &  0.6071 &  0.6361 &  0.1952 &  0.1908 &  0.0462 
          &  0.0365 &  0.0382 &  0.0118 &  0.0115 &  0.0028 \\ 
    0.065 & -0.0537 &  0.5805 &  0.0441 &  0.0329 &  0.0550 
          & -0.0029 &  0.0312 &  0.0024 &  0.0018 &  0.0030 \\ 
    0.087 & -0.1243 &  0.6248 &  0.1422 &  0.1366 &  0.0661 
          & -0.0059 &  0.0295 &  0.0067 &  0.0065 &  0.0031 \\ 
    0.118 & -0.3359 &  0.6516 &  0.0756 &  0.0689 &  0.0772 
          & -0.0133 &  0.0258 &  0.0030 &  0.0027 &  0.0031 \\ 
    0.166 &  1.3956 &  0.8851 &  0.0912 &  0.0252 &  0.0780 
          &  0.0418 &  0.0265 &  0.0027 &  0.0008 &  0.0023 \\ 
    0.239 & -1.2674 &  1.5039 &  0.3911 &  0.3781 &  0.1220 
          & -0.0230 &  0.0273 &  0.0071 &  0.0069 &  0.0022 \\ 
  \end{tabular}
\end{ruledtabular}
%%
%%  The source file ``deltaq.tex'' for this table is generated by running
%%  the following tool:
%%  ~/dqPaper/bin/deltaq2tex.pl -I 20,10,21,11,30 \
%%    -P ~/dqPaper/Data/Dq/FINAL/X6P_deltaqs.sys.all.out \
%%    -X ~/dqPaper/Data/Dq/FINAL/X6P_dqs_xDens.sys.all.out \
%%    ~/dqPaper/Latex/deltaq.tex
%%
\end{table*}

\begin{table*}
  \renewcommand{\arraystretch}{1.1}
  \caption{\label{tab:ubdb_results} The light sea flavor helicity
    asymmetry evolved to $Q_0^2=2.5\,\GeV^2$.  The systematic uncertainty
    due to the purities ($[\pm \mathit{Pur}]$) is included in the total
    systematic uncertainty ($\pm\mathrm{sys}$).}
  \begin{minipage}{0.6\textwidth}
\begin{ruledtabular}
  \begin{tabular}{q{4}|q{5}q{5}q{5}q{5}q{5}}
    \multicolumn{1}{c}{$\langle x \rangle$ \rule[-1.5ex]{0mm}{4.4ex}} &
    \multicolumn{1}{c}{$ x\cdot(\Delta{\bar{u}}-\Delta{\bar{d}}) $} &
    \multicolumn{1}{c}{$\pm \mathrm{stat}$} &
    \multicolumn{1}{c}{$\pm \mathrm{sys}$} &
    \multicolumn{1}{c}{$[\pm \mathit{Pur}]$} &
    \multicolumn{1}{c}{$\pm \mathrm{MC}$} \\ \hline
    0.033 &  0.0748  &  0.0653 &  0.0252 &  0.0242 &  0.0093\\
    0.048 &  0.1133  &  0.0670 &  0.0153 &  0.0132 &  0.0102\\
    0.065 & -0.0005  &  0.0639 &  0.0180 &  0.0174 &  0.0106\\
    0.087 & -0.0355  &  0.0655 &  0.0370 &  0.0367 &  0.0100\\
    0.118 & -0.1198  &  0.0606 &  0.0404 &  0.0399 &  0.0088\\
    0.166 &  0.0397  &  0.0709 &  0.0221 &  0.0211 &  0.0088\\
    0.239 &  0.0241  &  0.0885 &  0.0178 &  0.0077 &  0.0097\\
  \end{tabular}
\end{ruledtabular}
%%% JW: see and awk line in dqpaper.tex to generate the entries in
%%%    this table
  \end{minipage}
%%% a little awk script to generate the table: awk '{if ($1==60) printf"%5.3f & %7.4f  & %7.4f & %7.4f & %7.4f & %7.4f\\\\\n",$7,$13,$14,$15,$16,$17}' ~/dqPaper/Data/Dq/FINAL/X6X_dqs_xDens.sys.all.out 
\end{table*}

% --------------------------------------------------------------------------
%   The statistical and systematic correlations of the first moments
% --------------------------------------------------------------------------

\begin{table*}
  \renewcommand{\arraystretch}{1.1}
  \caption{\label{tab:first_mom_correlations} The statistical
    ($\rho_{\mathrm{stat}}$) and systematic ($\rho_{\mathrm{syst}}$)
    correlations of the first moments of the extracted helicity
    distributions in the measured $x$--range.}
  \vspace*{1ex}
  \begin{minipage}[t]{\linewidth}
    \hspace*{0.05\linewidth}
    \begin{minipage}[t]{0.4\linewidth}
      \begin{ruledtabular}
        \begin{tabular}{c||r|r|r|r|r}
          $\rho_{\mathrm{stat}}$ \rule[-1.5ex]{0mm}{3.8ex}
          & \multicolumn{1}{c|}{$\Delta u$}
          & \multicolumn{1}{c|}{$\Delta d$}
          & \multicolumn{1}{c|}{$\Delta \bar{u}$}
          & \multicolumn{1}{c|}{$\Delta \bar{d}$}
          & \multicolumn{1}{c}{$\Delta s$} \\ \hline \hline
          $\Delta u$ & %
           1.000 & -0.564 & -0.827 &  0.346 &  0.075 \\ \hline
          $\Delta d$ & %
          -0.564 &  1.000 &  0.301 & -0.758 &  0.032 \\ \hline
          $\Delta \bar{u}$ & %
          -0.827 &  0.301 &  1.000 & -0.385 & -0.257 \\ \hline
          $\Delta \bar{d}$ & %
           0.346 & -0.758 & -0.385 &  1.000 & -0.196 \\ \hline
          $\Delta s$ & %
           0.075 &  0.032 & -0.257 & -0.196 &  1.000 \\
        \end{tabular}
      \end{ruledtabular}
    \end{minipage}
    \hfill
    \begin{minipage}[t]{0.4\linewidth}
      \begin{ruledtabular}
        \begin{tabular}{c||r|r|r|r|r}
          $\rho_{\mathrm{syst}}$ \rule[-1.5ex]{0mm}{3.8ex}
          & \multicolumn{1}{c|}{$\Delta u$}
          & \multicolumn{1}{c|}{$\Delta d$}
          & \multicolumn{1}{c|}{$\Delta \bar{u}$}
          & \multicolumn{1}{c|}{$\Delta \bar{d}$}
          & \multicolumn{1}{c}{$\Delta s$} \\ \hline \hline
          $\Delta u$ &  1.000 & -0.732  &  0.577 & -0.176 &  0.388\\ \hline
          $\Delta d$ & -0.732 &  1.000  & -0.170 &  0.548 &  0.190\\ \hline
    $\Delta \bar{u}$ &  0.577 & -0.170  &  1.000 &  0.228 &  0.524\\ \hline
    $\Delta \bar{d}$ & -0.176 &  0.548  &  0.228 &  1.000 &  0.450\\ \hline
          $\Delta s$ &  0.388 &  0.190  &  0.524 &  0.450 &  1.000\\
        \end{tabular}
      \end{ruledtabular}
    \end{minipage}
    \hspace*{0.05\linewidth}
    \phantom{.}
  \end{minipage}
\end{table*}

% --------------------------------------------------------------------------
%   The statistical and systematic correlations of the second moments
% --------------------------------------------------------------------------

\begin{table*}
  \renewcommand{\arraystretch}{1.1}
  \caption{\label{tab:second_mom_correlations} The statistical
    ($\rho_{\mathrm{stat}}$) and systematic ($\rho_{\mathrm{syst}}$)
    correlations of the second moments of the extracted helicity
    distributions in the measured $x$--range.}
  \vspace*{1ex}
  \begin{minipage}[t]{\linewidth}
    \hspace*{0.05\linewidth}
    \begin{minipage}[t]{0.4\linewidth}
      \begin{ruledtabular}
        \begin{tabular}{c||r|r|r|r|r}
          $\rho_{\mathrm{stat}}$ \rule[-1.5ex]{0mm}{3.8ex}
          & \multicolumn{1}{c|}{$\Delta^{(2)} u$}
          & \multicolumn{1}{c|}{$\Delta^{(2)} d$}
          & \multicolumn{1}{c|}{$\Delta^{(2)} \bar{u}$}
          & \multicolumn{1}{c|}{$\Delta^{(2)} \bar{d}$}
          & \multicolumn{1}{c}{$\Delta^{(2)} s$} \\ \hline \hline
          $\Delta^{(2)} u$ & %
           1.000 & -0.803 & -0.522 &  0.277 &  0.048 \\ \hline
          $\Delta^{(2)} d$ & %
          -0.803 &  1.000 &  0.207 & -0.413 &  0.036 \\ \hline
          $\Delta^{(2)} \bar{u}$ & %
          -0.522 &  0.207 &  1.000 & -0.508 & -0.213 \\ \hline
          $\Delta^{(2)} \bar{d}$ & %
           0.277 & -0.413 & -0.508 &  1.000 & -0.197 \\ \hline
          $\Delta^{(2)} s$ & %
           0.048 &  0.036 & -0.213 & -0.197 &  1.000 \\
        \end{tabular}
      \end{ruledtabular}
    \end{minipage}
    \hfill
    \begin{minipage}[t]{0.4\linewidth}
      \begin{ruledtabular}
        \begin{tabular}{c||r|r|r|r|r}
          $\rho_{\mathrm{syst}}$ \rule[-1.5ex]{0mm}{3.8ex}
          & \multicolumn{1}{c|}{$\Delta^{(2)} u$}
          & \multicolumn{1}{c|}{$\Delta^{(2)} d$}
          & \multicolumn{1}{c|}{$\Delta^{(2)} \bar{u}$}
          & \multicolumn{1}{c|}{$\Delta^{(2)} \bar{d}$}
          & \multicolumn{1}{c}{$\Delta^{(2)} s$} \\ \hline \hline
      $\Delta^{(2)} u$ &  1.000 & -0.847 &  0.430 & -0.399 &  0.149 \\ \hline
      $\Delta^{(2)} d$ & -0.847 &  1.000 & -0.372 &  0.522 &  0.134 \\ \hline
$\Delta^{(2)} \bar{u}$ &  0.430 & -0.372 &  1.000 &  0.097 &  0.325 \\ \hline
$\Delta^{(2)} \bar{d}$ & -0.399 &  0.522 &  0.097 &  1.000 &  0.354 \\ \hline
      $\Delta^{(2)} s$ &  0.149 &  0.134 &  0.325 &  0.354 &  1.000 \\

        \end{tabular}
      \end{ruledtabular}
    \end{minipage}
    \hspace*{0.05\linewidth}
    \phantom{.}
  \end{minipage}
\end{table*}

% --------------------------------------------------------------------------
%   Include the bibliography, created via BibTeX
% --------------------------------------------------------------------------

\bibliography{dqpaper}   %  Produces the bibliography via BibTeX.

\end{document}

%% file: moments.tex
%% ----------------------------------------------------------------------------
%%  Table for moments of polarized quark distributions (LaTeX input)
%%  To be included into the source file "sect-VIII.tex" of the long
%%    delta-q paper (table \label{tab:moments})
%%
%%  This file was generated on Sun Nov 23 16:50:19 2003
%%  Modified manually on Sun Nov 23 18:36:29 MET 2003 to eliminate the
%%  extrapolations outside the measured range and the total integrals
%% ----------------------------------------------------------------------------
  \begin{ruledtabular}
    \begin{tabular}{lr} % \hline
      & \multicolumn{1}{c}{Moments in measured range} \\ \hline
      $\Delta u$ \rule{0mm}{3ex} & $ 0.601 \pm 0.039 \pm 0.049$ \\
      $\Delta \bar{u}$           & $-0.002 \pm 0.036 \pm 0.023$ \\
      $\Delta d$                 & $-0.226 \pm 0.039 \pm 0.050$ \\
      $\Delta \bar{d}$           & $-0.054 \pm 0.033 \pm 0.011$ \\
      $\Delta s$                 & $ 0.028 \pm 0.033 \pm 0.009$ \\
      \hline
      $\Delta u + \Delta\bar{u}$ & $ 0.599 \pm 0.022 \pm 0.065$ \\
      $\Delta d + \Delta\bar{d}$ & $-0.280 \pm 0.026 \pm 0.057$ \\
      \hline
      $\Delta u_{\mathrm{v}}$    & $ 0.603 \pm 0.071 \pm 0.040$ \\
      $\Delta d_{\mathrm{v}}$    & $-0.172 \pm 0.068 \pm 0.045$ \\
      \hline
      $\Delta\bar{u} - \Delta\bar{d}$ \rule{0mm}{2.5ex} &
      $ 0.048 \pm 0.057 \pm 0.028$ \\
      \hline
      $\Delta \Sigma$            & $ 0.347 \pm 0.024 \pm 0.066$ \\
      $\Delta q_3$               & $ 0.880 \pm 0.045 \pm 0.107$ \\
      $\Delta q_8$               & $ 0.262 \pm 0.078 \pm 0.045$ \\
      \hline
      $\Delta^{(2)} u$           & $ 0.142 \pm 0.009 \pm 0.011$ \\
      $\Delta^{(2)} \bar{u}$     & $-0.001 \pm 0.005 \pm 0.002$ \\
      $\Delta^{(2)} d$           & $-0.049 \pm 0.010 \pm 0.013$ \\
      $\Delta^{(2)} \bar{d}$     & $-0.003 \pm 0.004 \pm 0.001$ \\
      $\Delta^{(2)} s$           & $ 0.001 \pm 0.003 \pm 0.001$ \\
      \hline
      $\Delta^{(2)} u_{\mathrm{v}}$ & $ 0.144 \pm 0.013 \pm 0.011$ \\
      $\Delta^{(2)} d_{\mathrm{v}}$ & $-0.047 \pm 0.012 \pm 0.012$ \\
    \end{tabular}
  \end{ruledtabular}
%% ----------------------------------------------------------------------------
%%  End of file
%% ----------------------------------------------------------------------------